\documentclass[a4paper,11pt]{article}
\pdfoutput=1 

\usepackage{jheppub} 

\usepackage[T1]{fontenc} 
\usepackage{fontawesome}
\usepackage{pdflscape}
\usepackage[dvipsnames,table]{xcolor}
\usepackage[caption=false]{subfig}
\usepackage{epsfig}
\usepackage{makecell}
\usepackage{colortbl}
\usepackage[T1]{fontenc} 
\usepackage{hepnames}
\usepackage{adjustbox}
\usepackage{rotating}
\usepackage{multirow}
\usepackage{booktabs,siunitx,makecell}
\usepackage{bbold}
\usepackage{slashed}
\usepackage{bigints}
\usepackage[normalem]{ulem}
\usepackage{amsmath}
\usepackage{textcomp}
\usepackage{float}
\usepackage{subfig}
\usepackage{graphicx}
\usepackage{orcidlink}
\usepackage[table]{xcolor} 
\usepackage[toc,page]{appendix}
\usepackage[compat=1.1.0]{tikz-feynman} 
\usepackage{silence}
\tikzfeynmanset{warn luatex=false}
\newcommand{\stkout}[1]{\ifmmode\text{\sout{\ensuremath{#1}}}\else\sout{#1}\fi}
\definecolor{calpolypomonagreen}{rgb}{0.12, 0.3, 0.17}
\newcommand{\bea}{\begin{eqnarray}}
\newcommand{\eea}{\end{eqnarray}}
\newcommand{\beq}{\begin{equation}}
\newcommand{\eeq}{\end{equation}}
\usepackage{array} 
\usepackage[most]{tcolorbox}
\usepackage{xcolor}

\tcbset{highlight math style={
  colback=gray!20, colframe=gray!50!black,
  boxrule=0.5pt, sharp corners}}
\title{\textcolor{black}{A Study on Top Quark FCNC Interactions in SMEFT Framework}}
\author[a]{Subhajit Kala}
\author[a]{Soumitra Nandi}
\affiliation{Department of Physics, Indian Institute of Technology Guwahati, North Guwahati, Assam-781039, India.}

\emailAdd{s.kala@iitg.ac.in}
\emailAdd{soumitra.nandi@iitg.ac.in}

\abstract{We present a model-independent study of rare flavour-changing neutral current (FCNC) interactions of the top quark within the Standard Model Effective Field Theory (SMEFT). Matching a general top-FCNC parametrisation onto the SMEFT basis, we perform a global fit including low-energy flavour observables, electroweak precision data, Higgs measurements, collider limits on top-FCNC decays, and electric dipole moment constraints. Allowing for complex dipole operators, we derive stringent bounds on the real and imaginary parts of the top-FCNC couplings and, independently, obtain robust constraints on the corresponding SMEFT Wilson coefficients. We further provide predictions for rare top-FCNC branching ratios and CP asymmetries, and identify benchmark scenarios illustrating the complementary role of CP-violating observables in probing top-quark flavour dynamics.}


\DeclareUnicodeCharacter{202F}{\,}
\begin{document}
\maketitle
\flushbottom
\section{Introduction}\label{sec:Introduction}

The top quark, the heaviest particle in the Standard Model (SM), plays a crucial role in probing the dynamics of electroweak symmetry breaking and potential physics beyond the SM. Its large mass, close to the electroweak scale, suggests that the top quark may serve as a sensitive probe of new physics interactions. The top quark, being a cornerstone of particle physics, has decay and production modes that have been extensively studied at the Large Hadron Collider (LHC). Despite extensive studies, many of its characteristic properties remain unexplored, particularly in FCNC processes, such as $t\to c (u)$ decays. In the SM, such processes are forbidden at tree level and highly suppressed at higher orders due to the Glashow–Iliopoulos–Maiani (GIM) mechanism, leading to branching fractions of the order of $10^{-12}-10^{-14}$ \cite{TopQuarkWorkingGroup:2013hxj} for top decays such as $t \to u_j\,X$ with $X \in \{g, \gamma, Z, H\}$ and $u_j = u, c$, several orders of magnitude below the current experimental limits \cite{ATLAS:2024kxj}. Owing to the strong suppression of FCNCs in the SM, rare top-quark FCNC interactions provide a powerful probe of physics beyond the Standard Model. Consequently, any observable signal of top FCNC interactions at present or future colliders would provide an unambiguous indication of new physics. If the new physics (NP) scale is too high for direct production, its effects can still appear via virtual contributions, which can be systematically described within the effective field theory (EFT) framework. 

In this context, top-quark interactions have been widely studied within EFT approaches. Ref.~\cite{Bissmann:2019gfc} constrains flavour-conserving top dipole operators using flavour observables, particularly radiative $B$ decays, together with top-quark data. Ref.~\cite{Allwicher:2023shc} presents a comprehensive analysis of $\mathrm{U}(2)^5$-symmetric operators, treating the third generation as a singlet, and derives constraints from flavour, electroweak, and collider observables while consistently incorporating renormalisation group evolution (RGE). Similarly, refs.~\cite{Garosi:2023yxg,Bissmann:2020mfi} perform global analyses of flavour-conserving top operators using flavour, electroweak, and Higgs data within an RGE framework. More recently, ref.~\cite{Kala:2025srq} studies top charged-current interactions through a global fit combining flavour, electroweak, Higgs, and top observables, again connecting sectors via RGE. These works share a common strategy of operator matching, RGE evolution, and global fits.

In addition, several ultraviolet (UV) complete models, including two-Higgs-doublet models (2HDMs), supersymmetric scenarios, the Alternative Left--Right model, and frameworks with vector-like quarks or extended gauge sectors, predict significant enhancements of top-FCNC decays \cite{Cao:2007dk,Guasch:1999jp,Frank:2023fkc,Cai:2022xha,Fuyuto:2015gmk,Cho:2019stk,Hung:2017tts}. Simplified models \cite{Jueid:2024cge,Bhattacharya:2025mlg,Liu:2021crr,Chen:2022dzc,Chen:2023eof} have also been developed to parametrise such effects in a model-independent way, potentially bringing the corresponding branching ratios within experimental reach.

From an experimental perspective, searches for rare top decays and anomalous single-top production and top pair production modes have been carried out extensively at the LHC by the ATLAS~\cite{ATLAS:2023qzr, ATLAS:2022per, ATLAS:2024mih, ATLAS:2021amo} and CMS~\cite{CMS:2023bjm, CMS:2017wcz} collaborations, placing stringent bounds on a wide range of top FCNC channels. These searches are expected to improve significantly with the increased luminosity of the High-Luminosity LHC (HL-LHC). A recent study highlighting 
this prospective sensitivity can be found in ref.~\cite{Fuks:2025qgh}, 
indicating that future measurements will provide even stronger probes of 
rare top-quark interactions.

This work investigates how a wide range of observables, including low-energy flavour-changing charged-current (FCCC) and flavour-changing neutral-current (FCNC) processes, electroweak precision observables (EWPOs), collider constraints on top-FCNC decay rates, complementary high-energy measurements, and limits from electric dipole moment (EDM) experiments, constrain the effective couplings governing rare top-quark decays. We further derive bounds on the Wilson coefficients (WCs) of selected dimension-six SMEFT operators, focusing on dipole- and vector-type interactions that induce $t\to u_j X$ transitions at tree level. Previous studies have primarily concentrated on indirect constraints on four-fermion operators relevant to $t\to c\ell\ell$ transitions \cite{Malekhosseini:2018fgp,Aguilar-Saavedra:2008nuh,Bhattacharya:2023beo,Altmannshofer:2023bfk}, sometimes supplemented by low-energy FCNC observables such as $b\to s\ell^+\ell^-$ decays \cite{Bhattacharya:2023beo,Altmannshofer:2023bfk}. In contrast, by systematically combining information across multiple energy scales and explicitly allowing for complex SMEFT WCs, which are essential for probing CP-violating new-physics effects, we obtain significantly stronger and more comprehensive constraints on this class of operators than currently available in the literature. This multi-scale, multi-observable framework not only enhances sensitivity to rare top-quark flavour dynamics but also provides a more complete exploration of the underlying parameter space.

The paper is organised as follows. In section~\ref{sec:Effective_framework}, we introduce the effective couplings for top-FCNC interactions and the corresponding SMEFT WCs. Section~\ref{sec:methodology} outlines the analysis framework. In section~\ref{sec:List_of_obs}, we list all observables considered in this analysis. The results, including different fitting scenarios and the impact of direct collider bounds, are presented in section~\ref{sec:Result}. Finally, section~\ref{sec:summary} summarises our findings.
\section{Effective Framework}\label{sec:Effective_framework}
In the absence of direct evidence for new particles at current collider energies, an EFT approach provides a robust and model-independent framework to study potential NP effects. Our analysis is based on an effective Lagrangian that systematically captures possible new physics contributions to top FCNC vertices, like  $t \to u_j\, g$, $t \to u_j\,\gamma$, $t \to u_j\,Z$, and $t \to u_j\,H$.
The most general effective Lagrangian encompassing all these processes can be written as:
\begin{align}\label{eq:eff_vertex_top_FCNC}
    -\mathcal{L}_{\rm eff}=&\sum_{u_j=u,c}\bar{u}_j\Bigg[\frac{g_s}{2 m_t}T^A\sigma^{\mu\nu}\left(\xi_L^{u_jt}P_L+\xi_R^{u_jt}P_R\right)G^A_{\mu\nu}+\frac{e}{2 m_t}\sigma^{\mu\nu}\left(\lambda^{u_jt}_L P_L+\lambda_R^{u_jt} P_R\right)F_{\mu\nu}\nonumber\\
    &+\frac{g_W}{2c_W m_t}\sigma^{\mu\nu}\left(\kappa_L^{u_jt} P_L +\kappa_R^{u_jt}P_R\right)Z_{\mu\nu}-\frac{g_W}{2 c_W} \gamma^{\mu}\left(X^{u_jt}_LP_L + X_R^{u_j\,t}P_R\right)Z_{\mu}\\
   & -\frac{1}{\sqrt{2}}\left(\eta_L^{u_jt} P_L+\eta_R^{u_jt}P_R\right)H\bigg]t+\text{H.C.}\nonumber
\end{align}
Here $\sigma^{\mu\nu} = \frac{i}{2}[\gamma^{\mu}, \gamma^{\nu}]$, and $G_{\mu\nu}$, $F_{\mu\nu}$, and $Z_{\mu\nu}$ denote the field strength tensors of the gluon, photon, and $Z$ boson, respectively and $P_{L(R)}=(1\mp \gamma^5)/2$ corresponds to the chirality projection operators. Since FCNC processes are forbidden at tree level in the SM, all the couplings in eq.~\eqref{eq:eff_vertex_top_FCNC} vanish at leading order. However, such couplings can be generated at the one-loop level within the SM, or they may arise at tree level or one-loop in various beyond-the-Standard-Model (BSM) scenarios~\cite{Atwood:1996vj, Cakir:2010rs}. In this analysis, our objective is to constrain the strengths of the couplings associated with the operators introduced in the effective Lagrangian, considering a wide variety of observables to which these operators may contribute. Furthermore, we take this opportunity to place bounds on the WCs of the SMEFT operators that can also induce these FCNC interactions through tree-level matching, thereby effectively capturing the flavour structure of potential new physics effects. In the following paragraph, we discuss the SMEFT correspondence of the top-FCNC operators defined in the above Lagrangian.

\paragraph{\underline{\textbf{SMEFT Correspondence:}}}
The SMEFT framework is constructed by extending the SM Lagrangian with higher-dimensional operators $(d \geq 4)$, built solely from SM fields and invariant under the full SM gauge symmetries. The general SMEFT Lagrangian takes the form:
\begin{align}\label{eq:SMEFT_Lag}
    \mathcal{L}_{\rm SMEFT}=\mathcal{L}^{(4)}_{\rm SM}+\sum_{d>4}\frac{c_i^{(d)}}{\Lambda^{d-4}}\mathcal{O}_i^{(d)}
\end{align}
where $\mathcal{O}_i^{(d)}$ are the $d$-dimensional operators constructed from SM fields, and $c_i^{(d)}$ are the corresponding dimensionless WCs. It is worth mentioning that for each $i$-th operator, there will be another summation over the generation indices $ p$  and $ r$, so the dimension-six SMEFT Lagrangian can be rewritten as:
\begin{align}
    \mathcal{L}^{i\,(6)}_{\rm SMEFT}=\sum_{p,r}\left(\frac{c^i_{pr}}{\Lambda^2}\mathcal{O}^i_{pr}+\mathrm{h.c.}\right)
\end{align}
\indent{}We choose a few dimension-six top quark flavour-violating SMEFT operators in the representation of the Warsaw basis \cite{Grzadkowski:2010es} to match the effective couplings in eq.~\eqref{eq:eff_vertex_top_FCNC} at tree level. We have listed these operators in table~\ref{tab:SMEFT_Ops}. In these operators, one of the up-type quarks is from the third generation (top quark), while the other is taken from the first or second generation only. We assume that NP arises solely from this subset of SMEFT operators, with all other operators set to zero. For the rest of our analysis, we redefine the WCs as $\mathcal{C}_i \equiv c_i / \Lambda^2$, where $\Lambda$ is the NP scale. These rescaled coefficients have mass dimension $[\mathrm{GeV}^{-2}]$.
\begin{table}[t]
    \centering
    \renewcommand{\arraystretch}{1.5}
    \begin{tabular}{|c|c|c|c|}
        \hline
        \rowcolor{gray!20}
        \multicolumn{2}{|c|}{\textbf{Top dipole}} & \multicolumn{2}{c|}{\textbf{Top Higgs}} \\
        \hline
        $\mathcal{O}^{uG}_{pr}$ & $(\bar{q}_p \sigma^{\mu\nu}T^A u_r)\tilde{\phi}G_{\mu\nu}^A$ & $\mathcal{O}^{\phi q \,(1)}_{pr}$ & $(\phi^{\dagger}i \overleftrightarrow D_{\mu} \phi )(\bar{q}_p\gamma^{\mu} q_r)$  \\
        $\mathcal{O}^{uB}_{pr}$ & $(\bar{q}_p \sigma^{\mu\nu} u_r)\tilde{\phi}B_{\mu\nu}$ & $\mathcal{O}^{\phi q \,(3)}_{pr}$ & $(\phi^{\dagger}i \overleftrightarrow D_{\mu}^I \phi )(\bar{q}_p\tau^I\gamma^{\mu} q_r)$ \\
        $\mathcal{O}^{uW}_{pr}$ & $(\bar{q}_p \sigma^{\mu\nu} u_r)\tau^I\tilde{\phi}W_{\mu\nu}^I$ & $\mathcal{O}^{\phi u}_{pr}$ & $(\phi^{\dagger}i \overleftrightarrow D_{\mu} \phi )(\bar{u}_p\gamma^{\mu} u_r)$  \\
        & & $\mathcal{O}^{u\phi}_{pr}$ & $(\phi^{\dagger}\phi)(\bar{q}_p u_r \tilde{\phi})$ \\
        \hline
        
        \hline
    \end{tabular}
    \caption{List of dimension-six SMEFT operators that contribute to top quark FCNC interactions $t \to u_j\, X$ at tree level. Here, $\tau^I$ and $T^A$ denote the generators of the $\mathrm{SU}(2)_{\rm L}$ and $\mathrm{SU}(3)_{\rm C}$ gauge groups, respectively. The indices $p$ and $r$ are generation indices; in our analysis, we fix $p = 3$, while $r$ is allowed to take values $1$ or $2$, and vice versa through Hermitian conjugation.}
    \label{tab:SMEFT_Ops}
\end{table}

It is worth mentioning that, in order to match the general effective Lagrangian  eq.~\eqref{eq:eff_vertex_top_FCNC} expressed in the mass basis with the SMEFT operator basis eq.~\eqref{eq:SMEFT_Lag} defined in the flavour basis at the scale $(\Lambda)$. Hence, a flavour rotation of the relevant fields is required to find out the matching conditions. Since we are assuming that NP is mostly coupled to the top quark, it is logical to perform the analysis in $up-aligned$ basis, where we define all the right-handed fields in the mass basis, and the left-handed lepton doublet in the mass basis of charged leptons, while the left-handed quark doublet is initially defined in an arbitrary flavour basis. In this basis, quark Yukawa matrices can be expressed as
\begin{align}
  \Gamma_u=U_u \,\Gamma_u^{\rm diag}\,,\,\,\,\, \Gamma_d=U_d \,\Gamma_d^{\rm diag},\,\,\,\, q=\begin{pmatrix}
   U_{u_L}^{\dagger}\, u_L\\
    U_{d_L}^{\dagger}\, d_L
\end{pmatrix}
\end{align}
where $U_{u_L}$ and $U_{d_L}$ are unitary matrices, $\Gamma_u^{\rm diag}$ and $\Gamma_d^{\rm diag}$ are the diagonal Yukawa matrices, and $u_L$, $d_L$ are the mass eigenstates of the quark doublet $q$. After performing a flavour rotation $(q \to U_{u_L}\, q)$, which diagonalises the up-type Yukawa matrix, the quark doublet takes the form:
\begin{align}
    \Gamma_u=\Gamma_u^{\rm diag}\,,\,\,\,\Gamma_d=V_{\rm CKM}\Gamma_d^{\rm diag}\,,\,\,\, q=\begin{pmatrix}
        u_L\\
        V_{\rm CKM} d_L
    \end{pmatrix}
\end{align} 
where $V_{\rm CKM}=\left(U_{d_L}^{\dagger}U_{u_L}\right)$, and $V_{\rm CKM}$ is the Cabibbo-Kobayashi-Maskawa (CKM) matrix. Following this basis rotation, we obtain the tree-level matching condition of the SMEFT operators with the effective Lagrangian in eq.~\eqref{eq:eff_vertex_top_FCNC}, which are as given below  
\begin{align}\label{eq:tree_level_matching}
    &(\xi_L)_{pr}=\sqrt{2}  v\frac{m_t}{g_s} \left(\mathcal{C}^{uG}_{rp}\right)^* \,, \ \ \ \ \  (\xi_R)_{pr}=\sqrt{2} v \frac{m_t}{g_s}\mathcal{C}^{uG}_{pr}\,,\nonumber\\
   &(\lambda_L)_{pr}=\sqrt{2}  v\frac{m_t}{e}\left(s_W \left(\mathcal{C}^{uW}_{rp}\right)^*+c_W \left(\mathcal{C}^{uB}_{rp}\right)^*\right)\,, \ \  (\lambda_R)_{pr}=\sqrt{2}  v \frac{m_t}{e}\left(s_W \mathcal{C}_{pr}^{uW}+c_W\mathcal{C}^{uB}_{pr}\right)\,,\nonumber\\ 
   & (\kappa_L)_{pr}=\sqrt{2}  v \frac{c_W m_t}{g_W}\left(c_W \left(\mathcal{C}^{uW}_{rp}\right)^*-s_W\left(\mathcal{C}^{uB}_{rp}\right)^*\right)\,, \ \ (\kappa_R)_{pr}=\sqrt{2}  v \frac{c_W m_t}{g_W}\left(c_W\mathcal{C}^{uW}_{pr}-s_W\mathcal{C}^{uB}_{pr}\right)\,,\nonumber\\
   &(X_L)_{pr}= v^2 \left(\mathcal{C}^{\phi q (1)}_{pr}-\mathcal{C}^{\phi q (3)}_{pr}\right)\equiv v^2 \mathcal{C}^{\phi q(-)}_{pr}\,, \ \ \ \ (X_R)_{pr}= v^2 \mathcal{C}^{\phi u}_{pr}\,,\\
   &(\eta_L)_{pr}=\frac{3}{2}v^2 \left(\mathcal{C}^{u\phi}_{rp}\right)^*\,, \ \ \ \ (\eta_R)_{pr}=\frac{3}{2}v^2 \mathcal{C}^{u \phi}_{pr}\,.\nonumber
\end{align}
In eq.~\eqref{eq:tree_level_matching}, the flavour indices $p$ and $r$ correspond to the quark generation labels. The matching conditions for the left- and right-handed $top-charm$ effective couplings, as presented in eq.~\eqref{eq:eff_vertex_top_FCNC}, require the inclusion of both index configurations $(p,r) = (2,3)$ and $(3,2)$. Similarly, for the $top-up$ effective couplings in the left- and right-handed sectors, it is necessary to consider both $(p,r) = (1,3)$ and $(3,1)$ combinations. 

This requirement arises because the effective operators describing FCNC interactions are constructed to respect the hermiticity of the Lagrangian. Consequently, for any off-diagonal flavour structure involving two distinct generations, both orderings of the flavour indices must be included to account for the operator and its Hermitian conjugate. Neglecting one of the two configurations would lead to an incomplete description of the interaction and violate the symmetry properties of the effective theory\, \cite{Aguilar-Saavedra:2008nuh}. Eq.~\eqref{eq:tree_level_matching} represents the tree-level matching between the SMEFT operators and the effective Lagrangian given in eq.~\eqref{eq:eff_vertex_top_FCNC}, where a total $10$ effective couplings, together with $10$ Hermitian conjugate effective couplings (not shown explicitly), are matched with a set of SMEFT WCs mentioned in table.~\ref{tab:SMEFT_Ops}. Since vector-type SMEFT operators are Hermitian by construction, we take the corresponding top-FCNC vector couplings to be real, i.e., $X_L^* = X_L$ and $X_R^* = X_R$, while for the tensor $(\xi_{L(R)},\lambda_{L(R)},\kappa_{L(R)})$ and scalar $(\eta_{L(R)})$ type couplings, we retain their complex nature. Furthermore, this matching is performed at the top quark mass scale, $\mu_t$. The parameters $c_W$ and $s_W$ denote the cosine and sine of the weak mixing angle, respectively, and $v$ is the vacuum expectation value of the Higgs field, with $v \simeq 246\,\mathrm{GeV}$. 

In this work, we assume that NP effects arise exclusively from FCNC interactions involving up-type quarks, particularly the top quark. We want to emphasise that our primary goal is to constrain rare top FCNC couplings and map the effective operators in eq.~\eqref{eq:eff_vertex_top_FCNC} onto SMEFT operators that match at tree level. This SMEFT basis is chosen because its anomalous dimension matrix is known, enabling RGEs of WCs from the NP scale to the electroweak scale, $\mu_{\rm EW}$. Below $\mu_{\rm EW}$, the running to $\mu_b$ involves only QED and QCD corrections to low-energy operators, where SMEFT operators have no role. Therefore, even without SMEFT matching, constraints on top FCNC couplings can still be derived at both $\mu_{\rm EW}$ and $\mu_b$.

The SMEFT operators $\mathcal{O}^{\phi q,(1)}_{pr}$ and $\mathcal{O}^{\phi q,(3)}_{pr}$ can also generate down-type  FCNCs through the same WCs. These contributions occur at tree level and impact rare low-energy processes, which are subject to stringent experimental constraints. We analyse these effects separately. Although the WCs associated with $\mathcal{O}^{\phi q,(1)}_{pr}$ and $\mathcal{O}^{\phi q,(3)}_{pr}$ represent both up and down type FCNC interactions by single parameters in the SMEFT framework, their phenomenological implications vary across processes depending on the quark flavours involved. This variation arises from the decomposition of the $\mathrm{SU(2)}_L$ gauge currents into up-type and down-type components, leading to distinct observable signatures even when the underlying couplings are identical. 

In a model-independent EFT approach, constraints on WCs are extracted directly from experimental data and subsequently matched to NP scenarios. Therefore, the bounds obtained on up-type and down-type couplings are highly process-dependent and require careful consideration when mapped to fundamental model parameters. Conversely, model-dependent analyses may predict distinct FCNC patterns for up-type and down-type quarks. For example, certain NP frameworks \cite{Banerjee:2018fsx,Cacciapaglia:2019bqz,Cho:2019stk,Hou:2025bjy} induce FCNC effects exclusively in rare top-quark processes without generating $b\to s$ transitions. Moreover, if FCNC vertices arise only at the loop level, the effective couplings can deviate significantly from tree-level expectations. Accounting for these distinctions is essential for precision SMEFT fits and for the robust interpretation of experimental constraints.

Most of the SMEFT operators we consider, such as dipole and vector-like operators, can in principle mix with certain four-fermion operators (e.g., $\mathcal{O}^{\ell equ(3)}_{prst}$, $\mathcal{O}^{\ell q(1)}_{prst}$, $\mathcal{O}^{\ell q(3)}_{prst}$, $\mathcal{O}^{\ell u}_{prst}$, and $\mathcal{O}^{e u}_{prst}$ \cite{Alonso:2013hga,Jenkins:2013wua,Jenkins:2017jig}) under renormalisation group evolution (RGE). However, we have explicitly checked the size of these mixing contributions and found them to be extremely small, typically between $10^{-3}$ and $10^{-5}$ depending on the scenario and energy scale. This suppression occurs because the mixing of dipole operators into four-fermion operators is loop- and Yukawa-suppressed, while the reverse mixing is suppressed by gauge couplings and flavour structures. For instance, although the scalar-type operator $\mathcal{O}^{\ell equ(3)}_{prst}$ can mix into dipole operators via Higgs and gauge boson loops, the contribution is proportional to the corresponding Yukawa coupling and is therefore negligible for all but the top quark. 

Furthermore, these two operator classes impact largely distinct sets of observables. Dipole operators primarily drive radiative decays, magnetic moments, FCCC processes, and certain electroweak precision observables. In contrast, four-fermion operators dominantly contribute to contact interactions, scattering cross sections, and FCNC semileptonic rare decays like $b\to s\ell\ell$. Even when both classes influence the same observable, their contributions enter with different Lorentz structures and kinematic dependencies, making interference effects minimal. Consequently, mixing-induced effects do not meaningfully alter the phenomenology and can be safely ignored in our analysis without a loss of accuracy.

\subsection{Methodology} \label{sec:methodology}

Within the EFT framework, operators are defined at specific energy scales reflecting the relevant degrees of freedom. To systematically evaluate the impact of top-quark FCNC SMEFT operators defined at a new physics scale $\Lambda$, we employ a consistent sequence of RGE and one-loop matching steps. This allows us to connect high-scale couplings to a broad class of observables.

Specifically, our evaluation of the one-loop contributions induced by the top-FCNC effective vertices (defined in Eq.~\ref{eq:eff_vertex_top_FCNC}) relies on matching and running between two primary energy thresholds: the electroweak scale ($\mu_{\text{EW}}$) and the bottom-quark scale ($\mu_b$).

At the electroweak scale, $\mu_{\text{EW}}$, we obtain the corresponding effective couplings for high-energy observables via a matching procedure. Here, we evaluate anomalous $tbW$ interactions alongside various Higgs decay (production) processes, trilinear gauge couplings, the top chromo-magnetic dipole moment, and various $W$- and $Z$-pole observables. We also compute the loop-induced oblique parameters ($S, T$, and $U$) in terms of the top-FCNC couplings evaluated at this scale. Furthermore, to describe low-energy flavour observables, the SMEFT operators are matched onto the low-energy effective theory (LEFT, also referred to as WET) operator basis at this same electroweak threshold.

The LEFT WCs determined at $\mu_{\text{EW}}$ are subsequently evolved down to the bottom-quark scale, $\mu_b = m_b$. This provides the appropriate low-energy framework for applying experimental bounds from flavour data.

By determining the dependence of each observable on these effective couplings, our primary goal is to derive robust, global constraints on the corresponding SMEFT WCs. To implement the RGE running and one-loop matching consistently, we adopt the formalism of ref.~\cite{Dekens:2019ept} and closely follow the methodology developed in our earlier work~\cite{Kala:2025srq}, which provides a detailed flowchart of the full procedure. 

Specifically, starting from a chosen set of top-quark FCNC SMEFT operators defined at the new physics scale $\Lambda = 1\,\mathrm{TeV}$, we determine their impact on observables at different energy scales through the following sequence of RGE running and matching steps:

\begin{itemize}
    \item \underline{Running from $\Lambda \to \mu_{\rm EW}$:} We evolve the SMEFT WCs $\mathcal{C}_i$ from the new physics scale $(\Lambda)$ down to the electroweak scale $(\mu_{\rm EW}=m_Z\simeq 91.2 \,\mathrm{GeV})$. We perform this running at the leading-logarithmic (LL) approximation using the anomalous dimension matrix (ADM) of the SMEFT operators, taken from the literature~\cite{Jenkins:2013zja,Jenkins:2013wua,Alonso:2013hga}. It is important to mention, because the scale separation between the top quark, Higgs boson, $W$ boson, and $Z$ boson masses is relatively small. Consequently, we perform tree-level matching for the effective top-quark FCNC interactions and one-loop matching for the LEFT operator basis, EWPOs, and other high-energy observables at this common scale $\mu_{\rm EW}$. The underlying running equations connecting two different energy scales in LL approximation can be written as:
   \begin{align}\label{eq:RGE}
   \mathcal{C}_i(\mu)=\left(1+\frac{\gamma_{ii}}{16\pi^2}\log\left(\frac{\mu}{\Lambda}\right)\right)\mathcal{C}_i(\Lambda)+ \sum_{i\neq j}\frac{\gamma_{ij}}{16\pi^2}\log\left(\frac{\mu}{\Lambda}\right) \mathcal{C}_j(\Lambda)\,.
   \end{align} 
Taking the reference NP scale as $\Lambda = 1\,\mathrm{TeV}$, the SMEFT WCs at these two different scales are related by:
{\footnotesize
\begin{align}
    \begin{pmatrix}
    \mathcal{C}^{uB}_{23}\\
    \mathcal{C}^{uW}_{23}\\
    \mathcal{C}^{uG}_{23}\\
    \mathcal{C}^{\phi u}_{23}\\
    \mathcal{C}^{\phi q (1)}_{23}\\
    \mathcal{C}^{\phi q (3)}_{23}\\
    \mathcal{C}^{u\phi}_{23}
    \end{pmatrix}_{\mu_{\rm EW}}=\begin{pmatrix}
        0.888659 & 0.001755 & -0.025980 & 0 & 0 &0 & 0\\
        0.000585 & 0.921365 & -0.027466 & 0 & 0 & 0 & 0\\
        -0.019485 & -0.061798 & 1.04996 & 0 & 0 & 0 & 0 \\
        0 & 0 & 0 & 0.886147 & 0.000081 & 0 & 0\\
        0 & 0 & 0 & 0 & 0.908814 & 0.050908 & 0\\
        0 & 0 & 0 & 0 & 0.016969 & 0.955807 & 0 \\
        0.017356 & 0.007568 & -0.368257 & -0.010302 & 0 & -0.000030 & 0.977383
    \end{pmatrix}
    \begin{pmatrix}
    \mathcal{C}^{uB}_{23}\\
    \mathcal{C}^{uW}_{23}\\
    \mathcal{C}^{uG}_{23}\\
    \mathcal{C}^{\phi u}_{23}\\
    \mathcal{C}^{\phi q (1)}_{23}\\
    \mathcal{C}^{\phi q (3)}_{23}\\
    \mathcal{C}^{u\phi}_{23}
    \end{pmatrix}_{\mu_{\rm 1 TeV}}
    \label{Eq:SMEFT_Running}
\end{align}}
The RGE evolution of the WCs $\mathcal{C}_{13}$ is similar to that of $\mathcal{C}_{23}$; the only differences arise from the Yukawa-sector contributions, while the gauge-sector evolution is identical for both cases. The explicit expressions for the beta functions are provided in Appendix~\ref{Append:RGE_beta}.

    \item \underline{Running from $\mu_{\rm EW} \to \mu_b$:} The LEFT WCs, which dictate the low-energy observables, are further evolved from the electroweak scale ($\mu_{\rm EW}$) down to the $\mu_b \simeq 4.2\,\mathrm{GeV}$ scale relevant for $B$-decay processes. We perform this running using the ADM matrices provided in refs~\cite{Aebischer:2017gaw,Jenkins:2017jig}. This running is particularly important, as QCD corrections play a crucial role in the renormalisation of the LEFT operators. The underlying running equation responsible for the evolution of the LEFT couplings can be written as:
\begin{align}
    \frac{d\vec{L}_i}{d\log\mu}&=\frac{1}{16\pi^2}\sum_{j}\beta_{ij}\vec{L}_j=\frac{\alpha_s}{4\pi}\sum_j\beta_{ij}^s \vec{L}_j+\frac{\alpha_e}{4\pi}\sum_j \beta_{ij}^e \vec{L}_j\,.
\end{align}
Here, $\vec{L}_i$ represents the set of LEFT WCs associated with various low-energy processes, while $\alpha_s$ and $\alpha_e$ denote the strong and electromagnetic couplings, respectively. The matrices $\beta^s$ and $\beta^e$ are the ADMs arising from QCD and QED corrections to the LEFT operators. Considering only linear-order corrections from QED and QCD, the RGE takes a simplified form that connects the LEFT WCs at two different energy scales:
\begin{eqnarray}
\vec{L}_i(\mu)&=U_{ij}(\mu,\mu_0)\vec{L}_j=\Bigg[U_{ij}^s(\mu,\mu_0)+U_{ij}^e(\mu,\mu_0)\Bigg] \vec{L}_j(\mu_0)\,.
\end{eqnarray}
Here, $U_{ij}^s$ and $U_{ij}^e$ denote the evolution matrices arising from QCD and QED corrections, respectively. The values of these evolution matrices are taken from refs~\cite{Aebischer:2017gaw, Chetyrkin:1996vx, Buras:1998raa, Gonzalez-Alonso:2017iyc}. We introduce the relevant LEFT operator basis and corresponding WCs when discussing the low-energy processes in Section~\ref{sec:List_of_obs}. Furthermore, the numerical solutions of the RGEs for these WCs, obtained by running them from the electroweak scale $(\mu_{\rm EW})$ down to the low-energy scale $(\mu_b)$, are presented in the appropriate subsections.
\end{itemize}

Once the observables from the different energy scales are expressed in terms of either the top-FCNC effective couplings or the SMEFT WCs, a global $\chi^2$ fit is performed to constrain the corresponding parameters. The specific statistical techniques and details of this global $\chi^2$ analysis are discussed in the subsequent sections.

\section{Impact on various observables}
\label{sec:List_of_obs}
In the previous subsection~\ref{sec:methodology}, we outlined the methodology of our analysis. Building upon that foundation, in this section, we briefly outline the different classes of observables from various energy sectors that are sensitive to the top-FCNC effective couplings. These include low-energy flavour observables, EWPOs, and high-energy collider measurements, ensuring that the NP effects are probed across a broad range of processes and scales. A detailed theoretical framework explaining how these sectors are affected by NP interactions can be found in Appendix~\ref{Append:np_interactions}.

\subsection{Low Energy Observables}\label{subsec:Low_energy}
Low-energy observables associated with FCNC processes, such as radiative decays, semileptonic decays like $b \to d_j \ell^+ \ell^-$, rare and invisible decay modes, and FCCC processes, like $b \to c(u) \ell \nu$, are sensitive to the presence of top-quark FCNC interactions at the one-loop level. Since FCNC processes are rare and highly suppressed in the SM, they serve as excellent probes for new physics, providing stringent bounds on the corresponding NP couplings. Similarly, the FCCC processes are also useful in constraining the new physics information, given the wealth of improved measurements and the precision extractions of the observables associated with these processes.

\subsubsection{Observables related to FCNC Processes}\label{sec:FCNC}

In this section we summarise the FCNC observables included in our numerical analysis. The complete theoretical framework, including effective Hamiltonians, operator definitions, matching conditions, RG evolution, and the Feynman diagrams, is collected in Appendix~\ref{sec:appFCNC}.

\paragraph{\underline{\textbf{Radiative decays:}}}
Radiative decays $b\to s(d)\gamma$ provide some of the strongest indirect constraints on top-FCNC interactions. We include measurements of branching fractions and CP asymmetries from Belle, Belle II and HFLAV~\cite{Belle:2017hum,Belle-II:2024tru,Belle:2014sac,HeavyFlavorAveragingGroupHFLAV:2024ctg}. The list of branching-ratio observables used in the fit is given in Table~\ref{tab:radiative_data}.

\begin{table}[h!]
 	\centering
 	\renewcommand{\arraystretch}{1.5}
 	\setlength{\tabcolsep}{15pt}
 	\resizebox{\textwidth}{!}{%
 		\begin{tabular}{|c|c|c|}
 			\hline
 			\multicolumn{3}{|c|}{Radiative Decays} \\ 
 			\hline
            \rowcolor{gray!20}
 			$\text{Decay}$  & SM Value & Exp Value  \\
 			\hline
 			\hline
 			$\mathcal{B} (B^{+} \to K^{*+} \gamma)$     & $(4.60 \pm 1.28) \times 10^{-5}$ \cite{Ali:2007sj} &$ (3.76 \pm 0.16 ) \times  10^{-5}$ \cite{Belle:2017hum} \\
 			$\mathcal{B} (B^{0} \to K^{*} \gamma)$      & $(4.30 \pm 1.19 ) \times 10^{-5}$  \cite{Ali:2007sj} & $(3.96 \pm 0.16 ) \times  10^{-5}$ \cite{Belle:2017hum} \\
 			$\mathcal{B} (B^{+} \to \rho^{+} \gamma)$   & $(0.85 \pm 0.32 ) \times 10^{-6}$  \cite{Ali:2001ez} &$(1.31 \pm 0.23 ) \times 10^{-6} $ \cite{Belle-II:2024tru}   \\ 
 			$\mathcal{B} (B^{0} \to \rho^{0} \gamma)$   & $(0.49 \pm 0.17 ) \times 10^{-6}$  \cite{Ali:2001ez} & $(0.75 \pm 0.16 ) \times 10^{-6} $\cite{Belle-II:2024tru} \\
 			$\mathcal{B} (B^{0}_{s} \to \phi \gamma)$   & $(4.30 \pm 1.18) \times 10^{-5}$ \cite{Ali:2007sj}& $(3.60 \pm 0.86 ) \times  10^{-5} $  \cite{Belle:2014sac}  \\
 			$\mathcal{B} (B \to X_{s} \gamma)_{E_{\gamma}>1.6 \rm \, GeV } $          & $(3.40 \pm 0.17) \times 10^{-4}$ \cite{Misiak:2020vlo} & $ (3.49 \pm 0.19 ) \times 10^{-4} $ \cite{HeavyFlavorAveragingGroupHFLAV:2024ctg}  \\
 			\hline
 		\end{tabular}%
 	}
 	\caption{SM and Experimental branching ratios of the available radiative processes which we have taken into account. }
 	\label{tab:radiative_data}
 \end{table}

The Belle II Collaboration~\cite{Belle-II:2024kbc} reported
\begin{align}
	A_{\mathcal{CP}}(B^+\to K^{*+}\gamma)&=(-0.007\pm0.029\pm0.005),\\
	A_{\mathcal{CP}}(B^0\to K^{*0}\gamma)&=(-0.033\pm0.023\pm0.004).
\end{align}

The Belle Collaboration~\cite{Belle:2018iff} measured
\begin{align}
	A_{\mathcal{CP}}(B^+\to X_s\gamma)&=(0.0275\pm0.0184\pm0.0032),\\
	A_{\mathcal{CP}}(B^0\to X_s\gamma)&=(-0.009\pm0.018).
\end{align}

The Belle II Collaboration~\cite{Belle-II:2024uzp} reported
\begin{align}
	S_{K^*\gamma}=-0.00^{+0.27}_{-0.26}\pm0.03.
\end{align}

\paragraph{\underline{\textbf{Semileptonic $b\to d_i\ell\ell$ decays:}}}
We include a comprehensive set of observables associated with $B\to K^{(*)}\mu^+\mu^-$, $B\to\phi\mu^+\mu^-$ and $B \to \pi \mu\mu$ decays, including differential branching fractions, CP asymmetries and angular observables measured by LHCb, Belle, BaBar, CDF, ATLAS and CMS~\cite{LHCb:2013lvw,LHCb:2014cxe,LHCb:2014mit,LHCb:2014vgu,LHCb:2015svh,LHCb:2020gog,LHCb:2021zwz,LHCb:2022qnv,Belle:2009zue,Belle:2016fev,BaBar:2012mrf,CDF:2011tds,ATLAS:2018gqc,CMS:2017rzx}. We also include the LFU ratios $R_K$ and $R_{K^*}$ using the recent LHCb measurements~\cite{LHCb:2022qnv}. The corresponding theoretical expressions follow Refs.~\cite{Bobeth:2007dw,Altmannshofer:2008dz,Becirevic:2012fy,Descotes-Genon:2013vna,Capdevila:2016ivx,Biswas:2020uaq,Biswas:2022lhu}.

\paragraph{\underline{\textbf{Leptonic rare decays:}}}
We include the channels $B_s^0\to\mu^+\mu^-$, $B^0\to\mu^+\mu^-$ and $K_L\to\mu^+\mu^-$.

The latest measurements~\cite{CMS:2022mgd,LHCb:2021vsc} are
\begin{align}
	\mathcal{B}(B^{0}\to\mu^{+}\mu^{-})^{\rm Exp}&=\left(1.2^{+0.8}_{-0.7}\pm0.1\right)\times10^{-10},\\
	\mathcal{B}(B_{s}^{0}\to\mu^{+}\mu^{-})^{\rm Exp}&=\left(3.83^{+0.38\,+0.24}_{-0.36\,-0.21}\right)\times10^{-9}.
\end{align}

The corresponding SM predictions are~\cite{Beneke:2019slt}
\begin{align}
	\mathcal{B}(B^{0}\to\mu^+\mu^-)^{\rm SM}&=(1.03\pm0.05)\times10^{-10},\\
	\mathcal{B}(B_s^{0}\to\mu^+\mu^-)^{\rm SM}&=(3.66\pm0.14)\times10^{-9}.
\end{align}

For kaons,
\begin{align}
	\mathcal{B}(K_L\to\mu^+\mu^-)^{\rm Exp}&=(6.84\pm0.11)\times10^{-9},
\end{align}
with the SM short-distance prediction~\cite{Buras:2013rqa}
\begin{align}
	\mathcal{B}(K_L\to\mu^+\mu^-)_{\rm SD}^{\rm SM}=(0.79\pm0.12)\times10^{-9},
\end{align}
A conservative upper bound on the short-distance component has been derived in the literature (see, e.g., ref.~\cite{Isidori:2003ts}):
\begin{equation}
	(K_L \to \mu^+ \mu^-)_{\rm SD} < 2.5 \times 10^{-9}.
\end{equation}
This limit is obtained using the experimental constraint on the branching ratio of $K_L \to \gamma\gamma$ decays, which dominate the long-distance contribution to $K_L \to \mu^+ \mu^-$, for the details see the ref.~\cite{Isidori:2003ts}).

\paragraph{\underline{\textbf{Invisible decays:}}}
Invisible decays constitute another important class of channels for probing new physics. In this analysis, we have studied the effects on various $P \to P^{\prime} \nu \bar{\nu}$ channels, where $P$ denotes a pseudoscalar meson and $P^{\prime}$ denotes either a pseudoscalar or a vector meson.  Recently, the Belle collaboration \cite{Belle-II:2023esi} has reported the measurement of the di-neutrino channel $B^+ \to K^+ \nu \bar{\nu}$, for the first time. In our analysis, we have included the available data on the branching fractions $\mathcal{B}(B^+ \to K^+ \nu \bar{\nu})$, $\mathcal{B}(B^0 \to K^{*0} \nu \bar{\nu})$ and $\mathcal{B}(K^+ \to \pi^+ \nu \bar{\nu})$ which we have presented in table.~\ref{tab:Inv}, where the statistical uncertainties are also quoted. 
\begin{table}[h!]
	\centering
	\renewcommand{\arraystretch}{1.2}
	\setlength{\tabcolsep}{5pt}
	\resizebox{0.9\textwidth}{!}{
		\begin{tabular}{|c|c|c|}
			\hline
			\rowcolor{gray!20}
			\textbf{Branching Ratio} & \textbf{SM values} & \textbf{Experimental values}  \\
			\hline
			\hline
			$\mathcal{B}(B^+\to K^+\nu\bar{\nu})$ & $(5.22\pm 0.15 \pm 0.28)\times 10^{-6}$\,\,\cite{Becirevic:2023aov} & $(2.3\pm 0.5^{+0.5}_{-0.4})\times 10^{-5}$\,\, \cite{Belle-II:2023esi} \\
			\hline
			$\mathcal{B}(B^0\to K^{*\,0}\nu\bar{\nu})$ & $(9.47 \pm 1.28 \pm 0.57)\times 10^{-6}$\,\,\cite{Becirevic:2023aov} & \makecell[l]{$(3.8^{+2.9}_{-2.6})\times 10^{-5}$\,\,\cite{BaBar:2013npw} \\ $< 1.8 \times 10^{-5}$\,\,\cite{Belle:2017oht} } \\ \hline
			$\mathcal{B}(K^+\to \pi^+ \nu\bar{\nu})$ & $(8.60 \pm 0.42)\times 10^{-11}$\,\, \cite{Buras:2024ewl} & $(1.06^{+0.40}_{-0.34}\pm 0.09)\times 10^{-10}$\,\, \cite{NA62:2021zjw} \\
			\hline
	\end{tabular}}
	\caption{Updated SM predictions and experimental measurements of the branching ratios for invisible decays of $B$ and $K$ mesons.}
	\label{tab:Inv}
\end{table}

\paragraph{\underline{\textbf{Meson Mixing:}}}

In our analysis, we also investigate the impact of various neutral meson mixing processes. The contribution from $B_{s(d)}-\bar{B}_{s(d)}$ mixing is found to be numerically insignificant. This is because such mixing effects arise only at the two-loop level for the top FCNC NP operators, requiring double insertions of penguin-type topologies [see fig.~(\ref{fig:rad_1}--\ref{fig:rad_2})]. We have explicitly verified that these contributions do not lead to any substantial impact on the NP parameter space.

Furthermore, we examine the contributions to $D^0$--$\bar{D}^0$ mixing. In the SM, this process is highly suppressed due to the CKM hierarchy and is challenging to compute, as it is predominantly governed by long-distance (LD) effects. In our scenario, the mixing amplitude arises through box diagrams with \textit{quadruple insertions}, making the diagrams proportional to the product of couplings, such as $C_i C_j C_k C_l/\Lambda^8$. The $D^0$-$\bar{D}^0$ mixing might play an important role in constraining these product couplings, which is beyond the scope of the present study. However, we have explicitly checked that the allowed ranges obtained for individual couplings from other inputs are fully consistent with current data on the $D^0$-$\bar{D}^0$ mixing amplitude.

\subsubsection{Observables related to FCCC Processes}\label{sec:FCCC}
 The FCCC processes involve transitions between quarks of different flavours mediated by charged weak currents, typically through the exchange of $W^\pm$ bosons. In the SM, these processes occur at the tree level, making them experimentally accessible. Their coupling strengths provide direct measurements of the CKM matrix elements, offering a powerful avenue to test the flavour structure and internal consistency of the SM. In this work, we study the impact of top-FCNC couplings on semileptonic and leptonic decays of $B$, $B_s$, $K$, $D$, and $D_s$ mesons. We also investigate the indirect effects of these couplings on anomalous $t\to bW$ interactions. The complete effective-theory framework, operator basis, matching conditions, RG evolution, and relevant loop diagrams are presented in Appendix~\ref{sec:appFCCC}.
 
 \paragraph{\underline{\textbf{Leptonic and Semi-leptonic decays:}}} 
  The NP contributions modify the rates of semileptonic and leptonic decays and consequently affect the extraction of CKM matrix elements. This modification can be parameterised as 
  \begin{equation}
  	 \left|V_{u_j d_i}\right| \;\longrightarrow\; \left| V_{u_j d_i} \left( 1+C_{V_L}\pm C_{V_R} \right) \right|,
    \end{equation} 
where the $(+)$ sign applies to semileptonic decays and the $(-)$ sign applies to leptonic decays. Therefore, CKM elements extracted from processes of the type $P\to M\ell\nu_\ell$ can be used as sensitive probes of NP. In the SM, CKM elements are parameterised in terms of the Wolfenstein parameters $(A,\lambda,\rho,\eta)$, which must be determined simultaneously with the NP coefficients. In addition, we include branching-fraction measurements for a large number of semileptonic and leptonic decays. The complete list of observables and details of the fitting methodology can be found in Refs.~\cite{Biswas:2021pic,Kolay:2024wns,Kolay:2025jip}. Along with these inputs, we also include the updated measurements of the lepton-flavour-universality ratios 
\begin{equation} \label{eq:lfuv} 
	R(D^{(*)}) = \frac{ \Gamma(B\to D^{(*)}\tau\nu) }{ \Gamma(B\to D^{(*)}\mu\nu) }.
 \end{equation}
 The measured values are reported by Belle-II~\cite{Belle-II:2025yjp}. The numerical inputs used in the analysis, including CKM determinations, leptonic branching fractions, semileptonic observables and LFU measurements, are taken from Refs.~\cite{Biswas:2021pic,Kolay:2024wns,Kolay:2025jip,Belle-II:2025yjp}.

\paragraph{\underline{\textbf{Anomalous $tWb$ coupling:}}} The most general Lagrangian describing the effective $tWb$ vertex structure can be written as: 
\begin{equation}\label{eq:Wtb_Lagrangian}
\mathcal{L}_{tWb} =-\frac{g_2}{\sqrt{2}} \biggl( \bar{b} \gamma_{\mu} (V_L P_L +V_R P_R) t W^{\mu-}  + \bar{b} \frac{i \sigma_{\mu \nu} q^{\nu}}{m_W} (g_L P_L + g_R P_R) t W^{\mu-} \biggr) + \rm{h.c.}
\end{equation}
where $V_{L(R)}$, $g_{L(R)}$ are the anomalous $tWb$ coupling, where $V_L=V_{tb}^*(1+C_{V_L})$, $V_R=C_{V_R}$ in the SM $V_L=V_{tb}^*$ where other couplings are zero. The ATLAS collaboration~\cite{ATLAS:2024ojr,ATLAS:2024ppp,ATLAS:2023qrn} has recently provided bounds on the quantity $|f_L V_{tb}|$, which we incorporate in our analysis by assuming the SM value $V_{tb} = 1$. Whereas the bounds on other anomalous couplings at 95\% confidence level (CL) are obtained by ATLAS and CMS analyses \cite{CMS:2020ezf}. We do not obtain sizable contributions to $V_R$ and $g_{L,R}$ due to sensitivity to the light quark mass. Hence, they have negligible impact in the numerical analysis. 
\begin{figure}[t]
    \centering
\subfloat[]{
    \begin{tikzpicture}
        \begin{feynman}
        \vertex[](a1);
        \vertex[square dot, draw=black, fill=black, minimum size=4pt, inner sep=2.5pt,below left=0.8cm of a1](a2){};
        \vertex[below left=1.1cm of a2](a3){\(t\)};
        \vertex[above left=1cm of a1](a4);
        \vertex[above left=0.7cm of a4](a5){\(d_i\)};
        \vertex[right=1.cm of a1](a6){\(W\)};
        \diagram* {
            (a3) --[fermion,very thick,] (a2) --[fermion, very thick, edge label'=\(c/u\)] (a1) --[fermion,  edge label'=\(d_i\)] (a4) --[fermion] (a5),
            (a2) --[black,boson, very thick, edge label=\(\gamma/Z/g\)] (a4),
            (a1) --[ boson] (a6),
        };
        \end{feynman}
    \end{tikzpicture}
    }
     \subfloat[]{
    \begin{tikzpicture}
        \begin{feynman}
        \vertex[](a1);
        \vertex[square dot, draw=red, fill=red, minimum size=4pt, inner sep=2.5pt,below left=0.8cm of a1](a2){};
        \vertex[below left=1.1cm of a2](a3){\(t\)};
        \vertex[above left=1cm of a1](a4);
        \vertex[above left=0.7cm of a4](a5){\(d_i\)};
        \vertex[right=1.cm of a1](a6){\(W\)};
        \diagram* {
            (a3) --[fermion,very thick,] (a2) --[fermion, very thick, edge label=\(c/u\)] (a4)--[fermion] (a5),
            (a2) --[red,boson, very thick, edge label'=\(\gamma/Z\)] (a1),
            (a1) --[ boson,edge label'=\(W\)] (a4),
            (a1) --[boson](a6),
        };
        \end{feynman}
    \end{tikzpicture}
    }
\subfloat[]{
    \begin{tikzpicture}
        \begin{feynman}
        \vertex[](a1);
        \vertex[square dot, draw=violet, fill=violet, minimum size=4pt, inner sep=2.5pt,below left=0.8cm of a1](a2){};
        \vertex[below left=1.1cm of a2](a3){\(t\)};
        \vertex[above left=1cm of a1](a4);
        \vertex[above left=0.7cm of a4](a5){\(d_i\)};
        \vertex[right=1.cm of a1](a6){\(W\)};
        \diagram* {
            (a3) --[fermion,very thick,] (a2) --[fermion, very thick, edge label'=\(c/u\)] (a1) --[fermion,  edge label'=\(d_i\)] (a4) --[fermion] (a5),
            (a2) --[violet,scalar, very thick, edge label=\(H\)] (a4),
            (a1) --[ boson] (a6),
        };
        \end{feynman}
    \end{tikzpicture}
    }
    \subfloat[]{
    \begin{tikzpicture}
        \begin{feynman}
        \vertex[](a1);
        \vertex[square dot, draw=violet, fill=violet, minimum size=4pt, inner sep=2.5pt,below left=0.8cm of a1](a2){};
        \vertex[below left=1.1cm of a2](a3){\(t\)};
        \vertex[above left=1cm of a1](a4);
        \vertex[above left=0.7cm of a4](a5){\(d_i\)};
        \vertex[right=1.cm of a1](a6){\(W\)};
        \diagram* {
            (a3) --[fermion,very thick,] (a2) --[fermion, very thick, edge label=\(c/u\)] (a4)--[fermion] (a5),
            (a2) --[violet,scalar, very thick, edge label'=\(H\)] (a1),
            (a1) --[ boson,edge label'=\(W\)] (a4),
            (a1) --[boson](a6),
        };
        \end{feynman}
    \end{tikzpicture}
    }
 \caption{Effects of top FCNC operators on the $t \to W d_i$ form factor.} 
    \label{fig:top_Form_factor}
\end{figure}
In the presence of effective top FCNC interaction, the $tWb$ vertex will be modified at the one-loop level as depicted in fig.~\ref{fig:top_Form_factor}. These diagrams will contribute to all the $tWb$ anomalous vertex factors in eq.~ \ref{eq:Wtb_Lagrangian}. We have presented a detailed expression for $V_L$ in terms of the top FCNC couplings in eq.~\eqref{eq:Append_anomalous_Wtb} in Appendix~\ref{sec:appFCCC}. We have not provided expressions for $V_R$ and $g_{L,R}$, as they play an insignificant role in the numerical analysis. 
\subsection{Electric dipole moment of light quark}\label{sec:EDM}
Experimentally, EDMs are not measured directly for the light quarks, since they are confined inside the hadrons. Instead, constraints arise from the precision measurement of the composite systems, such as the neutron \cite{Abel:2020pzs}, proton EDMs, diamagnetic atom (e.g., Hg, Ra, Xe), or paramagnetic molecules (e.g., Tho, YbF, HfF) \cite{Grozin:2008nw, PhysRevLett.116.161601, Ema:2021jds}. These observables are sensitive to a combination of light-quark EDMs, chromo EDMs (CEDMs), and the CP-odd three gluon Weinberg operators, which allows us to place indirect bounds on quark-level CP violation. 
In our analysis, the EDMs of both light quarks ($u,d,s$) and heavy quarks ($c,b,t$) are affected by the top-FCNC couplings. The bounds from the top EDM are discussed in the subsequent section.  Generally, the low-energy effective Lagrangian for the hadronic EDM can be expressed as \cite{Gorbahn:2014sha}
\begin{align}\label{eq:lag_EDM}
	\mathcal{L}_{\rm eff}\supset +d_q(\mu) \frac{i}{2}\bar{q}\sigma^{\mu\nu}\gamma_5 qF_{\mu\nu}+\tilde{d}_q(\mu)\frac{i}{2}g_s(\mu) \bar{q}\sigma^{\mu\nu}T^a\gamma_5 q G^a_{\mu\nu}+w(\mu)\frac{1}{3}G^a_{\mu\sigma}G^{b,\sigma}_{\nu}\tilde{G}^{c,\mu\nu}
\end{align}
Here, $q$ runs over all quark flavours except the top quark. $F_{\mu\nu}$ and $G_{\mu\nu}^a$ denote the field strength tensors of QED and QCD, respectively, while $\tilde{G}^{a\,\mu\nu}=\tfrac{1}{2}\,\epsilon^{\mu\nu\alpha\beta}G^a_{\alpha\beta}$ is the dual QCD field strength tensor. The quantities $d_q(\mu)$ and $\tilde{d}_q(\mu)$ represent the EDM and chromo-EDM (cEDM) of quarks, respectively.  A detailed discussion describing the impact of our NP parameter space on the light quark EDM via neutron EDM can be found in Appendix~\ref{sec:appEDM}. Being a CP-violating observable, the neutron EDM directly impacts the imaginary part of the product of couplings, which can be realised from the loop topologies shown in fig.~\ref{fig:dipole_moment_light_quark}. This makes it an exceptionally sensitive probe for new physics. The current $90\%$ CL upper limit on the neutron EDM is~\cite{Abel:2020pzs}:
\begin{align}
	\left|\frac{d_n}{e}\right|<1.8\times 10^{-26}\,\mathrm{cm}\,.
\end{align}

As current experimental measurements provide only an upper limit on the neutron electric dipole moment, we incorporate this constraint into our global analysis using an appropriate symmetric two-sided Gaussian penalty function. The precise statistical formulation of this penalty within the $\chi^2$ likelihood is discussed in detail in the subsequent section.  
\subsection{Electroweak Precision Observables}\label{sec:EWPOs}

The considerable accuracy of electroweak precision measurements of the $W$ and $Z$-pole observables enables us to tightly constrain NP scenarios. In our framework, NP effects manifest as corrections to the $\gamma$ and $Z$ self-energies and modifications to the $Z\to f \bar{f}$ and $W \to u_i \bar{d}_j$ vertices. In this section, we briefly outline the EWPOs we are investigating, such as the oblique parameters $S$, $T$, and $U$. These oblique parameters are primarily generated by $\gamma$ and $Z$ boson self-energy corrections. The detailed theoretical derivations for these parameters can be found in Appendix~\ref{sec:appEWPO}. 

Apart from the oblique parameters, we evaluate a comprehensive set of $Z$- and $W$-pole observables affected by the modified $Z \to f \bar{f}$ and $W \to u_i \bar{d}_j$ vertices. The related observables included in our analysis are tabulated in Table~\ref{tab:W_pole}, \ref{tab:Z_pole} respectively. Detailed theoretical calculations describing these vertex modifications are also provided in Appendix~\ref{sec:appEWPO}.

\begin{table}[htb!]
\centering
\renewcommand{\arraystretch}{1.2}
	\setlength{\tabcolsep}{5pt}
	\resizebox{0.9\textwidth}{!}{
\begin{tabular}{|c|c|c|c|}
\hline
\rowcolor{gray!20}
\textbf{Observables} & \textbf{SM values} & \textbf{Experimental values} & \textbf{Defination} \\
\hline
\hline
$\Gamma_W \mathrm{(GeV)}$ & $2.088$ & $2.085 \pm 0.042$\,\, \cite{ParticleDataGroup:2024cfk} & $ \sum_f \Gamma(W\to ff^{\prime})$\\
\hline
$R_{Wc} $ &$0.50$ & $0.49 \pm 0.04$ \,\, \cite{ParticleDataGroup:2024cfk} & $\frac{\Gamma(W\to c s)}{\Gamma(W\to c s)+\Gamma(W\to u d)}$\\
\hline
\end{tabular}}
\caption{$W$ Pole Observables: List of observables affected by top quark FCNC operators.}
\label{tab:W_pole}
\end{table}

\begin{table}[htb!]
\centering
\renewcommand{\arraystretch}{1.5}
	\setlength{\tabcolsep}{5pt}
	\resizebox{0.9\textwidth}{!}{
\begin{tabular}{|c|c|c|c|}

\hline
\rowcolor{gray!20}
\textbf{Observables} & \textbf{SM values} & \textbf{Experimental values} & \textbf{Defination} \\
\hline
\hline
$\Gamma_Z$ (GeV) & $(2.4961 \pm 0.0010)$ & $(2.4955 \pm 0.0023)$ & $\sum_f \Gamma(Z\to f f)$\\
\hline
$\sigma_{\rm had}$ (nb) & $(41.484)$ & $(41.541\pm 0.037)$ & $\frac{12 \pi}{m_Z^2}\frac{\Gamma(Z\to e^+ e^-)\Gamma(Z\to q \bar{q})}{\Gamma_Z^2}$ \\
\hline
\hline
$A_e$ & $(0.1475 \pm 0.0010)$& $(0.1516 \pm 0.0021)$ & $\frac{\Gamma(Z\to e_L^+ e_L^-)-\Gamma(Z\to e_R^+ e_R^-)}{\Gamma(Z\to e^+ e^-)}$  \\
\hline
$A_{\mu}$ & $0.1472$ & $0.142 \pm 0.015$ &$\frac{\Gamma(Z\to \mu_L^+ \mu_L^-)-\Gamma(Z\to \mu_R^+ \mu_R^-)}{\Gamma(Z\to \mu^+ \mu^-)}$ \\
\hline
$A_{\tau}$ & $0.1472$ & $0.143 \pm 0.004$ & $\frac{\Gamma(Z\to \tau_L^+ \tau_L^-)-\Gamma(Z\to \tau_R^+ \tau_R^-)}{\Gamma(Z\to \tau^+ \tau^-)}$ \\
\hline
$A_b$ & $0.935$ & $0.923 \pm 0.020$\,\, &$\frac{\Gamma(Z\to b_L\bar{b}_L)-\Gamma(Z\to b_R\bar{b}_R)}{\Gamma(Z\to b\bar{b})}$\\
\hline
$A_s$ & $0.936$ & $0.90 \pm 0.09$\,\, & $\frac{\Gamma(Z\to s_L\bar{s}_L)-\Gamma(Z\to s_R\bar{s}_R)}{\Gamma(Z\to s\bar{s})}$\\
\hline
$A_c$ & $0.667$ & $0.670 \pm 0.027$\,\, &$\frac{\Gamma(Z\to c_L\bar{c}_L)-\Gamma(Z\to c_R\bar{c}_R)}{\Gamma(Z\to c\bar{c})}$\\
\hline
\hline
$A^{\rm FB}_{e}$ & $(1.606 \pm 0.006)\%$ & $(1.45 \pm 0.25)\%$ & $\frac{3}{4}A_e^2$\\
\hline
$A^{\rm FB}_{\mu}$ & $1.63\%$ & $(1.69 \pm 0.13)\%$& $\frac{3}{4}A_e A_{\mu}$\\
\hline
$A^{\rm FB}_{\tau}$ &$1.63\%$ & $(1.88 \pm 0.17)\%$& $\frac{3}{4}A_e A_{\tau}$\\
\hline
$A^{\rm FB}_b$ & $(10.34 \pm 0.07)\%$& $(9.92 \pm 0.16)\%$ \,\, &$\frac{3}{4}A_e A_b$\\
\hline
$A^{\rm FB}_s$ & $(10.35 \pm 0.07)\%$ & $(9.8\pm1.1)\%$\,\, &$\frac{3}{4}A_e A_s$\\
\hline
$A^{\rm FB}_c$ & $(7.35\pm 0.02)\%$& $(7.07 \pm 0.35)\%$ \,\, &$\frac{3}{4}A_e A_c$\\
\hline
\hline
$R_{e}$ & $20.736 \pm 0.010$ & $20.804 \pm 0.050$ & $\frac{\Gamma(Z \to q \bar{q})}{\Gamma(Z\to e^+ e^-)}$ \\
\hline
$R_{\mu}$ & $20.736 \pm 0.010$ & $20.784 \pm 0.034$ & $\frac{\Gamma(Z \to q \bar{q})}{\Gamma(Z\to \mu^+ \mu^-)}$\\
\hline
$R_{\tau}$ & $20.781 \pm 0.010$ & $20.764 \pm 0.045$ & $\frac{\Gamma(Z \to q \bar{q})}{\Gamma(Z\to \tau^+ \tau^-)}$\\
\hline
$R_b$ & $0.21581$ & $(0.21629 \pm 0.00066)$ & $\frac{\Gamma(Z\to b\bar{b})}{\sum_q (Z \to q \bar{q})}$\\
\hline
$R_c$ & $0.1722$ & $(0.1721 \pm 0.0030)$ & $\frac{\Gamma(Z\to c\bar{c})}{\sum_q (Z \to q \bar{q})}$\\
\hline
\end{tabular}}
\caption{$Z$ Pole Observables: List of observables affected by top quark FCNC operators. All experimental and SM values are taken from refs.~\cite{ALEPH:2005ab, Janot:2019oyi, dEnterria:2020cgt, SLD:2000jop, ParticleDataGroup:2024cfk}.}
\label{tab:Z_pole}
\end{table}


\subsection{Other relevant Observables}\label{sec:other_rel_obs}
In addition to low-energy observables and EWPOs, several other processes are also sensitive to top-quark FCNC interactions. For instance, various Higgs boson decays like $(H\to b\bar{b},WW^*,\,\gamma\gamma(Z))$ can receive one-loop level corrections from top-FCNC operators, potentially altering the decay width and coupling strength. Trilinear gauge couplings (TGC), which are tightly constrained by LEP and LHC data, can also be modified. Furthermore, these top FCNC operators can induce electric and chromomagnetic dipole and magnetic moments for various quark flavours, which could provide additional constraints on the new physics parameter space.
\subsubsection{Observables related to Higgs Physics}\label{sec:Higgs_obs}
In our analysis, we have incorporated the most recent measurements of Higgs boson production cross sections, decay branching ratios as reported by the ATLAS Collaboration~\cite{ATLAS:2022vkf}. Among the various Higgs decay channels, the processes $H \to b\bar{b}$ and $H \to WW^*$ receive contributions from top FCNC operators via single insertions at the one-loop level. On the other hand, the other visible-sector decays $H \to \gamma\gamma$ and $H \to \gamma Z$ and Higgs production via gluon fusion $(gg\to H)$ are affected only through double insertions of top FCNC operators.  In our analysis, we adopt the $\kappa$-framework to confront the NP scenario on these decay and production channels.
The ATLAS collaboration \cite{ATLAS:2022tnm,ATLAS:2022vkf} has recently reported measurements of various $\kappa_i$ coupling modifiers.
\begin{align}\label{eq:kappai}
    \kappa_{\gamma}=(1.02^{+0.08}_{-0.07})\,,\,\,\,\,\,\, \kappa_{Z\gamma}=1.38^{+0.31}_{-0.37}\,,\,\,\,\,\,\,\, \kappa_g=1.01^{+0.11}_{-0.09}\,.
\end{align}
  The detailed analytical expressions for the new physics modifications to these Higgs observables are provided in Appendix~\ref{sec:appHiggs}.
\subsubsection{Trilinear Gauge Couplings}
Trilinear gauge couplings (TGCs) describe the interactions among three gauge bosons, such as $WW\gamma$ and $WWZ$, and are a direct consequence of the non-Abelian structure of the electroweak gauge group in the SM \cite{Argyres:1992vv,Hagiwara:1986vm,Gounaris:1996rz}. Due to the high precision of TGC measurements at LEP \cite{ALEPH:2013dgf}, ATLAS \cite{ATLAS:2017pbb,ATLAS:2014ofc,ATLAS:2013way,ATLAS:2016zwm,ATLAS:2012mec} and CMS\cite{CMS:2015tmu,CMS:2013ryd,CMS:2016qth,CMS:2013ant}, particularly from diboson production channels, these observables provide stringent constraints on potential NP contributions. We include their effects in our analysis to constrain the parameter space of the relevant top FCNC effective operators. The general Lagrangian describing the triple gauge boson interactions involving $W^+W^-V$ ($V = \gamma, Z$) can be parameterized in a Lorentz-invariant form as:
\begin{align}\label{eq:TGC_couplings}
    \mathcal{L}_{WWV} &=i g_{WWV}\Bigg( g_1^V (W^+_{\mu \nu}W^{-\mu} W^{+\nu}-W^+_{\mu}W^-_{\nu}W^{\mu\nu})+ \kappa_V W^+_{\mu} W^-_{\nu} V^{\mu \nu} \nonumber\\
    &+\frac{\lambda_V}{M_W^2} W^+_{\lambda \mu}W^{-\mu}_{\nu} V^{\nu \lambda} + i g_4^V W_{\mu}^+ W^-_{\nu}(\partial^{\mu} V^{\nu}+ \partial^{\nu} V^{\mu}) -i g_5^V \epsilon^{\mu \nu \rho \sigma} (W_{\mu}^+ \stackrel{\leftrightarrow}{\partial}_{\rho} W_{\nu}^-) V_{\rho} \nonumber\\
    &+ \tilde{\kappa}_V W_{\mu}^+ W^-_{\nu} \tilde{V}^{\mu \nu} +\frac{\tilde{\lambda}_V}{M_W^2} W^+_{\lambda \mu} W^{-\mu} \tilde{V}^{\nu \lambda} \Bigg)\,.
\end{align}
Eq.~\eqref{eq:TGC_couplings} is parametrised by seven couplings. Among these, the couplings $g_1^V$, $\kappa_V$, and $\lambda_V$ conserve both charge $(C)$ and parity $(P)$ individually. The coupling $g_5^V$ violates $C$ and $P$ separately but preserves the combined $CP$ symmetry. On the other hand, the couplings $g_4^V$, $\tilde{\kappa}_V$, and $\tilde{\lambda}_V$ explicitly violate $CP$ and parameterise possible sources of $CP$ violation in the bosonic sector. Among these couplings, the SM at tree level predicts $g_1^Z=g_1^{\gamma}=\kappa_Z=\kappa_{\gamma}=1$, all the other couplings in eq.~\eqref{eq:TGC_couplings} are vanishing at tree level. The coupling $g_{WWV}$ takes the value $-i e$ for $g_{WW\gamma}$ and $-i e \cot{\theta_W}$ for $g_{WWZ}$, respectively.

\begin{figure}[t]
    \centering
\subfloat[]{
        \begin{tikzpicture}
          \begin{feynman}
            \vertex[](a1);
            \vertex[left=1.cm of a1](a2){\(W\)};
            \vertex[above right=1cm of a1](a3);
            \vertex[above right=1cm of a3](a4){\(W\)};
            \vertex[square dot, draw=red, fill=red, minimum size=4pt, inner sep=2.5pt,below right=1cm of a1](a5){};
            \vertex[below right=1.3cm of a5](a6){\(\gamma/Z\)};
            \diagram*{
                (a2) --[boson] (a1),
                (a3) --[boson](a4),
                (a5) --[red, very thick,boson](a6),
                (a5) --[fermion, very thick, edge label=\(t\)](a1) --[fermion, edge label=\(d_i\)](a3) --[fermion, very thick, edge label=\(u/c\)](a5),
            };
          \end{feynman}
        \end{tikzpicture}
    }
    \subfloat[]{
        \begin{tikzpicture}
          \begin{feynman}
            \vertex[](a1);
            \vertex[left=1.cm of a1](a2){\(W\)};
            \vertex[above right=1cm of a1](a3);
            \vertex[above right=1cm of a3](a4){\(W\)};
            \vertex[square dot, draw=red, fill=red, minimum size=4pt, inner sep=2.5pt,below right=1cm of a1](a5){};
            \vertex[below right=1.3cm of a5](a6){\(\gamma/Z\)};
            \diagram*{
                (a2) --[boson] (a1),
                (a3) --[boson](a4),
                (a5) --[red, very thick,boson](a6),
                (a5) --[fermion, very thick, edge label'=\(u/c\)](a3) --[fermion, edge label'=\(d_i\)](a1) --[fermion, very thick, edge label'=\(t\)](a5),
            };
          \end{feynman}
        \end{tikzpicture}
    }
    \caption{Feynman diagrams modifying $WW\gamma$ and $WWZ$ gauge couplings in the presence of top-FCNC operators.}
    \label{fig:WWV}
\end{figure}
These TGCs can be further parametrised in terms of momentum-dependent form factors. A detailed discussion of this parametrisation is provided in our previous work~\cite{Kala:2025srq}, and we follow the same methodology in the present analysis. The modification in different trilinear gauge vertices will be via the one-loop diagrams shown in fig.~\ref{fig:WWV}. In our analysis, the following TGCs will be impacted due to top-FCNC couplings
\begin{equation}\label{eq:tgci}
   g_i\in \{g_Z^1,\,\,\kappa_{\gamma},\,\, \lambda_{\gamma},\,\,\kappa_Z,\,\, \lambda_Z,\,\, g_Z^4,\,\, g^5_Z,\,\,\tilde{\kappa}_Z,\,\,\tilde{\lambda}_Z\,. \}
\end{equation}
The NP effects are parametrised through the shifts $\delta g_i$, such that
$g_i=g_i^{\rm SM}(1+\delta g_i)$. 

\subsubsection{Top Chromo-magnetic dipole moment}\label{sec:Top_CMDM}

\begin{figure}[h!]
    \centering
    \subfloat[]{
    \begin{tikzpicture}
        \begin{feynman}
        \vertex[](a1){\(t\)};
        \vertex[right=1cm of a1](a2);
        \vertex[right=1cm of a2](a3);
        \vertex[square dot,blue,right=1.cm of a3](a4){};
        \vertex[right=1cm of a4](a5){\(t\)};
        \vertex[below=1.3cm of a4](a6){\(g\)};
        \diagram* {
            (a1) --[fermion](a2) --[fermion, edge label'=\(d_i\)](a3) --[fermion, edge label'=\(u/c\)](a4) --[fermion](a5),
            (a4) --[thick,blue,gluon](a6),
            (a2) --[boson, half left, looseness=2, edge label = \(W\)](a3),
        };
        \end{feynman}
    \end{tikzpicture}
    }
\subfloat[]{
  \begin{tikzpicture}
      \begin{feynman}
          \vertex[](a1){\(t\)};
          \vertex[square dot,blue,right=1cm of a1](a2){};
          \vertex[right=1cm of a2](a3);
          \vertex[right=1cm of a3](a4);
          \vertex[right=1cm of a4](a5){\(t\)};
          \vertex[below =1.3 cm of a2](a6){\(g\)};
          \diagram* {
          (a1) --[fermion](a2) --[fermion,edge label'=\(u/c\)](a3) --[fermion, edge label'=\(d_i\)](a4) --[fermion](a5),
          (a3) --[boson, half left, looseness=2,edge label=\(W\)](a4),
          (a2) --[blue,thick,gluon](a6),
          };
      \end{feynman}
  \end{tikzpicture}}
    \caption{Magnetic dipole moment of top quarks affected by the top FCNC operators.}
    \label{fig:dipole_moment_top_quark}
\end{figure}
In the SM, the top quark chromo-magnetic dipole moment (CMDM) arises from radiative and QCD corrections to the $t\bar{t}g$ vertex and is predicted to be small. At the scale $q^2 = m_Z^2$, its SM value is estimated to be $\hat{\mu}_t^{\rm SM}(m_Z^2) = -0.0224$~\cite{Aranda:2020tox}, with a recent update including four-point vertex corrections yielding $\|\hat{\mu}_t^{\rm SM}(m_Z^2)\| = -0.0253$ \cite{Montano-Dominguez:2021zmy}. Experimentally, the CMS Collaboration reports $\hat{\mu}_t^{\rm Exp}(m_Z^2) = -0.024^{+0.013\,+0.016}_{-0.009\,-0.011}$~\cite{CMS:2019kzp}, which is in good agreement with the SM prediction. This makes the CMDM a sensitive and important observable for probing and constraining NP scenarios that may modify the top-gluon interaction.\\
The general Lagrangian that describes the effective $t\bar{t}g$ interaction can be written as:
\begin{align}\label{eq:top_EDM}
    \mathcal{L}_{t\bar{t}g}=-g_s \left(\bar{t}\gamma^{\mu}G_{\mu}t+i \frac{\hat{d}_t}{2 m_t}\bar{t}\sigma^{\mu\nu}\gamma_5 G_{\mu\nu}^a T^a t+\frac{\hat{\mu}_t}{2 m_t}\bar{t}\sigma^{\mu\nu}G_{\mu\nu}^a T^a t\right)
\end{align}
where $G_{\mu} \equiv G_{\mu}^a T^a$, where $G_{\mu}^a$ are the gluon fields and $T^a = \lambda^a/2$ $(a = 1, \cdots, 8)$ are the generators of $\mathrm{SU}(3)_C$. The gluon field strength tensor is defined as $G_{\mu\nu} = G_{\mu\nu}^a T^a$, where
$G_{\mu\nu}^a \equiv \partial_{\mu}G_{\nu}^a - \partial_{\nu}G_{\mu}^a - g_s f^{abc} G_{\mu}^b G_{\nu}^c,$ and $f^{abc}$ are the structure constants of $\mathrm{SU}(3)_C$. The CMDM and CEDM are defined as $\mu_t=g_s \hat{\mu}_t/m_t$ and $d_t= g_s \hat{d}_t/m_t$ respectively. In our analysis, top quark CMDM and CEDM will be impacted in the presence of NP interactions and the modifications will be via the one-loop level contributions as shown in the fig.~\ref{fig:dipole_moment_top_quark}.
Including the NP contributions, the effective CMDM or CEDM can be written as
\begin{align}
    \mu_t & =\frac{g_s}{m_t}\left(\hat{\mu}_t^{\rm SM}+\hat{\mu}_t^{\rm NP}\right),\\
    d_t & = \frac{g_s}{m_t}\left(\hat{d}_t^{\rm SM}+\hat{d}_t^{\rm NP}\right).
\end{align}
Here, $\hat{\mu}_t^{\rm NP}$ and $\hat{d}_t^{\rm NP}$ encode the NP effects, with its analytical form given in Appendix~\ref{sec:appTopCMDM} in eq.~\eqref{eq:Append_top_CMDM}. 

\subsection{Constraints from collider searches}\label{sec:appTopCMDM}

In this subsection, we derive bounds on the relevant effective parameters using current experimental upper limits on top-quark FCNC branching ratios. These limits originate from dedicated direct collider searches at the LHC, conducted by the ATLAS~\cite{ATLAS:2023qzr, ATLAS:2022per, ATLAS:2024mih, ATLAS:2021amo} and CMS~\cite{CMS:2023bjm, CMS:2017wcz} collaborations, which probe different top-FCNC processes $t \to u_j X$ ($X = g, \gamma, Z, H$) through single-top production channels and rare top-quark decays. The experimental upper bounds on these branching fractions can be straightforwardly translated into constraints on the corresponding top-quark FCNC WCs using eqs.~\eqref{eq:Branching_ratio_1}$-$\eqref{eq:Branching_ratio_4}. A summary of the most recent collider bounds is presented in table~\ref{tab:Branching_ratio_exp}. Crucially, in our subsequent global analysis, we explicitly incorporate these direct collider bounds alongside the indirect constraints from low-energy and electroweak precision measurements.
\begin{table}[h!]
	\centering
	\footnotesize
	\renewcommand{\arraystretch}{1.5}
	\begin{tabular}{|c|c|c|}
		\hline
		\rowcolor{gray!20}
		\textbf{Branching Ratio} 
		& \textbf{SM \cite{TopQuarkWorkingGroup:2013hxj}} 
		& \textbf{Experimental Bound} \\
		\hline\hline
		$\mathcal{B}(t \to \gamma c)$  & $5 \times 10^{-14}$   
		& \makecell[l]{$4.16 \times 10^{-5}$ (LH) \\ $4.16 \times 10^{-5}$ (RH) \cite{ATLAS:2022per}\\
			$1.51 \times 10^{-5}$ \cite{CMS:2023bjm}} \\ 
		\hline
		$\mathcal{B}(t \to \gamma u)$  & $4 \times 10^{-16}$   
		& \makecell[l]{$0.85 \times 10^{-5}$ (LH) \\ $1.2 \times 10^{-5}$ (RH) \cite{ATLAS:2022per}\\
			$0.95 \times 10^{-5}$ \cite{CMS:2023bjm}} \\ 
		\hline
		$\mathcal{B}(t \to Z c)$ & $1 \times 10^{-14}$   
		& \makecell[l]{$1.3 \times 10^{-4}$ (LH) \\ $1.2 \times 10^{-4}$ (RH) \cite{ATLAS:2023qzr}} \\ 
		\hline
		$\mathcal{B}(t \to Z u)$ & $7 \times 10^{-17}$   
		& \makecell[l]{$6.2 \times 10^{-5}$ (LH) \\ $6.6 \times 10^{-5}$ (RH) \cite{ATLAS:2023qzr}} \\ 
		\hline
		$\mathcal{B}(t \to H c)$ & $3 \times 10^{-15}$   
		& $3.4 \times 10^{-4}$ \cite{ATLAS:2024mih} \\ 
		\hline
		$\mathcal{B}(t \to H u)$& $2 \times 10^{-17}$   
		& $2.8 \times 10^{-4}$ \cite{ATLAS:2024mih} \\ 
		\hline
		$\mathcal{B}(t \to g c)$& $5 \times 10^{-12}$   
		& $3.7 \times 10^{-4}$ \cite{ATLAS:2021amo} \\ 
		\hline
		$\mathcal{B}(t \to g u)$& $4 \times 10^{-14}$   
		& $6.1 \times 10^{-5}$ \cite{ATLAS:2021amo} \\ 
		\hline
	\end{tabular}
	\caption{Summary of SM predictions and current experimental upper bounds on top-quark FCNC branching ratios.}
	\label{tab:Branching_ratio_exp}
\end{table}

The branching ratios of the various top-FCNC processes can be expressed in terms of the NP effective couplings as follows:
\begin{subequations}\label{eq:LHC_bound_to_NP_couplings}
	\begin{align}\label{eq:Branching_ratio_1}
		\mathcal{B}\left(t \to u_j  g\right)&= C_F \frac{m_t}{16 \pi \Gamma_t}\bigg[\left|\xi_L^{u_jt}\right|^2+\left|\xi_R^{u_jt}\right|^2\bigg]\,,
	\end{align}
	
	\begin{align}\label{eq:Branching_ratio_2}
		\mathcal{B}\left(t \to u_j  \gamma\right)&= e^2 \frac{m_t}{16 \pi \Gamma_t}\bigg[\left|\lambda_L^{u_jt}\right|^2+\left|\lambda_R^{u_jt}\right|^2\bigg]\,,
	\end{align}
	
	\begin{align}\label{eq:Branching_ratio_3}
		\mathcal{B}\left(t \to u_j Z\right)&=\frac{g_W^2 m_t}{128 \pi c_W^2 \Gamma_t}\left(1-\frac{M_Z^2}{m_t^2}\right)^2 \Bigg[\left(2+\frac{m_t^2}{M_Z^2}\right)\left(\left|X_L^{u_jt}\right|^2+\left|X_R^{u_jt}\right|^2\right) \nonumber\,,\\
		&+4\left(2+\frac{M_Z^2}{m_t^2}\right)\left(\left|\kappa_L^{u_jt}\right|^2+\left|\kappa_R^{u_jt}\right|^2\right)\,,\\&+6\left(X_L^{u_jt} \kappa_R^{*\,u_jt}+X_L^{*\,u_jt} \kappa_R^{u_jt}+X_R^{u_jt} \kappa_L^{*\,u_jt}+X_R^{*\,u_jt} \kappa_L^{u_jt}\right)\Bigg]\,,\nonumber
	\end{align}
	
	\begin{align}\label{eq:Branching_ratio_4}
		\mathcal{B}\left(t \to u_j H\right)&=\frac{m_t}{64 \pi \Gamma_t}\left(1-\frac{M_H^2}{m_t^2}\right)^2 \left(\left|\eta_L^{u_jt}\right|^2+\left|\eta_R^{u_jt}\right|^2\right)\,.
	\end{align}
	
\end{subequations}
Here, $\Gamma_t$ denotes the total width of the top quark, whose 
experimentally measured value is $\Gamma_t = 1.42^{+0.19}_{-0.15}~
\mathrm{GeV}$~\cite{ParticleDataGroup:2024cfk}, and $C_F = 4/3$ is the 
colour factor. Since current direct collider searches only provide upper bounds on the top-quark FCNC branching ratios, we account for these constraints in our global analysis by employing an appropriate half-Gaussian penalty function. The precise statistical implementation of this penalty in the $\chi^2$ likelihood is discussed in detail in the subsequent section.

 We can see from these expressions that the individual chiral components cannot be disentangled from current collider data in a model-independent way. Only in limiting scenarios, where one of the chiral couplings is set to zero, do the collider bounds translate into meaningful constraints on the remaining coupling. However, even in such cases, if the couplings are treated as complex, the experimental limits remain sensitive only to their absolute magnitude. These limitations underline the restricted constraining power of current collider bounds and motivate the need for complementary indirect probes. While these limits are valuable in constraining the overall size of the effective couplings, they are primarily sensitive only to the magnitude of the amplitudes and do not allow for a simultaneous and independent extraction of the real and imaginary components of the WCs.

\section{Combined Analysis Framework} \label{sec:Global_Analysis}

In the previous section \ref{sec:List_of_obs}, we discussed all the relevant observables included in our analysis. For convenience, we have provided a comprehensive summary of these observables in table~\ref{tab:observables}, grouped by their respective physical sectors. 
\begin{table}[t]
    \centering
    \small
    \renewcommand{\arraystretch}{1.3}
    \begin{tabular}{@{} >{\raggedright\arraybackslash}p{2.8cm} >{\raggedright\arraybackslash}p{4.2cm} >{\raggedright\arraybackslash}p{8cm} @{}}
        \toprule
        \rowcolor{gray!20}
        \textbf{Class} & \textbf{Category} & \textbf{Observables} \\
        \midrule
        
        \textbf{Low-Energy FCNC} 
        & $b \to s\ell\ell$ (with $\ell=\mu, e$) decays
        & Angular observables \cite{LHCb:2015svh, LHCb:2020lmf}, differential branching fractions $d\mathcal{B}/dq^2$~\cite{LHCb:2016ykl, CMS:2024syx}, LFUV ratios: $R_K$, $R_{K^*}$ \cite{LHCb:2022qnv}, CP Asymmetries \cite{LHCb:2014mit} \\
        \addlinespace
        & Radiative decays
        & $\mathcal{B}(B \to X_s\gamma)$~ \cite{LHCb:2014mit}, $\mathcal{B}(B_{s,d} \to V\gamma)$~\cite{Belle:2017hum,Belle-II:2024tru,Belle:2014sac,HeavyFlavorAveragingGroupHFLAV:2024ctg}, CP Asymmetries~\cite{Belle-II:2024kbc,Belle:2018iff,Belle-II:2024uzp}, $S_{K^*\gamma}$ \\
        \addlinespace
        & Rare/Invisible decays
        & $\mathcal{B}(B_{s,d} \to \mu^+\mu^-)$~\cite{CMS:2022mgd,LHCb:2021vsc}, $\mathcal{B}(K_L \to \mu^+\mu^-)$~\cite{ParticleDataGroup:2024cfk}, $\mathcal{B}(B \to K^{(*)}\nu\bar{\nu})$ ~\cite{Belle-II:2023esi}, $\mathcal{B}(K^+ \to \pi^+\nu\bar{\nu})$~\cite{NA62:2021zjw} \\
        \addlinespace
        & Neutral meson mixing amplitudes 
        & $\Delta M_{d(s)}$~\cite{HeavyFlavorAveragingGroupHFLAV:2024ctg} \\
        \midrule
        
        \textbf{Low-Energy FCCC} 
        & Semileptonic/Leptonic decays: $b\to c (u)\ell\nu$ ($\ell = \mu, \tau$)
        & $\mathcal{B}(P \to M\ell\nu)$, $\mathcal{B}(P \to \ell\nu)$~\cite{HeavyFlavorAveragingGroupHFLAV:2024ctg, ParticleDataGroup:2024cfk}, LFUV ratios: $R(D)$, $R(D^*)$~\cite{Belle-II:2025yjp} \\
        \midrule
        
        \textbf{$tWb$ Vertex} 
        & Anomalous couplings 
        & $|f_L V_{tb}|$~\cite{ATLAS:2024ojr,ATLAS:2024ppp,ATLAS:2023qrn}, $V_R$, $g_L$, $g_R$~\cite{CMS:2020ezf} \\
        \midrule
        
        \textbf{EW Precision} 
        & Global observables: W and Z-pole 
        & $S$, $T$, $U$. Other observables are listed in Tables \ref{tab:Z_pole} and \ref{tab:W_pole}, respectively~\cite{ALEPH:2005ab, Janot:2019oyi, dEnterria:2020cgt, SLD:2000jop, ParticleDataGroup:2024cfk} \\
        \midrule
        
        \textbf{Higgs Physics} 
        & Higgs decays/production 
        & $\Gamma(H\to b\bar{b}, WW^*, \gamma\gamma, Z\gamma$)~\cite{ParticleDataGroup:2024cfk,ATLAS:2023dnm,CMS:2022ley,CMS:2023vzh,ATLAS:2022tnm,ATLAS:2022vkf} \\
        \addlinespace
        & 
        & $\kappa_i$ variables: $\kappa_g$, $\kappa_\gamma$, $\kappa_{Z\gamma}$ (eq.~\ref{eq:kappai}) \\
        \midrule
        
        \textbf{Gauge \& Top} 
        & TGCs 
        & All the couplings listed in eq.~\ref{eq:tgci} \cite{ALEPH:2013dgf} \\
        \addlinespace
        & Dipole moments 
        & CMDM $(\mu_t)$, CEDM $(d_t)$~\cite{CMS:2019kzp} \\
        \midrule
        
        \textbf{EDM} 
        & Nucleon 
        & Neutron EDM $(d_n)$~\cite{Abel:2020pzs} \\
        \midrule
        \textbf{LHC Input} & Top FCNC & Branching fraction $(\mathcal{B})$~\cite{ATLAS:2023qzr, ATLAS:2022per, ATLAS:2024mih, ATLAS:2021amo, CMS:2023bjm, CMS:2017wcz}\\
        \bottomrule
    \end{tabular}
    \caption{Summary of observables included in the global analysis, grouped by physical sector.}
    \label{tab:observables}
\end{table}
To extract robust constraints on the NP parameters, we perform a global $\chi^2$ analysis. We denote the set of relevant top-FCNC effective couplings, evaluated at the electroweak scale ($\mu_{\rm EW}$), collectively as the vector $\vec{\tilde{g}}$. The total $\chi^2$ function is constructed as the sum of a standard contribution from precise measurements and a penalty contribution from experimental upper limits:
\begin{align} \label{eq:chi_sq_fun}
    \chi^2_{\text{total}}(\vec{\tilde{g}}) &= \chi^2_{\text{data}}(\vec{\tilde{g}}) + \chi^2_{\text{penalty}}(\vec{\tilde{g}})\,.
\end{align}

For standard measurements, the theoretical predictions, denoted $\mathcal{O}^{\rm theo}(\vec{\tilde{g}})$, are confronted with the experimental data using the covariance matrix:
\begin{align}\label{eq:chi_sq_data}
	\chi^2_{\text{data}}(\vec{\tilde{g}})
	&=
	\sum_{ij}
	\left(\mathcal{O}_i^{\rm theo}(\vec{\tilde{g}})-\mathcal{O}_i^{\rm exp}\right)
	\left(\sigma^2\right)^{-1}_{ij}
	\left(\mathcal{O}_j^{\rm theo}(\vec{\tilde{g}})-\mathcal{O}_j^{\rm exp}\right)\,,
\end{align}
where $\mathcal{O}^{\rm exp}$ are the central experimental values, and $(\sigma^2)_{ij}$ represents the elements of the total covariance matrix, constructed as the sum of the experimental and theoretical covariance matrices. 

For observables where only experimental upper bounds are available, we incorporate them into the likelihood using Gaussian penalty functions, following standard statistical methodologies for likelihood-based tests of NP \cite{Cowan:2010js, GAMBIT:2017yxo}. We divide these bounds into two distinct classes based on their physical domains, such that 
\begin{equation}
\chi^2_{\text{penalty}}(\vec{\tilde{g}}) = \chi^2_{\rm collider}(\vec{\tilde{g}}) + \chi^2_{\rm EDM}(\vec{\tilde{g}}).
\end{equation}
For rare top-quark FCNC decays ($t \to u_j X$), the SM background is effectively zero, as it is suppressed by at least ten orders of magnitude relative to current experimental sensitivities. Furthermore, the theoretical branching ratio is strictly positive-definite ($\mathcal{{B}}^{\text{theo}} \ge 0$). Therefore, we model these limits using a one-sided (half) Gaussian penalty function. By defining the variance such that the penalty reaches the requisite confidence level (CL) threshold ($\Delta\chi^2_{\text{CL}}$) exactly at the reported experimental upper limit ($\mathcal{B}^{\text{limit}}$), the constraint simplifies to:
\begin{align} \label{eq:chi_square_collider}
    \chi^2_{\rm collider}(\vec{\tilde{g}}) &= \sum_{k} \Delta\chi^2_{\text{CL},k} \left( \frac{\mathcal{B}_k^{\rm theo}(\vec{\tilde{g}}) - \mathcal{B}_k^{\text{SM}}}{\mathcal{B}_k^{\text{limit}}} \right)^2 \Theta\left(\mathcal{B}_k^{\rm theo}(\vec{\tilde{g}}) - \mathcal{B}_k^{\text{SM}}\right) \,,
\end{align}
where the Heaviside step function $\Theta(x)$ mathematically enforces that theoretical predictions falling at or below the SM expectation are not penalised as well as also take care of the fact $(\mathcal{B}^{\rm theo}\ge 0)$.

However, theoretical predictions for EDMs can be either positive or negative, as their sign depends on the phases of the imaginary WCs. Experimental constraints are therefore placed on the magnitude of the EDM ($\vert{}d^{\rm theo}\vert{} < d^{\text{limit}}$). Because the SM expectation for the neutron EDM is vanishingly small ($\sim 10^{-31} \, e\cdot\text{cm}$) compared to current bounds, we can safely set the central background value to zero. Consequently, the EDM contribution is implemented as a continuous, symmetric two-sided Gaussian penalty centered at the origin:
\begin{align} \label{eq:chi_square_edm}
    \chi^2_{\rm EDM}(\vec{\tilde{g}}) &= \sum_{m} \Delta\chi^2_{\text{CL},m} \left( \frac{d_m^{\rm theo}(\vec{\tilde{g}})}{d_m^{\text{limit}}} \right)^2 \,.
\end{align}

It is important to note that the scaling factor $\Delta\chi^2_{\text{CL}}$ differs between these two classes of observables because the respective collaborations quote their limits at different confidence levels. For top-quark FCNC branching ratios, limits are universally reported at the $95\%$ CL, corresponding to a threshold of $\Delta\chi^2_{\text{CL}} = 3.84$ (for one degree of freedom). Conversely, EDM upper limits are traditionally reported at the $90\%$ CL, which requires setting the threshold to $\Delta\chi^2_{\text{CL}} = 2.71$ in eq.~\eqref{eq:chi_square_edm}.

As will be discussed below, we extract the top-FCNC couplings and the SMEFT WCs by performing two separate $\chi^2$ minimisations. In both cases, the structure of the $\chi^2$ function remains identical to that defined above. However, the interpretation of the parameters differs between the two analyses. In one case, the parameters $\vec{\tilde{g}}_i$ correspond to the effective couplings defined in eq.~\ref{eq:eff_vertex_top_FCNC}, whereas in the other case they represent the SMEFT WCs $(\vec{\mathcal{C}})$.

By minimising the total $\chi^2_{\text{total}}$ function with respect to the effective couplings $\vec{\tilde{g}}$, we determine their best-fit values. We present our results in sec.~\ref{sec:Result}, showing all effective couplings and SMEFT WCs at different energy scales $(\Lambda_i)$, obtained by numerically solving the complete set of RGEs. All results are quoted at the $1 ~\sigma$ CL.

\section{Results}\label{sec:Result}
In this section, we present the numerical results of the global $\chi^2$ analysis framework defined in section~\ref{sec:Global_Analysis}. By confronting our theoretical predictions with the comprehensive set of observables summarised previously, we determine the allowed parameter space for both the anomalous effective couplings and the underlying SMEFT WCs.

In general, we assume that all dipole- and scalar-type effective couplings are complex quantities, thereby allowing for potential CP-violating phases. The extracted values for the various top-FCNC effective couplings are summarised in Tables~\ref{tab:top_gluon_FCNC_global}, \ref{tab:top_photon_FCNC_global}, \ref{tab:top_Z_FCNC_global}, and \ref{tab:top_Higgs_FCNC_Global}. These tables report the best-fit values and the corresponding allowed $1~\sigma$ ranges for the top-gluon, top-photon, top-$Z$, and top-Higgs FCNC interactions, respectively. To systematically explore the parameter space, our analysis considers several benchmark scenarios:
\begin{enumerate}
	\item \textbf{Single-Chirality Case:} Only one chiral structure (either left-handed or right-handed) is assumed to be present.
	\item \textbf{Dual-Chirality Case:} Both left- and right-handed couplings are allowed simultaneously. 
	\item \textbf{Real Coupling Case:} A special case where all couplings are assumed to be real-valued.
\end{enumerate}

All extracted coefficients and couplings are defined at the electroweak scale, $\mu_{\rm EW}$, which is the natural scale for processes involving the top quark and heavy electroweak bosons. However, for low-energy flavour observables, such as those relevant to $B$-meson decays, the appropriate scale is $\mu_b \sim m_b$. A consistent analysis requires translating the high-scale couplings to the low-energy scale or vice versa by using the RGE framework discussed earlier in section~\ref{sec:methodology}. This procedure ensures that predictions for low-energy observables, such as branching ratios of rare $B$ decays or meson mixing parameters, are computed using WCs evaluated at the correct scale. Without this step, one would risk introducing large theoretical uncertainties or inconsistencies between different sectors of the analysis. In summary, the connection between $\mu_{\rm EW}$ and $\mu_b$ through RGE evolution is essential for maintaining theoretical consistency.

\begin{table}[t!]
	\centering
	\footnotesize
	\renewcommand{\arraystretch}{1.5}
	\begin{tabular}{|c|c|c|c|}
		\hline
		\rowcolor{gray!20}
		\multicolumn{1}{|c|}{\textbf{Scenario}} & \multicolumn{1}{c|}{\textbf{Values $\times 10^2$}} &
		\multicolumn{1}{c|}{\textbf{Scenario }} & \multicolumn{1}{c|}{\textbf{Values $\times 10^2$}} \\
		\hline
		\hline
		$\mathrm{Re}(\xi_L^{ct})$, $\mathrm{Im}(\xi_L^{ct})$ 
		& $(0.06 \pm 1.98)$, $(-0.19 \pm 1.28)$
		& $\mathrm{Re}(\xi_L^{ut})$,$\mathrm{Im}(\xi_L^{ut})$ 
		& $(0.05 \pm 0.83)$, $(0.0 \pm 1.46)$ \\
		\hline
		\hline
		$\mathrm{Re}(\xi_R^{ct})$, $\mathrm{Im}(\xi_R^{ct})$ 
		& $(0.0 \pm 0.65)$, $(0.52 \pm 0.40)$ 
		& $\mathrm{Re}(\xi_R^{ut})$, $\mathrm{Im}(\xi_R^{ut})$  
		& $(-0.10 \pm 0.36 )$, $(0.0 \pm 0.79)$ \\
		\hline
		\hline
		$\mathrm{Re}(\xi_L^{ct})$ 
		& $(-0.12 \pm 2.00)$ 
		&
		$\mathrm{Re}(\xi_L^{ut})$ 
		& $(0.03 \pm 1.58)$ \\
		\hline
		\hline
		$\mathrm{Re}(\xi_R^{ct})$ 
		&  $(0.26 \pm 0.68)$ 
		&
		$\mathrm{Re}(\xi_R^{ut})$ 
		&  $(-0.10 \pm 0.39)$ \\
		\hline
		\hline
		$\mathrm{Re}\,(\xi_L^{ct})$, $\mathrm{Im}\,(\xi_L^{ct})$\,, & $(0.0 \pm 0.75)$, $(0.0 \pm 0.76)$,& $\mathrm{Re}\,(\xi_L^{ut})$, $\mathrm{Im}\,(\xi_L^{ut})$, & $(-0.06 \pm 0.55)$, $(0.03 \pm 0.60)$ , \\
		$\mathrm{Re}\,(\xi_R^{ct})$, $\mathrm{Im}\,(\xi_R^{ct})$ .& $(0.0\pm 0.68)$, $(0.56 \pm 0.38)$ . & $\mathrm{Re}\,(\xi_R^{ut})$, $\mathrm{Im}\,(\xi_R^{ut})$ . & $(-0.09 \pm 0.44)$, $(0.01 \pm 0.72)$ .\\
		\hline 
	\end{tabular}
	\caption{Extracted values of the complex top-gluon FCNC couplings $\xi_{L,R}^{u_j\,t}$ ($u_j=u,c$) from the global fit only, showing real and imaginary components under different scenarios.}
	\label{tab:top_gluon_FCNC_global}
\end{table}
\begin{table}[h!]
	\centering
	\footnotesize
	\renewcommand{\arraystretch}{1.5}
	\begin{tabular}{|c|c|c|c|}
		\hline
		\rowcolor{gray!20}
		\multicolumn{1}{|c|}{\textbf{Scenario}} & \textbf{Values $\times 10^2$} &
		\multicolumn{1}{c|}{\textbf{Scenario}} & \textbf{Values $\times 10^2$} \\
		\hline 
		\hline
		$\mathrm{Re}(\lambda_L^{ct})$, $\mathrm{Im}(\lambda_L^{ct})$ 
		& $(-0.17 \pm 0.81)$, $(-0.04 \pm 1.29)$
		& $\mathrm{Re}(\lambda_L^{ut})$, $\mathrm{Im}(\lambda_L^{ut})$
		& $(0.04 \pm 2.06)$, $(-0.01 \pm 3.19)$ \\
		\hline
		\hline
		$\mathrm{Re}(\lambda_R^{ct})$, $\mathrm{Im}(\lambda_R^{ct})$
		& $(-0.01 \pm 0.58)$, $(0.07 \pm 0.30)$
		& $\mathrm{Re}(\lambda_R^{ut})$, $\mathrm{Im}(\lambda_R^{ut})$
		& $(-0.17 \pm 0.47)$, $(-0.02 \pm 2.26)$ \\
		\hline
		\hline
		$\mathrm{Re}(\lambda_L^{ct})$
		& $(0.20 \pm 0.69)$
		&
		$\mathrm{Re}(\lambda_L^{ut})$
		& $(0.03 \pm 2.81)$\\
		\hline
		\hline
		$\mathrm{Re}(\lambda_R^{ct})$
		&  $(-0.08 \pm 0.17)$
		&
		$\mathrm{Re}(\lambda_R^{ut})$
		&  $(0.17 \pm 0.44)$ \\
		\hline
		\hline
		\makecell[l]{$\mathrm{Re}(\lambda_L^{ct})$, $\mathrm{Im}(\lambda_L^{ct})$ ,\\
			$\mathrm{Re}(\lambda_R^{ct})$, $\mathrm{Im}(\lambda_R^{ct})$.}
		&
		\makecell[l]{$(0.0 \pm 0.46)$, $(-0.03 \pm 0.45)$,\\
			$(-0.16 \pm 1.62)$, $(0.49 \pm 0.55)$.}
		&
		\makecell[l]{$\mathrm{Re}(\lambda_L^{ut})$, $\mathrm{Im}(\lambda_L^{ut})$ ,\\
			$\mathrm{Re}(\lambda_R^{ut})$, $\mathrm{Im}(\lambda_R^{ut})$.}
		&
		\makecell[l]{$(0.01 \pm 0.63)$, $(-0.03 \pm 0.62)$,\\
			$(-0.16 \pm 1.40)$, $(0.17 \pm 1.28)$.}\\
		\hline
	\end{tabular}
	\caption{Constraints on top-photon FCNC dipole couplings $\lambda_{L,R}^{qt}$ obtained from the global fit, showing real and imaginary components under different scenarios.}
	\label{tab:top_photon_FCNC_global}
\end{table}

\begin{table}[t!]
	\centering
	\footnotesize
	\renewcommand{\arraystretch}{1.5}
	\begin{tabular}{|c|c|c|c|}
		\hline
		\rowcolor{gray!20}
		\multicolumn{1}{|c|}{\textbf{Scenario}} & \textbf{Values $\times 10^2$} &
		\multicolumn{1}{c|}{\textbf{Scenario}} & \textbf{Values $\times 10^2$} \\
		\hline
		\hline
		
		$\mathrm{Re}(\kappa_L^{ct})$, $\mathrm{Im}(\kappa_L^{ct})$
		& $(0.21\pm 1.68)$, $(-0.07 \pm 2.54)$
		& $\mathrm{Re}(\kappa_L^{ut})$, $\mathrm{Im}(\kappa_L^{ut})$
		& $(0.17 \pm 1.03)$, $(0.03 \pm 1.71)$ \\
		\hline
		\hline
		$\mathrm{Re}(\kappa_R^{ct})$, $\mathrm{Im}(\kappa_R^{ct})$
		& $(0.22 \pm 1.48)$, $(0.11 \pm 2.20)$
		& $\mathrm{Re}(\kappa_R^{ut})$, $\mathrm{Im}(\kappa_R^{ut})$
		& $(-0.21 \pm 0.72)$, $(-0.05 \pm 1.33)$ \\
		\hline
		\hline
		$\mathrm{Re}(\kappa_L^{ct})$
		& $(0.09 \pm 3.48)$
		&
		$\mathrm{Re}(\kappa_L^{ut})$
		& $(0.17 \pm 1.03)$\\
		\hline
		\hline
		$\mathrm{Re}(\kappa_R^{ct})$
		&  $(0.18 \pm 1.68)$
		&
		$\mathrm{Re}(\kappa_R^{ut})$
		&  $(-0.27 \pm 0.93)$ \\
		\hline
		\hline
		$\mathrm{Re}(X_L^{ct})$
		& $(0.41 \pm 1.31)$
		&
		$\mathrm{Re}(X_L^{ut})$
		& $(0.45\pm 1.14)$ \\
		\hline
		\hline
		$\mathrm{Re}(X_R^{ct})$
		&  $(-0.22 \pm 2.63)$
		&
		$\mathrm{Re}(X_R^{ut})$
		&  $(-0.23\pm 2.46)$ \\
		\hline
		\hline
		\makecell[l]{$\mathrm{Re}(\kappa_L^{ct})$, $\mathrm{Im}(\kappa_L^{ct})$,\\
			$\mathrm{Re}(\kappa_R^{ct})$, $\mathrm{Im}(\kappa_R^{ct})$.}
		&
		\makecell[l]{$(0.06 \pm 5.99)$, $(0.02 \pm 7.30)$,\\
			$(0.03 \pm 7.16)$, $(-0.04 \pm 6.71)$}
		&
		\makecell[l]{$\mathrm{Re}(\kappa_L^{ut})$, $\mathrm{Im}(\kappa_L^{ut})$,\\
			$\mathrm{Re}(\kappa_R^{ut})$, $\mathrm{Im}(\kappa_R^{ut})$.}
		&
		\makecell[l]{ $(0.08 \pm 1.15)$,$(-0.16 \pm 0.99)$ \\
			$(-0.16 \pm 0.84)$, $(0.06 \pm 1.16) $}
		\\
		\hline
		\hline
		$\mathrm{Re}(X_L^{ct})$, $\mathrm{Re}(X_R^{ct})$
		& $(-0.43 \pm 1.29)$, $(-0.32 \pm 1.57)$
		& $\mathrm{Re}(X_L^{ut})$, $\mathrm{Re}(X_R^{ut})$
		& $(-0.09 \pm 1.13)$, $(-0.30 \pm 0.97)$ \\
		\hline
	\end{tabular}
	\caption{Fitted values of complex top–$Z$ boson dipole couplings $(\kappa_{L,R}^{u_j\,t})$ and vector couplings $(X_{L,R}^{u_j\,t})$ from the global fit, showing real and imaginary components under different scenarios.}
	\label{tab:top_Z_FCNC_global}
\end{table}
Apart from $X_L^{c(u)t}$, the collider limits on all left-chiral couplings are generally stronger than those obtained from the combined analysis. For $X_L^{c(u)t}$, however, the bounds from the combined analysis are comparable to the collider limits. For the right-chiral couplings, we obtain an improvement of a few orders of magnitude for $|\lambda_R^{ct}|$ relative to collider bounds, while the improvement for $|\lambda_R^{ut}|$ is marginal. The bound on $|\xi_R^{ct}|$ is also comparable to the collider constraint. For the remaining right-chiral couplings, collider bounds remain at least an order of magnitude stronger. We obtain similar observation for the one-operator scenarios with real coupling. It is also instructive to note that, for the couplings where the LHC bounds are stronger, their impact is largely confined to trimming the extremal regions of the parameter space. The bulk of the parameter space allowed by our global fit remains unaffected, with the collider-allowed regions lying well within it.
\subsection{Summary of the most sensitive probes}\label{subsec:top_coupling_fit}

While collider searches constrain only the modulus of the left- and right-chiral FCNC couplings through upper limits on decay rates, their inclusion as one-sided Gaussian likelihoods in the global $\chi^2$ analysis provides valuable complementary information. In particular, collider observables restrict the overall size of a coupling, $ |C|^2=[\mathrm{Re}(C)]^2+[\mathrm{Im}(C)]^2$, whereas low-energy observables and EWPOs are sensitive to specific combinations of its real and imaginary components. Consequently, the low-energy fit may exhibit parameter degeneracies or flat directions that allow relatively large values of individual coupling components while remaining consistent with the measured observables. By limiting the overall magnitude of the couplings, the collider constraints reduce these degeneracies and significantly shrink the allowed parameter space. The combined fit therefore yields more stringent bounds on both the real and imaginary parts of the couplings than those obtained from low-energy and electroweak precision data alone, even though collider observables do not directly distinguish between these components. 
	
In particular, the CP-violating imaginary components are largely inaccessible to collider observables, which predominantly probe the overall strength of the FCNC interactions through limits on decay branching fractions and are therefore only weakly sensitive to the underlying complex phases. The inclusion of flavour and electroweak precision observables in the global fit provides direct sensitivity to these phases, leading to stringent constraints on the imaginary parts of the couplings. Consequently, CP-violating contributions that would otherwise remain poorly constrained in a collider-only analysis are significantly restricted in the combined fit.
	
It is evident from tables~\ref{tab:top_gluon_FCNC_global}, \ref{tab:top_photon_FCNC_global}, \ref{tab:top_Z_FCNC_global}, and \ref{tab:top_Higgs_FCNC_Global} that the constraints on the real and imaginary parts of the couplings are found to be of comparable strength. This demonstrates that the current data set could provide a stable and robust global fit in the full complex parameter space. Furthermore, we can obtain strong and comparable bounds on all the left and right chiral complex and real couplings independently and combinely. In the below items, we highlight the most sensitive probes of each set of couplings.  
 
 \begin{table}[t!]
 	\centering
 	\footnotesize
 	\renewcommand{\arraystretch}{1.5}
 	\begin{tabular}{|c|c|c|c|}
 		\hline
 		\rowcolor{gray!20}
 		\multicolumn{1}{|c|}{\textbf{Scenario}} & \textbf{Values $\times 10^2$} &
 		\multicolumn{1}{c|}{\textbf{Scenario}} & \textbf{Values $\times 10^2$} \\
 		\hline
 		\hline
 		
 		$\mathrm{Re}(\eta_L^{ct})$, $\mathrm{Im}(\eta_L^{ct})$
 		& $(-0.91 \pm 6.38)$, $(1.59 \pm 4.92)$
 		&
 		$\mathrm{Re}(\eta_L^{ut})$, $\mathrm{Im}(\eta_L^{ut})$
 		& $(1.28 \pm 4.71)$, $(1.30 \pm 4.68)$ \\
 		\hline
 		\hline
 		$\mathrm{Re}(\eta_R^{ct})$, $\mathrm{Im}(\eta_R^{ct})$
 		& $(-0.70 \pm 6.26)$, $(1.66 \pm 4.61)$
 		&
 		$\mathrm{Re}(\eta_R^{ut})$, $\mathrm{Im}(\eta_R^{ut})$
 		& $(-1.21 \pm 4.08)$, $(0.0\pm 8.74)$ \\
 		\hline
 		\hline
 		$\mathrm{Re}(\eta_L^{ct})$
 		& $(-1.67 \pm 4.43)$	
 		&
 		$\mathrm{Re}(\eta_L^{ut})$
 		& $(-0.89 \pm 6.82)$\\
 		\hline
 		\hline
 		$\mathrm{Re}(\eta_R^{ct})$
 		&  $(-0.83\pm 7.92)$
 		&
 		$\mathrm{Re}(\eta_R^{ut})$
 		&  $(-1.21 \pm 4.08)$ \\
 		\hline
 		\hline
 		\makecell[l]{$\mathrm{Re}(\eta_L^{ct})$, $\mathrm{Im}(\eta_L^{ct})$,\\
 			$\mathrm{Re}(\eta_R^{ct})$, $\mathrm{Im}(\eta_R^{ct})$}
 		&
 		\makecell[l]{$(0.11 \pm 7.34)$, $(0.24 \pm 7.30)$,\\
 			$(1.72 \pm 4.22)$, $(0.09 \pm 7.34)$}
 		&
 		\makecell[l]{$\mathrm{Re}(\eta_L^{ut})$, $\mathrm{Im}(\eta_L^{ut})$,\\
 			$\mathrm{Re}(\eta_R^{ut})$, $\mathrm{Im}(\eta_R^{ut})$}
 		&
 		\makecell[l]{$(0.08 \pm 6.44)$, $(-0.52 \pm 6.17)$,\\
 			$(-1.27 \pm 4.07)$, $(-0.88 \pm 5.65)$}
 		\\
 		\hline
 	\end{tabular}
 	\caption{Global fit constraints on top–Higgs FCNC couplings $\eta_{L,R}^{qt}$ under various scenarios.}
 	\label{tab:top_Higgs_FCNC_Global}
 \end{table}
 
\begin{itemize}
    \item \textbf{$\xi_L^{u_j t},~\xi_R^{u_j t}$:} From Table~\ref{tab:top_gluon_FCNC_global}, it is evident that the constraints on the real and imaginary parts of the right-chiral couplings are nearly an order of magnitude stronger than those on the left-chiral couplings. The bounds on $|\xi_L^{u_j t}|$ are driven mainly by direct collider searches, with the most stringent limits arising from the experimental upper bounds on the FCNC branching fractions $\mathcal{B}(t\to u_j g)$. In contrast, the much stronger constraints on $\xi_R^{u_j t}$ originate predominantly from low-energy flavour observables, particularly radiative $B$-meson decays, where the renormalisation-group mixing of the chromomagnetic dipole operator $O_8$ into the electromagnetic dipole operator $O_7$ significantly enhances the sensitivity to right-handed couplings. Notably, $\mathrm{Im}(\xi_R^{ct})$ exhibits a preference for a non-zero value at the $1\sigma$ level, driven primarily by the direct CP asymmetry $A_{\mathcal{CP}}(B^+\to X_s\gamma)$. While currently subleading, contributions to the effective Higgs-gluon coupling ($\delta_{gg}$) and anomalous $Wtb$ interactions provide complementary constraints and may become increasingly relevant as experimental precision improves.

    \item \textbf{$\lambda_L^{u_j t},~\lambda_R^{u_j t} :$} From Table~\ref{tab:top_photon_FCNC_global}, we observe that the pattern of constraints on the top-photon FCNC couplings is qualitatively similar to that obtained for the top-gluon FCNC interactions, exhibiting a pronounced chiral hierarchy between the left- and right-handed sectors. In particular, the right-handed couplings, $|\lambda_R^{u_j t}|^2$, are constrained at the level of $\mathcal{O}(10^{-6})$, while the corresponding left-handed couplings, $|\lambda_L^{u_j t}|^2$, remain at the level of $\mathcal{O}(10^{-5})$, indicating an approximately one-order-of-magnitude stronger sensitivity to the former. The dominant constraints on $\lambda_R^{u_j t}$ arise from flavour observables, especially the inclusive radiative decay branching fraction $\mathcal{B}(B \to X_s \gamma)$, with additional sensitivity provided by exclusive radiative $B$-meson decays and loop-induced modifications of the effective Higgs-to-diphoton coupling. In contrast, these low-energy observables are only weakly sensitive to the left-handed couplings, leading to comparatively weaker indirect bounds. Consequently, direct collider searches become essential for constraining $\lambda_L^{u_j t}$, with the current upper limits on the FCNC branching ratio $\mathcal{B}(t \to u_j \gamma)$ playing a crucial role in restricting the allowed parameter space and significantly reducing the corresponding uncertainty bands. This behaviour highlights the complementarity between flavour and collider probes, where flavour observables predominantly constrain the right-handed top-photon FCNC couplings, while collider measurements provide the leading sensitivity to the left-handed sector.

    \item \textbf{$\kappa_L^{u_j t},~\kappa_R^{u_j t} :$} The bounds reported in table~\ref{tab:top_Z_FCNC_global} on both $\kappa_L^{u_j t}$ and $\kappa_R^{u_j t}$ are predominantly driven by the current collider limits on the FCNC branching fraction $\mathcal{B}(t\to u_j Z)$. These direct measurements significantly reduce the allowed parameter space and lead to constraints that are considerably stronger than those obtained from the presently available indirect probes. Nevertheless, a pronounced chiral dependence is observed in the sensitivity of low-energy and electroweak precision observables. The right-handed top-$Z$ dipole couplings, $\kappa_R^{u_j t}$, receive additional and potentially powerful constraints from electroweak precision $Z$-pole measurements, the loop-induced $\delta_{\gamma Z}$ coupling, and various FCCC observables. Although these indirect bounds are currently subdominant compared to the collider limits, future improvements in both theoretical predictions and experimental precision could substantially enhance their constraining power, particularly for the right-handed sector. In contrast, the left-handed couplings, $\kappa_L^{u_j t}$, exhibit only limited sensitivity to these observables and therefore remain predominantly constrained by direct collider searches. Furthermore, the combined analysis places meaningful and often stringent constraints on the imaginary components of $\kappa_{L,R}^{u_j t}$, providing information that would remain largely inaccessible in a collider-only analysis.

    \item \textbf{$X_L^{u_j t},~X_R^{u_j t} :$} At the effective coupling level, indirect measurements provide weak constraints, and direct collider limits on $\mathcal{B}(t \to u_j Z)$ dominate by a factor of ten or more. However, this hierarchy inverts in the SMEFT framework. Because left-handed vector interactions originate from $\text{SU}(2)_L$ doublets, gauge invariance directly links them to down-type FCNCs. Consequently, the corresponding SMEFT WCs face far more severe constraints from rare $B$-meson decays (such as $b \to s(d)\ell^+\ell^-$ and $b \to s(d)\nu\bar{\nu}$) than from direct collider searches, as detailed in the subsequent section.

    \item \textbf{$\eta_L^{u_j t},~\eta_R^{u_j t} :$} The top-Higgs FCNC couplings are best probed directly by the LHC experiments through anomalous single-top plus Higgs ($tH$) production and FCNC top-quark decays in $t\bar{t}$ events. For these scalar interactions, direct collider measurements completely dominate the global fit, yielding bounds that are approximately two orders of magnitude more stringent than those derived from low-energy observables and Higgs decay.
    
    \item \textbf{Multioperator scenarios:} In all the tables, we have mentioned earlier, we have presented the results of the multioperator scenarios as well. For all the top-FCNC couplings, we have extracted simultaneously the left- and right-handed complex couplings. These scenarios will contribute to the neutron EDM $(d_n)$ and imposes important constraints. The neutron EDM is highly sensitive to CP-violating phases via it's sensitivity to the imaginary components of the products of the left- and right-handed couplings. In a couple of scenarios, this bound helps to improve the constraints on the imaginary component of the left-handed couplings. As is clearly evident for the top-gluon ($\xi_L^{u_j t}$), top-photon ($\lambda_L^{u_j t}$), and top-$Z$ ($\kappa_L^{u t}$) dipole interactions, the constraints on their imaginary components are stronger in the multi-parameter fits compared to the scenarios where only a single complex coupling is fitted at a time. Furthermore, in the multi-parameter analysis, it is essential to examine the correlations among the fitted parameters, as these correlations strongly influence the allowed parameter ranges and the overall stability of the fit. For this reason, we explicitly report the correlation matrices for each multi-parameter scenario in Appendix~\ref{Append:correlation_mat}.   
\end{itemize}
 
 Overall, these results highlight the necessity of a global EFT framework for studying top-quark FCNC interactions. Although present collider measurements set the dominant bounds on many effective couplings, indirect probes remain indispensable for accessing their chiral and CP properties, particularly through their sensitivity to right-handed interactions and complex phases. Future improvements in flavour, electroweak precision, Higgs, and EDM measurements, together with increased collider statistics, are expected to substantially strengthen these constraints and further reduce the available parameter space for physics beyond the SM.

\subsection{Extraction of SMEFT Wilson Coefficients}\label{subsec:SMEFT_Fit}
In this subsection, we present the extracted values of the SMEFT WCs corresponding to the operators listed in table~\ref{tab:SMEFT_Ops}, obtained through the minimisation of the $\chi^2$ function defined in eq.~\ref{eq:chi_sq_fun}. For this part of the analysis, the $\vec{\tilde{g}}$ are replaced by the appropriate SMEFT WCs defined at the electroweak scale. The set of observables and the associated inputs used in this analysis are summarised in table~\ref{tab:observables}.
	
The motivation for performing this separate analysis can be directly inferred from the tree-level matching relation in eq.~\eqref{eq:tree_level_matching}. In a purely phenomenological framework based on effective couplings, the couplings are treated as independent parameters, and no correlations are imposed a priori. In contrast, within the SMEFT framework, even a single operator contributes simultaneously to multiple effective top-FCNC couplings, thereby inducing non-trivial correlations among them, even in single-operator scenarios (for example $\mathcal{C}^{uB}, \mathcal{C}^{uW}$). Such correlations are intrinsically absent in a purely effective-coupling approach.
	
Furthermore, although one could, in principle, attempt to invert eq.~\eqref{eq:tree_level_matching} to extract SMEFT WCs from the effective couplings, this procedure is neither unique nor practically efficient. The presence of complex WCs necessitates a consistent treatment of their real and imaginary components, while in multi-operator scenarios, the system becomes highly coupled, leading to significant correlations and potential degeneracies among the SMEFT WCs. Consequently, a reliable and meaningful extraction of the WCs requires a dedicated global fit performed directly within the SMEFT framework, despite the increased computational complexity.

To express the observables used in this analysis in terms of the SMEFT WCs, we employ the mapping provided in eq.~\eqref{eq:tree_level_matching}. This relation establishes a direct correspondence between the coefficients in the effective basis~\eqref{eq:eff_vertex_top_FCNC} and the SMEFT WCs. The extracted values of the WCs are summarised in table~\ref{tab:SMEFT_complex_parameter}. As discussed earlier, all NP effective couplings in our analysis are treated as complex quantities, and this feature is consistently carried over to the corresponding SMEFT WCs. For completeness, we also present the results for the real-coupling scenarios in table~\ref{tab:SMEFT_Real_parameter} and \ref{tab:SMEFT_C3i}, where the SMEFT coefficients are reported at different NP scales~$\Lambda$.
\begin{table}[t!]
    \centering
    \footnotesize
    \renewcommand{\arraystretch}{2.}
    \begin{tabular}{|c|c|c|c|}
    \hline
\multicolumn{4}{|c|}{\textbf{Complex SMEFT Couplings} $(\mu_{\rm EW})$} \\
    \hline
    \rowcolor{gray!20}
       \textbf{Scenarios}  & \textbf{Value}~$(10^{7}\times\mathrm{GeV}^{-2})$ &  \textbf{Scenarios}  & \textbf{Value}~$(10^{7}\times\mathrm{GeV}^{-2})$  \\
\hline
\hline
$\mathrm{Re}(\mathcal{C}^{uG}_{23})\,, \mathrm{Im}(\mathcal{C}^{uG}_{23})$ & $(0.01 \pm 1.41),~(1.15 \pm 0.78)$ &$\mathrm{Re}(\mathcal{C}^{uG}_{13})\,, \mathrm{Im}(\mathcal{C}^{uG}_{13})$& $(-0.21 \pm 0.80),~(-0.03 \pm 1.46)$ \\
\hline
$\mathrm{Re}(\mathcal{C}^{uB}_{23})\,, \mathrm{Im}(\mathcal{C}^{uB}_{23})$ & $(-0.0 \pm 0.76),~(0.01 \pm 0.14)$ &$\mathrm{Re}(\mathcal{C}^{uB}_{13})\,, \mathrm{Im}(\mathcal{C}^{uB}_{13})$ & $(-0.10 \pm 0.34),~(-0.09 \pm 0.37)$\\
\hline
$\mathrm{Re}(\mathcal{C}^{uW}_{23})\,, \mathrm{Im}(\mathcal{C}^{uW}_{23})$ & $(0.01 \pm 0.26),~(0.0 \pm 0.45)$ &$\mathrm{Re}(\mathcal{C}^{uW}_{13})\,, \mathrm{Im}(\mathcal{C}^{uW}_{13})$& $(-0.15 \pm 0.53),~(0.0 \pm 0.80)$ \\
\hline
$\mathrm{Re}(\mathcal{C}^{u\phi }_{23})\,, \mathrm{Im}(\mathcal{C}^{u\phi}_{23})$ & $(0.81 \pm 9.87),~(0.30 \pm 16.98)$ &$\mathrm{Re}(\mathcal{C}^{u\phi }_{13})\,, \mathrm{Im}(\mathcal{C}^{u\phi}_{13})$ & $(-1.68 \pm 3.87),~(0.35 \pm 7.32)$ \\
\hline
\hline
$\mathrm{Re}(\mathcal{C}^{uG}_{32})\,, \mathrm{Im}(\mathcal{C}^{uG}_{32})$ & $(0.11 \pm 9.99),~(0.0 \pm 17.44)$ &$\mathrm{Re}(\mathcal{C}^{uG}_{31})\,, \mathrm{Im}(\mathcal{C}^{uG}_{31})$& $(0.0 \pm 5.75),~(-0.05 \pm 3.34)$ \\
\hline
$\mathrm{Re}(\mathcal{C}^{uB}_{32})\,, \mathrm{Im}(\mathcal{C}^{uB}_{32})$ & $(0.05 \pm 0.99),~(0.01 \pm 1.68)$ &$\mathrm{Re}(\mathcal{C}^{uB}_{31})\,, \mathrm{Im}(\mathcal{C}^{uB}_{31})$ & $(-0.02 \pm 1.16),~(-0.03 \pm 0.98)$\\
\hline
$\mathrm{Re}(\mathcal{C}^{uW}_{32})\,, \mathrm{Im}(\mathcal{C}^{uW}_{32})$ & $(0.02 \pm 5.49),~(-0.05 \pm 3.65)$ &$\mathrm{Re}(\mathcal{C}^{uW}_{31})\,, \mathrm{Im}(\mathcal{C}^{uW}_{31})$& $(0.06 \pm 1.47),~(0.08 \pm 1.24)$ \\
\hline
$\mathrm{Re}(\mathcal{C}^{u\phi }_{32})\,, \mathrm{Im}(\mathcal{C}^{u\phi}_{32})$ & $(-0.25 \pm 36.01),~(-0.03 \pm 62.11)$ &$\mathrm{Re}(\mathcal{C}^{u\phi }_{31})\,, \mathrm{Im}(\mathcal{C}^{u\phi}_{31})$ & $(0.52 \pm 14.66),~(0.16 \pm 22.83)$ \\
\hline
\hline
$\mathrm{Re}(\mathcal{C}^{uB}_{23})\,, \mathrm{Re}(\mathcal{C}^{uW}_{23})$ & $(-0.13 \pm 1.12),~(0.15 \pm 2.03)$ &$\mathrm{Re}(\mathcal{C}^{uB}_{13})\,, \mathrm{Re}(\mathcal{C}^{uW}_{13})$ & $(0.10 \pm 0.39),~(-0.37 \pm 0.61)$\\
\hline
$\mathrm{Re}(\mathcal{C}^{uB}_{32})\,, \mathrm{Re}(\mathcal{C}^{uW}_{32})$ & $(0.10 \pm 4.70),~(-0.16 \pm 3.52)$ &$\mathrm{Re}(\mathcal{C}^{uB}_{31})\,, \mathrm{Re}(\mathcal{C}^{uW}_{31})$ & $(-0.04 \pm 0.87),~(0.17 \pm 1.42)$\\
\hline
\end{tabular}
\caption{Fit results for top–quark FCNC processes. The SMEFT WCs 
are treated as complex, with the shorthand notation $\mathcal{C}$ denoting 
$c/\Lambda^{2}$. All coefficients are evaluated at the electroweak 
scale $\mu_{\rm EW}$.}
\label{tab:SMEFT_complex_parameter} 
\end{table}

Table~\ref{tab:SMEFT_complex_parameter} summarises the constraints on the complex SMEFT WCs relevant for top-quark FCNC interactions at the electroweak scale. A clear pattern emerges in which the coefficients associated with the right-handed dipole operators, $\mathcal{C}^{uG}_{i3}$, $\mathcal{C}^{uB}_{i3}$ and $\mathcal{C}^{uW}_{i3}$, are generally constrained relatively strongly than their chirality-flipped counterparts, $\mathcal{C}^{uG}_{3i}$, $\mathcal{C}^{uB}_{3i}$ and $\mathcal{C}^{uW}_{3i}$. Among all operators, the most stringent limits are obtained for the electroweak dipole coefficients $\mathcal{C}^{uB}_{i3}$ and $\mathcal{C}^{uW}_{i3}$, which are constrained at the level of a few $10^{-8}\,\mathrm{GeV}^{-2}$, reflecting the powerful sensitivity of flavour and electroweak precision observables. By contrast, the chirality-flipped coefficients $\mathcal{C}^{uG}_{3i}$, $\mathcal{C}^{uB}_{3i}$, and $\mathcal{C}^{uW}_{3i}$ receive comparatively weaker indirect constraints, and their bounds are largely determined by the present LHC limits on FCNC top-quark production and decay processes. The fit also demonstrates that both the real and imaginary parts of the WCs are constrained with comparable precision, underscoring the important role played by CP-sensitive observables in the global analysis. In particular, the imaginary parts of the $i3$ dipole coefficients, including $\mathrm{Im}(\mathcal{C}^{uG}_{i3})$, $\mathrm{Im}(\mathcal{C}^{uB}_{i3})$, and $\mathrm{Im}(\mathcal{C}^{uW}_{i3})$, are found to be very tightly constrained, highlighting the sensitivity of CP-violating flavour observables included in the fit. Notably, a mild preference for a non-zero value of $\mathrm{Im}(\mathcal{C}^{uG}_{23})$ is observed, in agreement with the behaviour identified at the effective-coupling level. In contrast, the scalar operators $\mathcal{C}^{u\phi}_{ij}$ remain comparatively weakly constrained, with allowed ranges typically one to two orders of magnitude larger than those of the dipole operators. Finally, the simultaneous fits involving $\mathcal{C}^{uB}_{ij}$ and $\mathcal{C}^{uW}_{ij}$ reveal no significant degeneracies, indicating that the available flavour, precision, and collider data are already capable of effectively disentangling the contributions of these electroweak dipole operators.

The operators $\mathcal{O}^{\phi q\,(1)}_{pr}$ and 
$\mathcal{O}^{\phi q\,(3)}_{pr}$, which involve the $\mathrm{SU}(2)_{\rm L}$ doublet 
$q_L$, induce both up- and down-type FCNC interactions. Since they contribute 
at tree level to down-type FCNC processes like $b\to s(d)\ell\ell$ decays, the corresponding bounds can be extremely restrictive, reaching the level of $\mathcal{O}(10^{-10})$,
\begin{align}\label{eq:cphiq13}
    \mathcal{C}^{\phi q\,(-)}_{32}(\mu_{\rm EW}) = (-0.83 \pm 1.96)\times 10^{-10},\quad 
\mathcal{C}^{\phi q\,(-)}_{31}(\mu_{\rm EW}) = (0.23 \pm 4.95)\times 10^{-10}.
\end{align}
However, if these coefficients were extracted by considering only up-type 
FCNC couplings, their values would be
\begin{align}\label{eq:cphiq13up}
\mathcal{C}^{\phi q\,(-)}_{32}(\mu_{\rm EW}) = (0.67 \pm 2.16)\times 10^{-7}, 
\qquad 
\mathcal{C}^{\phi q\,(-)}_{31}(\mu_{\rm EW}) = (0.57 \pm 1.34)\times 10^{-7}.
\end{align}
The bounds are derived from their one-loop level contributions to various observables we considered in our analysis. As discussed earlier, certain NP models can affect top-quark FCNC interactions and down-type FCNC interactions in different ways. These different types of interactions may influence distinct sets of observables, and even when they impact the same observables, their contributions can vary significantly, as illustrated in the example above. Therefore, it is crucial to determine independent constraints on the couplings associated with top-quark and down-type FCNC interactions.

It is worth noting that in our analysis, the coefficients $\mathcal{C}^{\phi q(1)}$ and $\mathcal{C}^{\phi q(3)}$ always appear in the 
specific combination $\mathcal{C}^{\phi q(1)} - \mathcal{C}^{\phi q(3)}$. As a result, it is not possible to disentangle them individually from the fit. Therefore, in eq.~\eqref{eq:cphiq13} we extract their values through the 
combination $\mathcal{C}^{\phi q(-)} \equiv \mathcal{C}^{\phi q(1)} - 
\mathcal{C}^{\phi q(3)}.$

\begin{figure}[t!]
    \centering
    \includegraphics[width=1.0\linewidth]{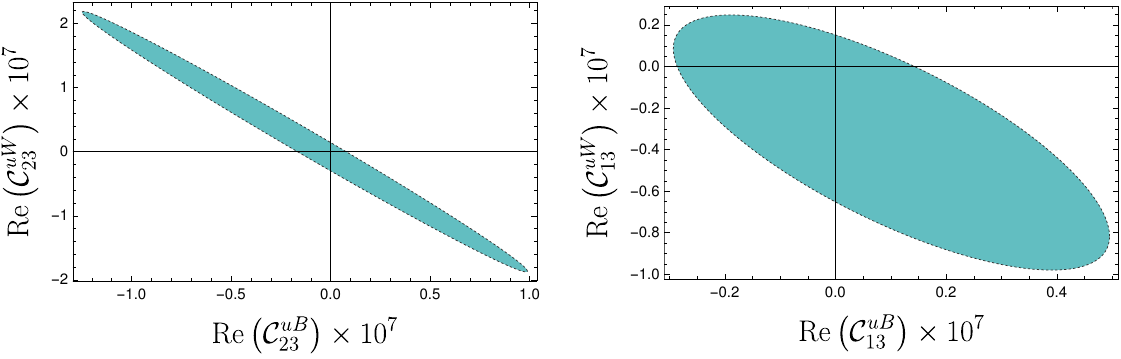}
    \caption{Correlations among the SMEFT WCs contributing to the top FCNC processes $t \to u_j \,\gamma(Z)$ are shown. The pastel blue ellipse represents $1\sigma$ errors. SMEFT WCs are defined at the reference scale $\mu_{\rm EW}$.}
     \label{fig:correl_CuBW}
\end{figure}

Unlike the vector combination $\mathcal{C}^{\phi q(-)}$, the dipole operators $\mathcal{C}^{uB}$ and $\mathcal{C}^{uW}$ enter the $tu_j\gamma$ and $tu_jZ$ interactions through specific linear combinations, as shown in eq.~\eqref{eq:tree_level_matching}. Consequently, these two WCs exhibit a pronounced negative correlation when varied simultaneously in the global fit, as shown in fig.~\ref{fig:correl_CuBW}. This behaviour originates from the fact that $\mathcal{C}^{uB}$ and $\mathcal{C}^{uW}$ are significantly more sensitive to radiative $B$-meson decays through the $tu_j\gamma$ interaction than to the $b\to s(d)\ell^+\ell^-$ processes mediated by the $tu_jZ$ interaction. Since the global fit is overwhelmingly dominated by the particular linear combination entering the photon vertex in eq.~\eqref{eq:tree_level_matching}, the two WCs are forced to satisfy an approximately inverse relation, giving rise to the strong anti-correlation observed in fig.~\ref{fig:correl_CuBW}. The corresponding results of the simultaneous fit are presented in the last row of table~\ref{tab:SMEFT_complex_parameter}.
 The simultaneous fit yields a noticeably more relaxed parameter space compared to the individual parameter fits, reflecting the strong interplay between these two dipole couplings. It is worth noting that we 
present only the real-parameter fit for this operator pair; in the complex scenario, the number of free parameters increases and the strong 
correlations between them prevent the fit from converging to a stable minimum.

\begin{table}[t!]
    \centering
    \footnotesize
    \renewcommand{\arraystretch}{1.5}
    \begin{tabular}{|c|c|c|c|c|c|}
    \hline
    \multicolumn{6}{|c|}{\textbf{Real-Parameter Scenario}} \\
    \hline
    \rowcolor{gray!20}
    \textbf{Coupling} $(\mathrm{GeV}^{-2})$ 
      & $\mu_{b}$ 
      & $\mu_{t}$  
      & $\mu=1~\mathrm{TeV}$ 
      & $\mu=5~\mathrm{TeV}$ 
      & $\mu=10~\mathrm{TeV}$  \\
    \hline
       \hline
      $\mathcal{C}^{uG}_{23}\times 10^7$ & $(-0.58 \pm 1.63)$ & $(-0.50 \pm 1.41)$ & $(-0.48 \pm 1.36)$ & $(-0.46 \pm 1.31)$ & $(-0.46 \pm 1.30)$ \\
      \hline
      $\mathcal{C}^{uB}_{23}\times 10^8$ & $(0.25 \pm 0.43)$ & $(0.01 \pm 1.40)$  & $(-0.09 \pm 1.81)$ & $(-0.17 \pm 2.16)$ & $(-0.21 \pm 2.31)$  \\
      \hline
      $\mathcal{C}^{uW}_{23} \times 10^8$ & $(-0.42 \pm 1.17)$ & $(-0.87 \pm 2.42)$  & $(-1.05 \pm 2.92)$ & $(-1.20 \pm 3.34)$ & $(-1.26 \pm 3.52)$ \\
      \hline
     $\mathcal{C}^{\phi q\,(1)}_{23}\times 10^{10}$ & $(-0.58 \pm 1.86)$ & $(-0.58 \pm 2.05)$ & $(-0.58 \pm 2.11)$ & $(-0.58 \pm 2.17)$ & $(-0.58 \pm 2.19)$ \\
      \hline
       $\mathcal{C}^{\phi q\,(3)}_{23}\times 10^{10}$ & $(0.58 \pm 1.87)$ & $(0.58\pm 2.04)$ & $(0.58 \pm 2.09)$ & $(0.58 \pm 2.12)$ & $(0.58 \pm 2.13)$ \\
      \hline
      $\mathcal{C}^{\phi u}_{23} \times 10^7 $ & $(-0.21 \pm 4.82)$ & $(-0.27 \pm 6.28)$ & $(-0.30 \pm 6.87)$ & $(-0.32 \pm 7.39)$ & $(-0.33 \pm 7.60)$ \\
      \hline
      $\mathcal{C}^{u\phi}_{23} \times 10^6$ & $(0.05 \pm 0.88)$ &  $(0.09 \pm 0.91)$ & $(0.11 \pm 0.91)$  & $(0.11 \pm 0.90)$ & $(0.11 \pm 0.89)$ \\
      \hline
      \hline
      $\mathcal{C}^{uG}_{13} \times 10^7$ & $(-0.24 \pm 0.84)$ & $(-0.22 \pm 0.76)$  & $(-0.22 \pm 0.75)$ & $(-0.22 \pm 0.74)$ & $(-0.22 \pm 0.74)$ \\
      \hline
      $\mathcal{C}^{uB}_{13}\times 10^8$ & $(0.93 \pm 1.80)$ & $(1.07 \pm 2.75)$  & $(1.12 \pm 3.14)$ & $(1.17 \pm 3.48)$ & $(1.18 \pm 3.62)$ \\
      \hline
      $\mathcal{C}^{uW}_{13}\times 10^8$ & $(-1.62 \pm 3.07)$ & $(-2.19 \pm 4.38)$  &  $(-2.41 \pm 4.90)$ & $(-2.60 \pm 5.33)$ & $(-2.67 \pm 5.51)$  \\
      \hline
      $\mathcal{C}^{\phi q\,(1)}_{13}\times 10^{10}$ & $(0.27 \pm 4.57)$ & $(0.27 \pm 5.02)$ & $(0.27 \pm 5.18)$ & $(0.27 \pm 5.32)$ & $(0.27 \pm 5.37)$ \\
      \hline
      $\mathcal{C}^{\phi q\,(3)}_{13}\times 10^{10}$ & $(-0.27 \pm 4.59)$ & $(-0.27\pm 5.00)$ & $(-0.27 \pm 5.12)$ & $(-0.27 \pm 5.20)$ & $(-0.27 \pm 5.23)$ \\
      \hline
      $\mathcal{C}^{\phi u}_{13}\times 10^7$ & $(-0.22 \pm 2.38)$ & $(-0.28 \pm 3.10)$ & $(-0.31 \pm 3.40)$ & $(-0.33 \pm 3.65)$ & $(-0.34 \pm 3.75)$ \\
      \hline
      $\mathcal{C}^{u\phi}_{13}\times 10^7$ & $(-1.28 \pm 4.40)$ & $(-1.32 \pm 4.51)$ & $(-1.30 \pm 4.46)$ & $(-1.28 \pm 4.38)$ & $(-1.27 \pm 4.33)$  \\
      \hline
    \end{tabular}
    \caption{Fit results for the top--FCNC SMEFT WCs in the real--coupling scenario. The scale dependence of the coefficients is shown for different values of the NP scale $(\Lambda)$.}
    \label{tab:SMEFT_Real_parameter}
\end{table}
The results shown in table~\ref{tab:SMEFT_complex_parameter} can be straightforwardly translated into bounds on the top-FCNC effective couplings using the matching relations between the two operator bases given in Eq.~\ref{eq:tree_level_matching}. In what follows, we present a few representative limits that exhibit significant improvements compared to those obtained directly in the effective-coupling framework and listed in table~\ref{tab:top_Z_FCNC_global} (we have shown only upper limits):
\begin{align}
	|\lambda^R_{tc}|^2 &<  3.680 \times 10^{-6} \,, & |\lambda^R_{tu}|^2 &< 1.925 \times 10^{-6} \,, \\[10pt]
	|\kappa^R_{tc}|^2 &< 1.857 \times 10^{-6} \,, & |\kappa^R_{tu}|^2 &<  1.137 \times 10^{-5} \,, \\[10pt]
	|X^L_{tc}|^2 &<  1.192 \times 10^{-10} \,, & |X^L_{tu}|^2 &<   8.339 \times 10^{-11} \,.
\end{align}
The bounds on $|\kappa_R^{qt}|^2$ are derived from the constraints on the complex WCs $\mathcal{C}^{qB}_{2(1)3}$ and $\mathcal{C}^{qW}_{2(1)3}$, while those on $|X_L^{qt}|^2$ follow from the limits on $\mathcal{C}^{q\phi}_{2(1)3}$. 


The origin of these improvements can be understood from the correlations imposed by the SMEFT framework. In the effective-coupling approach, the top-photon and top-$Z$ FCNC interactions are treated as independent parameters. Consequently, the coupling $|\lambda_R^{qt}|$ is strongly constrained by flavour observables, most notably the radiative decay $B\to X_s\gamma$, whereas $|\kappa_R^{qt}|$ is constrained mainly by collider measurements and a limited set of rare FCNC observables, resulting in comparatively weaker bounds. In the SMEFT framework, however, the electroweak dipole operators $\mathcal{C}^{qB}_{2(1)3}$ and $\mathcal{C}^{qW}_{2(1)3}$ simultaneously generate both top-photon and top-$Z$ FCNC interactions after electroweak symmetry breaking. As a result, the stringent constraints from radiative $B$-meson decays, which strongly restrict the WCs through their contributions to top-photon FCNC processes, are automatically propagated to the corresponding top-$Z$ interactions. This SMEFT correlation leads to an improvement in the bounds on $|\kappa_R^{qt}|^2$ relative to those obtained in the effective-coupling framework and illustrates the power of combining flavour and collider information within a gauge-invariant EFT description.

The improvement in the bounds on $|X_L^{qt}|$ arises from a different mechanism, which was discussed in detail around eqs.~\eqref{eq:cphiq13} and \eqref{eq:cphiq13up}. Owing to the $\mathrm{SU}(2)_L$ gauge structure of the SMEFT, the operators responsible for left-handed top-$Z$ FCNC interactions are directly linked to flavour-changing processes in the down-type quark sector. Consequently, rare $B$-meson decays impose stringent constraints on the corresponding WCs, yielding limits that are relatively stronger than those obtained from direct collider searches alone. Taken together, these results demonstrate how the SMEFT framework can translate the precision of low-energy flavour measurements into improved constraints on top-quark FCNC interactions. We therefore expect these stronger bounds on the relevant WCs to yield correspondingly tighter predictions for the associated top-FCNC branching fractions, a point that will be explored in the next subsection.

\paragraph{\underline{\textbf{Scale dependency of the couplings:}}}
Table~\ref{tab:SMEFT_Real_parameter} presents the results for the SMEFT WCs in the real-coupling scenarios. We display these results separately because several of the observables discussed earlier are particularly sensitive to the imaginary parts of the NP couplings. To isolate the effects free from CP-violating contamination, we therefore examine this special case in which all couplings are taken to be real. The table also shows 
the RGE of these coefficients for different choices of the NP scale $\Lambda$. Notably, we explicitly perform the RGE running only for the $\mathcal{C}_{i3}$ couplings; however, the evolution of the $\mathcal{C}_{3i}$ couplings can be derived in a completely analogous manner. The resulting SMEFT WCs for $\mathcal{C}_{3i}$ evaluated at the electroweak scale, $\mu_{\rm EW}$, are provided in Table~\ref{tab:SMEFT_C3i} of Appendix~\ref{Append:SMEFT_couplings}. As mentioned earlier, to perform the RGE of the 
SMEFT WCs, we use the ADMs
derived in refs.~\cite{Jenkins:2013zja,Jenkins:2013wua,Alonso:2013hga}. The evolution is obtained by solving the full set of seven coupled differential equations, with initial conditions given by the values of the SMEFT 
coefficients at the matching scale $(\mu_{\rm EW})$. The details of the ADMs, 
together with the corresponding initial conditions of SM input parameters, are provided in 
Appendix~\ref{Append:RGE_beta}. It is worth mentioning that, for running the WCs, $\mathcal{C}^{\phi q(1)}$ and $\mathcal{C}^{\phi q(3)}$, we assign identical 
initial conditions as mentioned in eq.~\eqref{eq:cphiq13} to both coefficients but with a relative minus sign. This choice is justified in the single-parameter scenarios (only one parameter is present at a time), since these operators always appear in a specific combination hard to disentangle from the fit.
\begin{figure}[h!]
    \centering
    \includegraphics[scale=0.82]{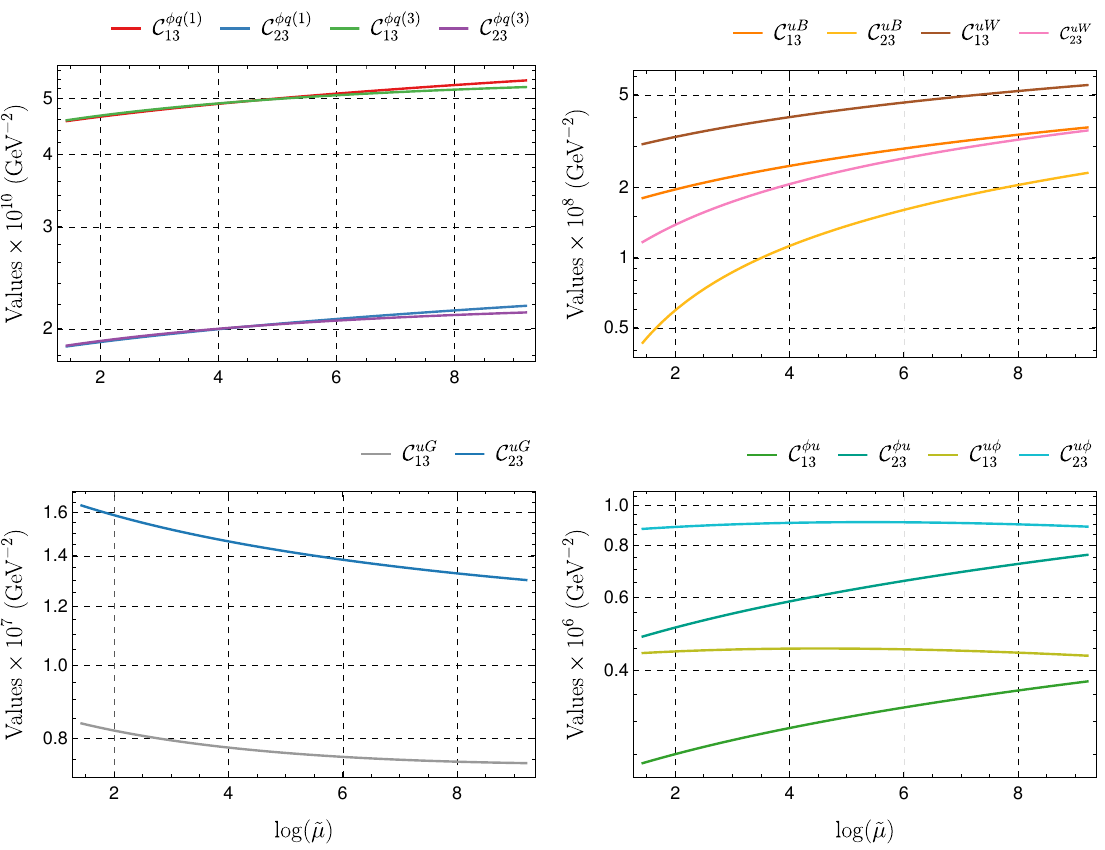}
     \caption{Evolution of the SMEFT couplings $(\mathcal{C}_i/\Lambda^2)$ over the 
energy range up to $10~\mathrm{TeV}$. We use the shorthand notation $(\mathcal{C}_i \equiv \mathcal{C}_i/\Lambda^2)$ throughout. The quantity $\tilde{\mu}$ denotes the normalised energy scale, defined as $\tilde{\mu} = \mu / 1~\mathrm{GeV}$.}

     \label{fig:SMEFT_RGE}
\end{figure}\\
In fig.~\ref{fig:SMEFT_RGE}, we show the extrapolated evolution of the SMEFT 
couplings over the energy range from the scale $\mu_b$ up to the NP scale of $10~\mathrm{TeV}$.
We observe that the coefficients $\mathcal{C}^{uG}$ 
and $\mathcal{C}^{u\phi}$ decrease as the scale $\mu$ is increased, 
indicating that their running is dominated by the strong coupling $g_s$. In 
contrast, the remaining coefficients $\mathcal{C}^{uB}$, $\mathcal{C}^{uW}$, 
$\mathcal{C}^{\phi q(1)}$, $\mathcal{C}^{\phi q(3)}$, and 
$\mathcal{C}^{\phi u}$ grow with increasing energy scale, reflecting their 
ADM structure and the dominance of electroweak contributions in their 
evolution. As evident from fig.~\ref{fig:SMEFT_RGE}, the WCs
$\mathcal{C}^{uB}$ and $\mathcal{C}^{uW}$ exhibit a substantial change as 
they evolve from the scale $\mu_b$ up to $\mu_{\Lambda} = 10~\mathrm{TeV}$.

\subsection{Predictions for branching ratios }\label{subsec:pred_topfcnc_smeft}
\begin{table*}[t]
	\footnotesize
	\centering
	\begin{tabular}{@{} l l c @{\quad} l l c @{}}
		\toprule
		\rowcolor{gray!20}
		\multicolumn{1}{c}{\textbf{Process}} & \multicolumn{1}{c}{\textbf{Scenario}} & \multicolumn{1}{c}{\textbf{Our Prediction}} &
		\multicolumn{1}{c}{\textbf{Process}} & \multicolumn{1}{c}{\textbf{Scenario}} & \multicolumn{1}{c}{\textbf{Our Prediction}} \\
		\midrule
		
		\multirow{2}{*}{$\mathcal{B}(t\to c g)$}
		& $\mathrm{Re}\,\mathcal{C}^{uG}_{23}\neq0$
		& $< 4.12 \times 10^{-4}$  
		& \multirow{2}{*}{$\mathcal{B}(t\to u g)$}
		& $\mathrm{Re}\,\mathcal{C}^{uG}_{13}\neq0$
		& $< 0.94 \times 10^{-4}$ \\
		
		& $\mathrm{Re}\,\mathcal{C}^{uG}_{23}, \mathrm{Im}\,\mathcal{C}^{uG}_{23}\neq0$
		& $< 5.08 \times 10^{-4}$
		&
		& $\mathrm{Re}\,\mathcal{C}^{uG}_{13}, \mathrm{Im}\,\mathcal{C}^{uG}_{13}\neq0$
		& $< 0.93 \times 10^{-4}$ \\
		
		\midrule
		
		\multirow{5}{*}{$\mathcal{B}(t\to c \gamma)$}
		& $\mathrm{Re}\,\mathcal{C}^{uB}_{23}\neq0$
		& $< 2.66 \times 10^{-7}$
		& \multirow{5}{*}{$\mathcal{B}(t\to u \gamma)$}
		& $\mathrm{Re}\,\mathcal{C}^{uB}_{13}\neq0$
		& $< 1.32 \times 10^{-5}$ \\
		
		& $\mathrm{Re}\,\mathcal{C}^{uW}_{23}\neq0$
		& $< 2.59 \times 10^{-6}$
		&
		& $\mathrm{Re}\,\mathcal{C}^{uW}_{13}\neq0$
		& $< 1.28 \times 10^{-5}$ \\
		
		& $\mathrm{Re}\,\mathcal{C}^{uB}_{23},\mathrm{Re}\,\mathcal{C}^{uW}_{23}\neq0$
		& $< 2.63 \times 10^{-6}$
		&
		& $\mathrm{Re}\,\mathcal{C}^{uB}_{13},\mathrm{Re}\,\mathcal{C}^{uW}_{13}\neq0$
		& $< 1.31 \times 10^{-5}$ \\
		
		& $\mathrm{Re}\,\mathcal{C}^{uB}_{23}, \mathrm{Im}\,\mathcal{C}^{uB}_{23}\neq0$
		& $<  2.87 \times 10^{-7}$
		&
		& $\mathrm{Re}\,\mathcal{C}^{uB}_{13}, \mathrm{Im}\,\mathcal{C}^{uB}_{13}\neq0$
		& $< 3.63 \times 10^{-6}$ \\
		
		& $\mathrm{Re}\,\mathcal{C}^{uW}_{23}, \mathrm{Im}\,\mathcal{C}^{uW}_{23}\neq0$
		& $< 3.56 \times 10^{-7}$
		&
		& $\mathrm{Re}\,\mathcal{C}^{uW}_{13}, \mathrm{Im}\,\mathcal{C}^{uW}_{13}\neq0$
		& $< 3.03 \times 10^{-6}$ \\
		
		\midrule
		
		\multirow{5}{*}{$\mathcal{B}(t\to c Z)$}
		& $\mathrm{Re}\,\mathcal{C}^{uB}_{23}\neq0$
		& $< 4.59 \times 10^{-8}$
		& \multirow{5}{*}{$\mathcal{B}(t\to u Z)$}
		& $\mathrm{Re}\,\mathcal{C}^{uB}_{13}\neq0$
		& $< 2.29 \times 10^{-6}$ \\
		
		& $\mathrm{Re}\,\mathcal{C}^{uW}_{23}\neq0$
		& $< 4.93 \times 10^{-6}$
		&
		& $\mathrm{Re}\,\mathcal{C}^{uW}_{13}\neq0$
		& $< 2.43 \times 10^{-5}$ \\
		& $\mathrm{Re}\,\mathcal{C}^{uB}_{23},\mathrm{Re}\,\mathcal{C}^{uW}_{23}\neq0$
		& $< 1.65 \times 10^{-4}$
		&
		& $\mathrm{Re}\,\mathcal{C}^{uB}_{13},\mathrm{Re}\,\mathcal{C}^{uW}_{13}\neq0$
		& $< 0.92 \times 10^{-4}$ \\
		
		& $\mathrm{Re}\,\mathcal{C}^{uB}_{23}, \mathrm{Im}\,\mathcal{C}^{uB}_{23}\neq0$
		& $< 4.96 \times 10^{-8}$
		&
		& $\mathrm{Re}\,\mathcal{C}^{uB}_{13}, \mathrm{Im}\,\mathcal{C}^{uB}_{13}\neq0$
		& $< 6.27 \times 10^{-7}$ \\
		
		& $\mathrm{Re}\,\mathcal{C}^{uW}_{23}, \mathrm{Im}\,\mathcal{C}^{uW}_{23}\neq0$
		& $< 6.78 \times 10^{-7}$
		&
		& $\mathrm{Re}\,\mathcal{C}^{uW}_{13}, \mathrm{Im}\,\mathcal{C}^{uW}_{13}\neq0$
		& $< 5.77 \times 10^{-6}$ \\
		
		& $\mathrm{Re}\,\mathcal{C}^{\phi q(-)}_{23}\neq0$
		& $< 1.48 \times 10^{-10}$
		&
		& $\mathrm{Re}\,\mathcal{C}^{\phi q(-)}_{13}\neq0$
		& $< 1.66 \times 10^{-10}$ \\
		
		\midrule
		
		\multirow{2}{*}{$\mathcal{B}(t\to c H)$}
		& $\mathrm{Re}\,\mathcal{C}^{u\phi}_{23}\neq0$
		& $< 4.78 \times 10^{-4}$  
		& \multirow{2}{*}{$\mathcal{B}(t\to u H)$}
		& $\mathrm{Re}\,\mathcal{C}^{u\phi}_{13}\neq0$
		& $< 3.63 \times 10^{-4}$ \\
		& $\mathrm{Re}\,\mathcal{C}^{u\phi}_{23}, ~\mathrm{Im}\,\mathcal{C}^{u\phi}_{23}\neq0$
		& $< 4.89 \times 10^{-4}$  
		&
		& $\mathrm{Re}\,\mathcal{C}^{u\phi}_{13},~\mathrm{Im}\,\mathcal{C}^{u\phi}_{13}\neq0$
		& $< 3.95 \times 10^{-4}$ \\
		
		\bottomrule
	\end{tabular}
	\caption{95\% CL upper limits on branching ratios for top FCNC decay modes evaluated in different SMEFT scenarios.}
	\label{tab:BR-pred}
\end{table*}
In the previous subsections~\ref{subsec:top_coupling_fit} and \ref{subsec:SMEFT_Fit}, we presented the bounds on various top effective couplings and their corresponding SMEFT WCs, obtained collectively from low-energy flavour observables, EWPOs, direct collider and other relevant measurements. Here, using these bounds, we predict the branching fractions of various top-FCNC decays and the FCNC decays of the Z- and H-bosons.

\paragraph{Top-FCNC decays:}
We present the predicted branching fractions for radiative top-FCNC decays in Table~\ref{tab:BR-pred}, allowing for direct comparison with the current experimental upper limits and the corresponding SM predictions listed in Table~\ref{tab:Branching_ratio_exp}. The SM expectations for these processes are extremely suppressed, lying in the range $\mathcal{O}(10^{-12})$–$\mathcal{O}(10^{-17})$, and remain far below present experimental sensitivities.

Having established the constraints on the effective couplings and SMEFT WCs, we now turn to their implications for top-quark FCNC branching fractions. This step is particularly important because, in several SMEFT scenarios, the strongest bounds on the relevant WCs do not originate from direct searches for FCNC top-quark processes, but rather from low-energy flavour and electroweak precision observables. These indirect probes can impose restrictions that are relatively stronger than those obtained from collider measurements alone. Consequently, the resulting predictions for the FCNC top-quark branching fractions can differ from those obtained in collider measurements. In the following, we translate the fitted constraints on the SMEFT WCs into predictions for the branching fractions of the various FCNC top-quark decay modes and quantify the impact of these indirect constraints on the discovery reach of future collider experiments.

By translating our indirect constraints into predictions for top-quark FCNC decays, we find that the allowed branching fractions in several channels are substantially tighter than current direct collider limits. In particular, for scenarios involving the dipole operators $\mathcal{O}^{uB}_{23}$ or $\mathcal{O}^{uW}_{23}$ with either real or complex WCs, the predicted maximum for $\mathcal{B}(t\to c \gamma)$ lies approximately one to two orders of magnitude below present experimental bounds. For the same operators with complex coefficients, the predicted limit on $\mathcal{B}(t\to c Z)$ is restricted to values two to three orders of magnitude below current collider limits.

For the first-generation couplings, an analysis of $\mathcal{O}^{uB}_{13}$ with complex coefficients yields predicted bounds that are roughly one order of magnitude stronger for $\mathcal{B}(t\to u \gamma)$ and two orders of magnitude stronger for $\mathcal{B}(t\to u Z)$ relative to direct measurements. For $\mathcal{O}^{uW}_{13}$ with complex coefficients, the improvements are comparatively milder, tightening the limits on $\mathcal{B}(t\to u \gamma)$ and $\mathcal{B}(t\to u Z)$ by approximately one order of magnitude. However, the predicted branching ratios $\mathcal{B}(t\to u_j H)$ lie within the reach of current experimental sensitivities.

Finally, the most dramatic enhancement in constraining power is observed in scenarios involving the left-handed vector operator $\mathcal{O}^{\phi q(-)}_{2(1)3}$. In these cases, the severe restrictions imposed by low-energy observables force the predicted branching fractions for $t \to c(u) Z$ to lie approximately six orders of magnitude below current experimental collider limits.

\begin{table*}[t!]
\centering
\footnotesize
\begin{tabular}{l c c c}
\toprule
\rowcolor{gray!20}
\textbf{Scenario} &
$\mathcal{B}(t\to c e^+ e^-)$ &
$\mathcal{B}(t\to c \mu^+ \mu^-)$ &
$\mathcal{B}(t \to c \nu \bar{\nu})$ 
\\
\midrule
via $t\to c \gamma$ & $< 2.45 \times 10^{-8}$ & $< 1.30 \times 10^{-8}$ & $-$ \\
via $t\to c Z$ & $< 1.13 \times 10^{-7}$ & $< 1.13 \times 10^{-7}$ & $< 8.86 \times 10^{-10}$ \\
via $t\to c H$ & $< 0.94 \times 10^{-16}$ & $< 4.01 \times 10^{-12}$ & $-$ \\

\midrule

\rowcolor{gray!20}
\textbf{Scenario} &
$\mathcal{B}(t\to u e^+ e^-)$ &
$\mathcal{B}(t\to u \mu^+ \mu^-)$ &
$\mathcal{B}(t \to u \nu \bar{\nu})$ \\
\midrule
via $t \to u\gamma$ & $< 1.22 \times 10^{-7}$ & $< 6.45 \times 10^{-8}$ & $-$\\
via $t \to uZ$ & $< 6.27 \times 10^{-8}$ & $< 6.27 \times 10^{-8}$ & $< 4.93 \times 10^{-10}$\\
via $t \to uH$ & $< 0.76 \times 10^{-16}$ & $< 3.25 \times 10^{-12}$ & $-$\\

\bottomrule
\end{tabular}
\caption{95\% CL upper limits on branching ratios for different three-body top-FCNC processes.}
\label{tab:BR-tcll-95CL}
\end{table*}
In addition to the two-body FCNC modes, we have also studied the three-body decays $t \to u_j \ell^+ \ell^-$, which arise through off-shell photon, 
$Z$-, and $H$-boson exchange. Depending on the choice of intermediate mediator, different SMEFT operators contribute, leading to distinct 
predictions for the branching fractions of the various $t \to u_j \ell^+ \ell^-$ and $t \to u_j \nu\bar{\nu}$ channels. The general 
expression for the differential decay width of a sequential three-body decay is given by
\begin{align}
    d\Gamma_{A\to b,q^*(c,d)}=\frac{\left|\mathcal{M}\right|^2}{512 \pi^3 m_A^3 q^2}\lambda^{1/2}(m_A^2,m_b^2,q^2)\,\lambda^{1/2}(q^2,m_c^2,m_d^2)dq^2 d\mathrm{cos}\theta
\end{align}
where $\lambda$ is the Källén function and $q^\ast$ denotes the off-shell propagator, which in our case corresponds to $\gamma^\ast$, $Z^\ast$, or $H^\ast$. The expression for differential deay width can be written as
\begin{subequations}\label{eq:diff_decay}
\begin{align}
    \frac{d\Gamma(q^2)}{dq^2}\Big|_{t\to u_j\gamma^*(\ell\ell)}&=\frac{e^2 v^2 m_t^3}{96\pi^3 q^2}\left(1-\frac{q^2}{m_t^2}\right)^2\left(2+\frac{q^2}{m_t^2}\right)\left|c_W~\mathcal{C}^{uB}_{pq}+s_W~\mathcal{C}^{uW}_{pq}\right|^2\,,
    \end{align}
    
    \begin{align}
    \frac{d\Gamma(q^2)}{dq^2}\Big|_{t\to u_jZ^*(\ell \ell)}&=\frac{g_Z^2 m_t^3 v^2(g_{Z_L}^2+g_{Z_R}^2)}{1536\pi^3 (q^2-M_Z^2)^2}\left(1-\frac{q^2}{m_t^2}\right)^2\left\{8q^2\left(2+\frac{q^2}{m_t^2}\right)\left|c_W~\mathcal{C}^{uW}_{pq}-s_W~\mathcal{C}^{uB}_{pq}\right|^2\right.\nonumber\\
    &\left.+ g_Z^2v^2 \left(1+\frac{2 q^2}{m_t^2}\right)\left|\mathcal{C}^{\phi q(-)}_{pq}\right|^2\right\}\,,
    \end{align}
    
    \begin{align}
    \frac{d\Gamma(q^2)}{dq^2}\Big|_{t\to u_jH^*(\ell\ell)}&=\frac{9 m_t m_{\ell}^2 v^2 q^2}{1024 \pi^3(q^2-M_H^2)^2}\left(1-\frac{q^2}{m_t^2}\right)^2\left|\mathcal{C}^{u\phi}_{pq}\right|^2\,.
\end{align}
\end{subequations}
where $g_Z$, $g_{Z_L}$, and $g_{Z_R}$ denote the couplings associated with 
the neutral--current interactions of the $Z$ boson with fermions. They are 
given by
$$g_Z = \frac{e}{s_W c_W}, \qquad 
g_{Z_L} = I_W^{(3)} - Q_f\, s_W^2, \qquad 
g_{Z_R} = -\, Q_f\, s_W^2 ,$$
with $I_W^{(3)}$ being the weak isospin of the fermion and $Q_f$ its 
electric charge. Since the decay $t \to u_j \nu\bar{\nu}$ proceeds only 
via an off-shell $Z$ boson, its differential decay rate has the same structure as 
$t \to u_j \ell^+\ell^-$ in the massless limit, with the proper replacement of $Z\nu\bar{\nu}$ coupling. In 
table~\ref{tab:BR-tcll-95CL}, we summarise the predicted branching fractions for the three-body FCNC modes in the different scenarios considered. To the best 
of our knowledge, no experimental upper limits for these channels currently exist. The corresponding SM predictions are extremely small, as reported in 
refs.~\cite{Kala:2025srq, Frank:2006ku}:
\begin{align}
    \mathcal{B}(t\to c e^+ e^-)&=8.48 \times 10^{-15}\,,& \mathcal{B}(t \to c \mu^+ \mu^-)&=9.55\times 10^{-15}\,,\nonumber\\
    \mathcal{B}(t\to u e^+ e^-)&=6.81 \times 10^{-17}\,,& \mathcal{B}(t \to u \mu^+ \mu^-)&=7.68\times 10^{-17}\,,\\
    \mathcal{B}(t \to c~\nu\bar{\nu})&=2.99\times 10^{-14}\,, & \mathcal{B}(t \to u~\nu \bar{\nu})&=2.40\times 10^{-16}\,.\nonumber
\end{align}
 In our analysis, we have predicted branching ratio several order magnitude larger than SM predictions for all these channels, which will provide a 
well motivated target range for future experimental searches.

\subsection*{$Z(H)$ mediated FCNC interaction:}
The top-FCNC interaction vertices can also contribute to the FCNC decays of 
the $Z$ and $H$ bosons. In the SM, the processes 
$Z(H)\to d_i \bar d_j$ $(i\neq j)$ occur only at the one-loop level and are therefore extremely suppressed, unlike other SM FCNC transitions. 
The corresponding SM predictions for the branching ratios are 
$\mathcal{B}(Z (H)\to b\bar{s}(\bar{d})) \sim \mathcal{O}(10^{-7}(10^{-9}))$ ~\cite{PhysRevLett.57.1514, Aranda:2020tqw}, however, at collider, no FCNC decays of the $Z$ or $H$ 
bosons have been observed so far.\\
In our analysis, the top-FCNC interactions contributing to the rare $Z$ FCNC decays, with a similar topology as shown in 
fig.~\ref{fig:Z_pole}. For $\mathcal{B}(Z \to b\bar{s})$, the loop amplitude contains a CKM factor 
of $\mathcal{O}(1)$ when the $tcZ$ coupling is inserted, resulting in the 
dominant contribution to this decay. Similarly, for 
$\mathcal{B}(Z \to b\bar{d})$, the loop arising from the $tuZ$ coupling also 
carries a CKM factor of $\mathcal{O}(1)$. The same pattern holds for the Higgs FCNC decays,
the branching ratio $\mathcal{B}(H \to b\bar{s})$ is dominated by the $tcH$ 
coupling, while $\mathcal{B}(H \to b\bar{d})$ receives its dominant 
contribution from the $tuH$ coupling. However, our NP predictions for these modes remain extremely small and effectively indistinguishable from the SM expectations. Consequently, if any 
FCNC decay of the $Z$ or $H$ boson were to be observed at colliders, such a signal would not mainly originate from top–FCNC interactions.

\subsection{Predictions of a few CP asymmetries}
We now present our predictions for the CP-violating observables in the radiative FCNC decay of the top quark. In this analysis, the source of CP violation arises from the interference between amplitudes with relative phase differences. In our framework, the weak phase originates from the complex effective top FCNC couplings, which encapsulate the possible NP effects. On the other hand, there are additional phases generated through one-loop corrections to the decay amplitudes as shown in fig.~\ref{fig:top_CP_asym}, where loop kinematics give rise to absorptive parts once the intermediate states can go on-shell. Such absorptive contributions are essential to generate a non-vanishing CP asymmetry. In addition, we have taken into account the SM one-loop contributions to these radiative top-FCNC decays. The resulting interplay between the weak phases associated with the complex NP couplings and the CKM-induced weak phase from the SM diagrams, together with the phase arising from the absorptive part of the loop dynamics (sometimes called strong phase), constitutes a direct source of CP violation in radiative FCNC top decays.
\begin{figure}[htb!]
    \centering
\subfloat[]{
    \begin{tikzpicture}
        \begin{feynman}\label{fig:top_rad_1}
        \vertex[square dot, draw=red, fill=red, minimum size=5pt, inner sep=2pt](a1){};
        \vertex[ above right=1cm of a1](a2);
        \vertex[ above right=2cm of a1](a3){$u_j$};
        \vertex[ above left=1cm of a1](a4);
        \vertex[ above left=2cm of a1](a5){$t$};
        \vertex[below=1.0cm of a1](a6){\(\gamma\)};
        
        \diagram* {
            (a5) --[fermion] (a4) --[fermion, very thick, edge label'=\(t\)] (a1) --[fermion, very thick, edge label'=\(u_j\)] (a2) --[fermion] (a3),
            (a2) --[boson, edge label'=\(X\)] (a4),
            (a1) --[red, boson, very thick] (a6),
        };
        \end{feynman}
    \end{tikzpicture}
    }
\subfloat[]{
    \begin{tikzpicture}
        \begin{feynman}\label{fig:top_rad_2}
        \vertex[](a1);
        \vertex[ square dot, draw=black, fill=black, minimum size=5pt, inner sep=2pt, above right=0.95 cm of a1](a2){};
        \vertex[ above right=1.6cm of a1](a3){$u_j$};
        \vertex[ above left=1cm of a1](a4);
        \vertex[ above left=1.5cm of a1](a5){$t$};
        \vertex[below=0.8cm of a1](a6){\(\gamma\)};
        
        \diagram* {
            (a5) --[fermion] (a4) --[fermion, edge label'=\(t\)] (a1) --[fermion, very thick, edge label'=\(t\)] (a2) --[fermion, very thick] (a3),
            (a2) --[boson, very thick, edge label'=\(X\)] (a4),
            (a1) --[ boson] (a6),
        };
        \end{feynman}
    \end{tikzpicture}
    }
\subfloat[]{
    \begin{tikzpicture}
        \begin{feynman}\label{fig:top_rad_3}
        \vertex[](a1);
        \vertex[  above right=1.1cm of a1](a2);
        \vertex[ above right=1.5cm of a1](a3){$u_j$};
        \vertex[square dot, draw=black, fill=black, minimum size=5pt, inner sep=2pt, above left=0.9 cm of a1](a4){};
        \vertex[ above left=1.6 cm of a1](a5){$t$};
        \vertex[below=0.8cm of a1](a6){\(\gamma\)};
        
        \diagram* {
            (a5) --[fermion, very thick] (a4) --[fermion, very thick, edge label'=\(u_j\)] (a1) --[fermion,  edge label'=\(u_j\)] (a2) --[fermion] (a3),
            (a2) --[boson, very thick, edge label'=\(X\)] (a4),
            (a1) --[ boson] (a6),
        };
        \end{feynman}
    \end{tikzpicture}
    }
   \caption{
One-loop Feynman diagrams contributing to the radiative decay 
$t \to u_j \gamma$, with $X = \gamma,\, Z,\, g,\, H$ denoting neutral intermediate particles. 
These diagrams also contribute to the FCNC processes 
$t \to u_j g$. 
Self-energy corrections to the external legs of the effective top--photon/gluon 
vertices are not shown explicitly.
}

    \label{fig:top_CP_asym}
\end{figure}
\paragraph{} The most general form of the polarised $t \to u_j\,\gamma_{\pm}$ amplitudes can be written as,
\begin{align}
    i\mathcal{M}(t \to u_j+\gamma_{\pm})=i \bar{u}(p_f)\,\Gamma^{\mu}_{fi,\gamma}(q^2)\,u(p_i)\,\varepsilon^*_{\pm,\mu}(q)\,,
\end{align}
where $u(p_i)$ and $u(p_f)$ are the Dirac spinors of the initial top and final charm (down) states, respectively, and $\varepsilon$ denotes the photon polarisation vectors having momentum $q=(p_i-p_f)$, $\gamma_+$ and $\gamma_-$ represent the positively and negatively polarised photons, respectively. Here we have followed the same notation as used in ref.~\cite{Balaji:2020qjg}. The effective vertex can be written as
\begin{align}
    \Gamma^{\mu}_{fi,\,\gamma}(q^2)=i \sigma^{\mu\nu}q_{\nu}\left(f^L_{fi,\gamma}P_L+f^R_{fi,\gamma}P_R\right)\,.
\end{align}
where the form factors $f^L_{fi,\gamma}$ and $f^R_{fi,\gamma}$ consist of CKM elements and the complex NP couplings and one-loop integral loop factors and can be further parametrised into
\begin{subequations}\label{eq:top_asym_FF1}
\begin{align}
    f^L_{fi,\gamma}&=\sum_{\alpha}V_{t\alpha}^*V_{u_j\alpha}m_{u_j}\mathcal{F}_{\alpha}^L+\sum_{l_{\rm NP}}\mathcal{C}_{l_{\rm NP}}\mathcal{K}^L_{l_{\rm NP}}\,,\\
    f^R_{fi,\gamma}&=\sum_{\alpha}V_{t\alpha}^*V_{u_j\alpha}m_t\mathcal{F}_{\alpha}^R+\sum_{l_{\rm NP}}\mathcal{C}_{l_{\rm NP}}\mathcal{K}^R_{l_{\rm NP}}\,.
\end{align}
\end{subequations}
where the first term in eq.~\eqref{eq:top_asym_FF1} corresponds to the one-loop SM contribution to the $t \to u_j \gamma$ decay, while the second term arises solely from the new-physics sector. In the SM term, $\alpha~(\alpha = d, s, b)$ denotes the internal down-type quarks, and $\mathcal{F}^{L,R}$ represent the purely loop-induced kinematic functions obtained from the SM contribution, whose explicit expressions can be found in ref.~\cite{Balaji:2020qjg}. In contrast, the NP contribution involves $\mathcal{C}^{L,R}_{l_{\rm NP}}$, which contain the complex NP couplings, and $\mathcal{K}^{L,R}$, which denote the corresponding loop kinematic functions. 
\paragraph{} 
Under a CP transformation, each particle is replaced by its corresponding antiparticle, and all additive quantum numbers, including the electric charge, are reversed, while the spin remains unchanged. Therefore, the polarised amplitudes of CP-conjugated processes can be written as
\begin{align}
 i\mathcal{M}(\bar{t}\to\bar{u}_j+\gamma_{\mp})=i\mathcal{M}^{CP}(t \to u_j+\gamma_{\pm})=i \bar{v}(p_i)\bar{\Gamma}^{\mu}_{if,\gamma}(q^2) v(p_f)\varepsilon^*_{\mp,\mu}(q)\, 
\end{align}
where $v(p_i)$ and $v(p_f)$ are the Dirac spinors associated with the initial and final anti-fermion states, respectively. The effective vertex function is parametrised in the same manner as introduced earlier.
\begin{align}
        \bar{\Gamma}^{\mu}_{if,\,\gamma}(q^2)=i \sigma^{\mu\nu}q_{\nu}\left(\bar{f}^L_{if,\gamma}P_L+\bar{f}^R_{if,\gamma}P_R\right)\,.
    \end{align}
    
\begin{subequations}\label{eq:top_asym_FF2}
    \begin{align}
        \bar{f}^L_{if,\gamma}&=\sum_{\alpha}V_{t\alpha}V_{u_j\alpha}^*m_{t}\mathcal{F}_{\alpha}^R+\sum_{l_{\rm NP}}\mathcal{C}_{l_{\rm NP}}^{\,*}\mathcal{K}^R_{l_{\rm NP}}\,,\\
    \bar{f}^R_{if,\gamma}&=\sum_{\alpha}V_{t\alpha} V_{u_j\alpha}^* m_{u_j}\mathcal{F}_{\alpha}^L+\sum_{l_{\rm NP}}\mathcal{C}_{l_{\rm NP}}^{\,*}\mathcal{K}^L_{l_{\rm NP}}\,.
    \end{align}
\end{subequations}
It is important to note that, among the various loop diagrams shown in fig.~\ref{fig:top_CP_asym}, only the diagram fig.~\ref{fig:top_rad_3} gives rise to an absorptive part.
In a similar manner, one can compute the polarised amplitudes for the $t \to u_j g_{\pm}$ processes following the same procedure.
\paragraph{} The polarised CP asymmetries $\Delta_{\rm CP\,,\pm}$ between $t \to u_j\,\gamma_{\pm}$ and its CP conjugate process $\bar{t}\to\bar{u}_j\gamma_{\mp}$ are defined respectively, as \cite{Balaji:2019fxd}
\begin{subequations}\label{eq:polarised_CP_asym}
    \begin{align}
        \Delta_{\rm CP,+}&=\frac{\Gamma(t \to u_j \gamma_+)-\Gamma(\bar{t}\to \bar{u}_j\gamma_-)}{\Gamma(t \to u_j\gamma)+\Gamma(\bar{t} \to \bar{u}_j\gamma)}=\frac{\left| f^{L}_{fi} \right|^{2} - \left| \bar{f}^{R}_{if} \right|^{2}}
{\left| f^{L}_{fi} \right|^{2} + \left| f^{R}_{fi} \right|^{2}
 + \left| \bar{f}^{L}_{if} \right|^{2} + \left| \bar{f}^{R}_{if} \right|^{2}} \,,\\
        \Delta_{\rm CP,-}&=\frac{\Gamma(t \to u_j \gamma_-)-\Gamma(\bar{t}\to \bar{u}_j\gamma_+)}{\Gamma(t \to u_j\gamma)+\Gamma(\bar{t} \to \bar{u}_j\gamma)}=\frac{\left| f^{R}_{fi} \right|^{2} - \left| \bar{f}^{L}_{if} \right|^{2}}
{\left| f^{L}_{fi} \right|^{2} + \left| f^{R}_{fi} \right|^{2}
 + \left| \bar{f}^{L}_{if} \right|^{2} + \left| \bar{f}^{R}_{if} \right|^{2}} \,,
    \end{align}
\end{subequations}
A photon–polarization–independent CP asymmetry can also be obtained by summing over the two polarized asymmetries $\Delta_{\rm CP,+}$ and $\Delta_{\rm CP,-}$, which gives
\begin{align}\label{eq:direct_CP_asym}
    \Delta_{\rm CP}&=\frac{\Gamma(t \to u_j \gamma_+)-\Gamma(\bar{t}\to \bar{u}_j\gamma_-)+\Gamma(t \to u_j \gamma_-)-\Gamma(\bar{t}\to \bar{u}_j\gamma_+)}{\Gamma(t \to u_j\gamma)+\Gamma(\bar{t} \to \bar{u}_j\gamma)}\nonumber\,,\\
 &=\frac{\Gamma(t \to u_j\gamma)-\Gamma(\bar{t}\to \bar{u}_j\gamma)}{\Gamma(t \to u_j\gamma)+\Gamma(\bar{t} \to \bar{u}_j\gamma)}=\frac{\left| f^{L}_{fi} \right|^{2}+\left| f^{R}_{fi} \right|^{2} - \left| \bar{f}^{R}_{if} \right|^{2}- \left| \bar{f}^{L}_{if} \right|^{2}}
{\left| f^{L}_{fi} \right|^{2} + \left| f^{R}_{fi} \right|^{2}
 + \left| \bar{f}^{L}_{if} \right|^{2} + \left| \bar{f}^{R}_{if} \right|^{2}}\,.
\end{align}
Using the parametrisations of the form factors given in eqs.~\eqref{eq:top_asym_FF1} and~\eqref{eq:top_asym_FF2}, the two polarised CP asymmetries can be further expressed in terms of the particle masses, loop functions, CKM matrix elements, and complex NP couplings. The resulting expression can then be written as
\begin{subequations}
\begin{align}\label{eq:top_cp_plus}
    \Delta_{\rm CP,+}&= -\frac{\makecell[l]{m_{u_j}^2\,\displaystyle\sum_{\alpha,\,\beta}\mathcal{J}_{\alpha\beta}\,\mathrm{Im}\left(\mathcal{F}^L_{\alpha}\mathcal{F}^{L\,*}_{\beta}\right)+2m_{u_j}\sum_{\alpha,\,l_{\rm NP}}\mathrm{Im}\left(V_{t\alpha}V_{u_j\alpha}^*\mathcal{C}_{l_{\rm NP}}\right)\mathrm{Im}\left(\mathcal{K}^L_{l_{\rm NP}}\mathcal{F}^{L\,*}_{\alpha}\right)\\ \quad \displaystyle+\sum_{l_{\rm NP}\neq l_{\rm NP}^{\prime}}\mathrm{Im}\left(\mathcal{C}_{l_{\rm NP}}\mathcal{C}^{\,*}_{l^{\prime}_{\rm NP}}\right)\mathrm{Im}\left(\mathcal{K}^L_{l_{\rm NP}}\mathcal{K}^{L\,*}_{l^{\prime}_{\rm NP}}\right)}}{\mathcal{D}}
    \end{align}
    \begin{align}\label{eq:top_cp_minus}
    \Delta_{\rm CP,-}&= -\frac{\makecell[l]{m_{t}^2\,\displaystyle\sum_{\alpha,\,\beta}\mathcal{J}_{\alpha\beta}\,\mathrm{Im}\left(\mathcal{F}^R_{\alpha}\mathcal{F}^{R\,*}_{\beta}\right)+2m_{t}\sum_{\alpha,\,l_{\rm NP}}\mathrm{Im}\left(V_{t\alpha}V_{u_j\alpha}^*\mathcal{C}_{l_{\rm NP}}\right)\mathrm{Im}\left(\mathcal{K}^R_{l_{\rm NP}}\mathcal{F}^{R\,*}_{\alpha}\right)\\ \quad \displaystyle+\sum_{l_{\rm NP}\neq l_{\rm NP}^{\prime}}\mathrm{Im}\left(\mathcal{C}_{l_{\rm NP}}\mathcal{C}^{\,*}_{l^{\prime}_{\rm NP}}\right)\mathrm{Im}\left(\mathcal{K}^R_{l_{\rm NP}}\mathcal{K}^{R\,*}_{l^{\prime}_{\rm NP}}\right)}}{\mathcal{D}}
\end{align}
\end{subequations}
with the denominator given by
\begin{align}\label{eq:top_asym_deno}
    \mathcal{D}=& \displaystyle\sum_{a,\beta} \mathcal{R}_{\alpha\beta}\left\{m_t^2\,\mathrm{Re}\left(\mathcal{F}^R_{\alpha}\mathcal{F}^{R\,*}_{\beta}\right)+m_{u_j}^2\,\mathrm{Re}\left(\mathcal{F}^L_{\alpha}\mathcal{F}^{L\,*}_{\beta}\right)\right\}\nonumber\,,\\
    &+2\sum_{\alpha,\,l_{\rm NP}}\mathrm{Re}\left(V_{t\alpha}V_{u_j\alpha}^*\mathcal{C}_{l_{\rm NP}}\right)\left\{m_t\,\mathrm{Re}\left(\mathcal{F}^{R\,*}_{\alpha}\mathcal{K}^R_{l_{\rm NP}}\right)+m_{u_j}\mathrm{Re}\left(\mathcal{F}^{L\,*}_{\alpha}\mathcal{K}^L_{l_{\rm NP}}\right)\right\}\nonumber\,,\\
    &+\sum_{l_{\rm NP}\,,l^{\prime}_{\rm NP}}\mathrm{Re}\left(\mathcal{C}_{l_{\rm NP}}\mathcal{C}^{\,*}_{l^{\prime}_{\rm NP}}\right)\left\{\,\mathrm{Re}\left(\mathcal{K}^L_{l_{\rm NP}}\mathcal{K}^{L\,*}_{l^{\prime}_{\rm NP}}\right)+\,\mathrm{Re}\left(\mathcal{K}^R_{l_{\rm NP}}\mathcal{K}^{R\,*}_{l^{\prime}_{\rm NP}}\right)\right\}\,.
\end{align}
where $\alpha,\beta={d,s,b}$ and 
\begin{align}
    \mathcal{J}_{\alpha\beta}&=\mathrm{Im}\left(V_{t\alpha}^*V_{u_j\alpha}V_{t\beta}V^*_{u_j\beta}\right)\,,& \mathcal{R}_{\alpha\beta}=\mathrm{Re}\left(V_{t\alpha}^*V_{u_j\alpha}V_{t\beta}V^*_{u_j\beta}\right)\,.
\end{align}
It is evident from eqs.~\eqref{eq:top_cp_plus} and~\eqref{eq:top_cp_minus} that a non-vanishing CP asymmetry requires the interference of at least two amplitudes possessing different weak and strong phases. In our analysis, the weak phase originates from the complex NP effective couplings, the Jarlskog invariant $(\mathcal{J}_{\alpha\beta})$, and the products of CKM matrix elements with these complex couplings, which together quantify the strength of CP violation. 

\begin{table}[htb!]
\centering
\begin{tabular}{l c c }
\toprule
\rowcolor{gray!20}
\textbf{Decay Channel} &
$\Delta_{\mathrm{CP},+}$ &
$\Delta_{\mathrm{CP},-}$  
\\
\midrule
$t\to c~\gamma$ & $(0.01 \pm 1.08)\times 10^{-3}$ & $(0.01 \pm 4.67)\times 10^{-2}$ \\
$t\to u~\gamma$ & $(-0.38 \pm 1.79)\times 10^{-9}$ & $(-0.26 \pm 3.63)\times 10^{-6}$ \\
$t\to c~g$ & $(0.0\pm 0.80)\times 10^{-4}$ & $(0.13 \pm 9.88)\times 10^{-3}$ \\
$t\to u~g$ & $(-0.56 \pm 4.86)\times 10^{-9}$ & $(-0.15 \pm 2.72)\times 10^{-6}$ \\
\bottomrule
\end{tabular}
\label{tab:cp_asym}
\caption{Results for the CP asymmetries for different top FCNC radiative decay channels $t\to u_j ~\gamma$ and $t\to u_j ~g$.}
\end{table}

Since the left-chiral effective couplings are generally less constrained than their right-chiral counterparts across the various top FCNC processes, and because we have already discussed that their allowed parameter space can be further restricted using collider bounds, which we have included in our global analysis through the penalty function, we tighten the left-chiral 
tensor and scalar couplings by incorporating the current collider limits, 
while keeping the remaining couplings at their global-fit values. In table~\ref{tab:cp_asym}, we summarise our predictions for the CP asymmetries in different radiative top FCNC decays. The results clearly exhibit the 
hierarchy $\Delta_{\rm CP,+} < \Delta_{\rm CP,-}$, which follows directly from angular momentum conservation and the $V\!-\!A$ structure of the weak 
interaction, indicate that the photons (or gluons) emitted in top FCNC 
radiative decays are predominantly left-handed. Moreover, the inclusion of complex  NP couplings leads to a substantial enhancement of the CP-violating observables. The resulting CP asymmetries are significantly larger than the corresponding 
SM expectations~\cite{Aguilar-Saavedra:2002lwv}, with 
enhancements of at least $\mathcal{O}(10^3)\times \Delta_{\rm CP}^{\rm SM}$ for 
the $t \to c\gamma$ and $t \to c g$ channel which can be tested in future precision experiments.

\section{Summary}\label{sec:summary}

In this work, we have presented a comprehensive global analysis of FCNC interactions involving the top quark within both an effective-coupling framework and the SMEFT. The analysis combines direct collider searches with a wide range of indirect probes, including flavour observables, electroweak precision measurements, Higgs observables, and stringent CP-violating bounds from the neutron EDM, thereby providing a unified assessment of the current constraints on anomalous top-quark interactions.

At the effective-coupling level, we have derived stringent constraints on the complex top-gluon, top-photon, top-$Z$, and top-Higgs FCNC couplings. A recurring feature of the analysis is the strong complementarity between collider and low-energy observables. While collider measurements generally provide the dominant sensitivity to several top-FCNC branching fractions, flavour and precision observables play a crucial role in constraining specific chiral structures and CP-violating phases. In particular, the top-gluon and top-photon right-handed dipole interactions are found to be subject to relatively stronger constraints than their left-handed counterparts owing to their enhanced contributions to radiative and rare $B$-meson decays. While other constraints remain broadly consistent with, and in some cases complementary to, existing collider bounds. The global fit also imposes meaningful constraints on the imaginary parts of the FCNC couplings, demonstrating the importance of CP-sensitive observables, which probe information inaccessible to collider measurements that are primarily sensitive to the overall magnitude of the couplings.

The SMEFT analysis reveals several features that are hidden in a model-independent effective-coupling description. Gauge invariance correlates different FCNC interactions and connects top-quark observables to flavour processes in the down-type quark sector, allowing low-energy measurements to place stringent constraints on several WCs. The strongest effects are observed for the electroweak dipole operators, which are constrained at the level of $\mathcal{O}(10^{-7})$-$\mathcal{O}(10^{-8})$ by radiative and rare $B$-meson decays. These bounds are directly translated into substantially improved limits on top-photon and top-$Z$ FCNC interactions, while simultaneously providing sensitivity to CP-violating phases and chiral structures that are largely inaccessible to collider measurements. Furthermore, the renormalisation-group evolution of the WCs up to multi-TeV scales exhibits characteristic patterns dictated by the gauge structure, highlighting the SMEFT as a consistent framework for connecting low- and high-energy probes of new physics.

A clear hierarchy emerges between operators with flavour structures $i3$ and $3i$: the former are tightly constrained by flavour and electroweak precision observables, leading to stringent bounds on both their real and imaginary components, whereas the latter are constrained predominantly by current LHC searches for FCNC top-quark production and decay. In contrast, the top-Higgs operator remains comparatively weakly constrained, at the level of $\mathcal{O}(10^{-6})$. Translating the fitted WCs into predictions for FCNC top-quark branching fractions shows that, particularly for the $t\to q\gamma$ and $t\to qZ$ channels associated with the operators $\mathcal{O}^{uB}_{23}$, $\mathcal{O}^{uW}_{23}$, and $\mathcal{O}^{\phi q(-)}_{23}$, the SMEFT framework yields substantially stronger limits than the effective-coupling approach. Depending on the specific interaction, the resulting branching-fraction bounds are typically two to three orders of magnitude below the current experimental sensitivities. These improvements arise from SMEFT correlations that efficiently propagate the stringent constraints from low-energy flavour observables into the top-quark sector, demonstrating that indirect measurements can probe new-physics effects at a level that is competitive with, and in some cases exceeds, the reach of direct collider searches.

 The fitted WCs also lead to sharp predictions for a variety of FCNC top-quark decay channels. In particular, the rare decays $t \to u_j \ell^+\ell^-$ and $t \to u_j \nu\bar{\nu}$ are predicted to lie in the range $10^{-9}$$-$$10^{-6}$, providing well-motivated targets for future collider searches. By contrast, FCNC decays of the $Z$ and Higgs bosons remain highly suppressed within the scenarios considered here, such that any experimental observation of these modes would point to additional sources of new physics beyond the top-FCNC interactions analysed in this work. We further find that complex WCs can generate sizeable CP asymmetries in radiative top decays, offering a complementary probe of CP-violating new-physics effects. For the dipole and vector operators constrained in our analysis, the predicted branching fractions for $t\to q\gamma$ and $t\to qZ$ are typically several orders of magnitude below current experimental limits, thereby establishing theoretically clean and experimentally relevant benchmarks for future high-luminosity collider programmes.

Overall, our study highlights the necessity of a global and gauge-invariant EFT framework for interpreting top-quark FCNC interactions. The simultaneous use of flavour, electroweak precision, Higgs, EDM, and collider data not only yields the most stringent current constraints on the relevant couplings and WCs, but also provides unique sensitivity to their chiral and CP properties. Future improvements in flavour physics, CP-violation measurements, electroweak precision observables, and high-luminosity collider searches will further strengthen these constraints and enhance the potential of top-quark FCNC processes as probes of physics beyond the SM.

\section*{Acknowledgments}
SK would like to thank Dr.~Lipika Kolay and Dr.~Utsab Dey for several useful 
and insightful discussions.

\appendix
\section{RGEs: beta functions}\label{Append:RGE_beta}
In this section, we present the beta functions employed in our analysis, 
together with the values of the relevant SM input parameters 
used in the RGE.

\begin{subequations}\label{eq:beta_lambda2EW}
\begin{align}
    &[\beta^{uG}]_{rs}=-\frac{1}{36}\left(81 g_2^2+19g_1^2+204g_s^2\right)[\mathcal{C}^{uG}]_{rs}+6 g_2 g_s[\mathcal{C}^{uW}]_{rs}+\frac{10}{3}g_1 g_s[\mathcal{C}^{uB}]_{rs}\\
    &+2[\Gamma_u\Gamma_u^{\dagger}\mathcal{C}^{uG}]_{rs}-2[\Gamma_d\Gamma_d^{\dagger}\mathcal{C}^{uG}]_{rs}+[\mathcal{C}_{uG}\Gamma_u^{\dagger}\Gamma_u]+\gamma_H^{(Y)}[\mathcal{C}^{uG}]_{rs}+[\gamma_q^{(Y)}\mathcal{C}^{uG}]_{rs}+[\mathcal{C}^{uG}\gamma_u^{(Y)}]_{rs}\,,\nonumber
\end{align}

\begin{align}
    [\beta^{uW}]_{rs}&=-\frac{1}{36}(33 g_2^2+ 19 g_1^2 -96 g_s^2)[\mathcal{C}^{uW}]_{rs} + \gamma_H^{(Y)}[\mathcal{C}^{uW}]_{rs}+[\gamma_q^{(Y)} \mathcal{C}^{uW}]_{rs} +[\mathcal{C}^{uW} \gamma_u^{(Y)}]_{rs}\nonumber\\
    &+[\mathcal{C}^{uW}\Gamma^{\dagger}_u \Gamma_u]_{rs}+2[\Gamma_d \Gamma^{\dagger}_d \mathcal{C}^{uW}]_{rs}-\frac{1}{6}g_1 g_2 [\mathcal{C}^{uB}]_{rs}+\frac{8}{3} g_2 g_s[\mathcal{C}^{uG}]_{rs}
    \end{align}
    
    \begin{align}
      &[\beta^{uB}]_{rs}=-\frac{1}{36}(81 g_2^2-313 g_1^2 -96 g_s^2)[\mathcal{C}^{uB}]_{rs} + \gamma_H^{(Y)}[\mathcal{C}^{uB}]_{rs}+[\gamma_q^{(Y)} \mathcal{C}^{uB}]_{rs} +[\mathcal{C}^{uB} \gamma_u^{(Y)}]_{rs}\nonumber\\
    &-\frac{1}{2}g_1 g_2 [\mathcal{C}^{uW}]_{rs}-2[\Gamma_d\Gamma_d^{\dagger}\mathcal{C}^{uB}]_{rs}+2[\Gamma_u\Gamma_u^{\dagger}\mathcal{C}^{uB}]_{rs}+[\mathcal{C}^{uB}\Gamma_u^{\dagger}\Gamma_u]_{rs}+\frac{40}{9} g_1 g_s[\mathcal{C}^{uG}]_{rs}
\end{align}

\begin{align}
    [\beta^{\phi q\,(1)}]_{rs} &= \frac{1}{3}g_1^2[\mathcal{C}^{\phi q\,(1)}]_{rs} + \frac{3}{2} \bigg([ \Gamma_d \Gamma_d^{\dagger} \mathcal{C}^{\phi q\,(1)} ]_{rs} + [ \Gamma_u \Gamma_u^{\dagger} \mathcal{C}^{\phi q\,(1)} ]_{rs} + [ \mathcal{C}^{\phi q\,(1)} \Gamma_d \Gamma_d^{\dagger} ]_{rs} + [ \mathcal{C}^{\phi q\,(1)} \Gamma_u \Gamma_u^{\dagger} ]_{rs}\nonumber \\
    &\quad + 3[ \Gamma_d \Gamma_d^{\dagger} \mathcal{C}^{\phi q\,(3)} ]_{rs} - 3[ \Gamma_u \Gamma_u^{\dagger} C^{\phi q\,(3)} ]_{rs} + 3[ C^{\phi q\,(3)} \Gamma_d \Gamma_d^{\dagger} ]_{rs} - 3[ C^{\phi q\,(3)} \Gamma_u \Gamma_u^{\dagger} ]_{rs} \bigg)\nonumber\\
    &\quad +2 \gamma_H^{(Y)}[C^{\phi q\,(1)}]_{rs}+[\gamma_q^{(Y)} C^{\phi q\,(1)}]_{rs}+[C^{\phi q\,(1)} \gamma_q^{(Y)}]_{rs}
    \end{align}

    \begin{align}
    [\beta^{\phi q\,(3)}]_{rs}&=-\frac{17}{3}g_2^2[\mathcal{C}^{\phi q\,(3)}]_{rs}+\frac{1}{2}\Bigg(3[\Gamma_d \Gamma_d^{\dagger}\mathcal{C}^{\phi q\,(1)}]_{rs}-3[\Gamma_u \Gamma_u^{\dagger}\mathcal{C}^{\phi q\,(1)}]_{rs}+3[\mathcal{C}^{\phi q\,(1)}\Gamma_d \Gamma_d^{\dagger}]_{rs}\nonumber\\
   &\quad - 3[\mathcal{C}^{\phi q\,(1)}\Gamma_u \Gamma_u^{\dagger}]_{rs}+[\Gamma_d \Gamma_d^{\dagger}\mathcal{C}^{\phi q\,(3)}]+[\Gamma_u \Gamma_u^{\dagger}\mathcal{C}^{\phi q\,(3)}]_{rs}+[\mathcal{C}^{\phi q\,(3)}\Gamma_d \Gamma_d^{\dagger}]_{rs}+ [\mathcal{C}^{\phi q\,(3)}\Gamma_u \Gamma_u^{\dagger}]_{rs} \Bigg)\nonumber\\
   &\quad +2 \gamma_H^{(Y)}[\mathcal{C}^{\phi q\,(3)}]_{rs}+[\gamma_q^{(Y)} \mathcal{C}^{\phi q\,(3)}]_{rs}+[\mathcal{C}^{\phi q\,(3)} \gamma_q^{(Y)}]_{rs}
\end{align}

\begin{align}
    [\beta^{u\phi}]_{rs}&=-\Bigg(\frac{35}{12}g_1^2+\frac{27}{4}g_2^2+8 g_s^2\Bigg)[\mathcal{C}^{u\phi}]_{rs}-g_1(5g_1^2-3g_2^2)[\mathcal{C}^{uB}]_{rs}+ g_2(5 g_1^2-9g_2^2)[\mathcal{C}^{uW}]_{rs}\nonumber\\
    &\quad -(3 g_2^2-g_1^2)[\Gamma_u \mathcal{C}^{\phi u}]_{rs}+4 g_1^2[\mathcal{C}^{\phi q\,(1)}\Gamma_u]_{rs}- 4 g_1^2[\mathcal{C}^{\phi q\,(3)}\Gamma_u]_{rs}-5 g_1[\mathcal{C}^{uB}\Gamma_u^{\dagger}\Gamma_u+\Gamma_u \Gamma_u^{\dagger}\mathcal{C}^{uB}]_{rs}\nonumber\\
    &\quad -3g_2[\mathcal{C}^{uW}\Gamma_u^{\dagger}\Gamma_u-\Gamma_u \Gamma_u^{\dagger}\mathcal{C}^{uW}]_{rs}-12 g_2[\Gamma_d \Gamma_d^{\dagger}\mathcal{C}^{uW}]_{rs}-16 g_s[\mathcal{C}^{uG}\Gamma_u^{\dagger}\Gamma_u+\Gamma_u \Gamma_u^{\dagger}\mathcal{C}^{uG}]_{rs}\nonumber\\
    &\quad +\lambda\left(12[\mathcal{C}^{u\phi}]_{rs}-2[\mathcal{C}^{\phi q\,(1)}\Gamma_u]_{rs}+6[\mathcal{C}^{\phi q\,(3)}\Gamma_u]+2[\Gamma_u \mathcal{C}^{\phi u}]_{rs}\right)\\
    &\quad -2[\mathcal{C}^{\phi q\,(1)}\Gamma_u \Gamma_u^{\dagger}\Gamma_u]_{rs}+6[\mathcal{C}^{\phi q\,(3)}\Gamma_d \Gamma_d^{\dagger}\Gamma_u]_{rs}+2[\Gamma_u \Gamma_u^{\dagger}\Gamma_u \mathcal{C}^{\phi u}]_{rs}+4[\mathcal{C}^{u\phi}\Gamma_u^{\dagger} \Gamma_u]_{rs}\nonumber\\
    &\quad +5[\Gamma_u \Gamma_u^{\dagger}\mathcal{C}^{u\phi}]_{rs}-2[\Gamma_d \Gamma_d^{\dagger}\mathcal{C}^{u\phi}]_{rs}+3 \gamma_H^{(Y)}[\mathcal{C}^{u\phi}]_{rs}+[\gamma_q^{(Y)} \mathcal{C}^{u\phi}]_{rs}+[\mathcal{C}^{u\phi}\gamma_u^{(Y)}]_{rs}\nonumber
\end{align}
\end{subequations}
Here, $\gamma^i$ denote the wavefunction renormalisation constants of the 
corresponding fields, and $\Gamma_i$ represent the quark Yukawa matrices. In eq.~\eqref{eq:beta_lambda2EW} wavefunction renormalisation terms are
\begin{align}
&\gamma_H^{(Y)}=\frac{3}{4}(g_1^2+3 g_2^2)+\rm{Tr}\left(3\Gamma_u\Gamma_u ^{\dagger}+3\Gamma_d \Gamma_d^{\dagger}+\Gamma_e\Gamma_e^{\dagger}\right) \,,& \nonumber\\
    &[\gamma_q^{(Y)}]_{rs} =\frac{1}{2}[\Gamma_u \Gamma_u^{\dagger}+\Gamma_d \Gamma_d^{\dagger}] \,,& \nonumber\\
    &[\gamma_u^{(Y)}]_{rs} =[\Gamma_u^{\dagger} \Gamma_u]_{rs}\,, & \\
   & [\gamma_d^{(Y)}]_{rs} =[\Gamma_d^{\dagger} \Gamma_d]_{rs}\,,& \nonumber\\
    &[\gamma_e^{(Y)}]_{rs} = [ \Gamma_e^{\dagger} \Gamma_e]_{rs}\,.& \nonumber
\end{align} 
\textbf{Input Parameters:} All SM input parameters at the electroweak scale, used in our analysis, are as follows:
\begin{align}
    g_1=0.3576\,,\,\,g_2=0.6515\,,\,\,g_s=1.220\,,\,\,\lambda=0.2813\,,\\
    m_u=1.27\times 10^{-3}\,,\,\,m_d=2.7\times 10^{-3}\,,\,\,m_s=5.51\times 10^{-2}\,,\,\,m_c=0.635\,,\,\,m_b=2.85\,,\nonumber
\end{align}
All the light quark masses are given in units of GeV. Different scales used in this works are
\begin{align}
    \mu_{\rm had}=2\mathrm{GeV}\,,\,\mu_b=4.18\mathrm{GeV}\,,\, \mu_{\rm EW} = 91.1876~\mathrm{GeV}\,,\,\mu_t=172.69\mathrm{GeV}\,,\,\mu_{\Lambda}=\mathcal{O}\mathrm{(TeV)}\,.\nonumber
\end{align}

\section{Theoretical Framework: Connecting NP Interactions to Observables}
\label{Append:np_interactions}
In this appendix, we present the detailed derivation of our theoretical framework, explicitly demonstrating how NP interactions modify various observables at the loop level. To bridge the different physical sectors, we outline the proper running matrices required to connect these observables and provide the explicit analytical forms of the loop contributions. The loop calculations are cross-verified using the automated \textsc{Mathematica} package \texttt{Package-X}~\cite{Patel:2016fam} and \texttt{FeynCalc}~\cite{Mertig:1990an}. The loop factors are expressed as functions of the renormalisation scale $\mu$. They can subsequently be evolved to other energy scales using the appropriate RGEs, thereby allowing a consistent connection to observables defined at different scales, as discussed earlier in section~\ref{sec:methodology}. It is worth noting that the loop functions are written in terms of ratios of heavy masses, defined as $x_i \equiv m_i^2/M_W^2$, where $m_i$ denotes the mass of the $i$th particle (quark, gauge boson, or scalar). In some cases, mass ratios of the form $x_{i/j} \equiv m_i^2/m_j^2$ are also used. We define all the LEFT couplings relevant to low-energy observables as $L_i$, following the operator basis notation of ref.~\cite{Jenkins:2017jig}. Furthermore, for clarity and consistency with the existing literature, we also provide the correspondence between this LEFT basis and most commonly used operator bases for the study of low-energy observables. 
\subsection{FCNC Processes}\label{sec:appFCNC}
In the SM, FCNC processes are loop suppressed, making them highly sensitive to the new physics. In this subsection, we study observables associated with decays of $B$ and $K$ mesons and use their experimental measurements to constrain the top FCNC effective couplings.
\paragraph{\underline{\textbf{Radiative decays:}}}

\begin{figure}[t]
	\centering
	\subfloat[]{
		\begin{tikzpicture}
			\begin{feynman}\label{fig:rad_1}
				\vertex[square dot, draw=red, fill=red, minimum size=4pt, inner sep=2pt](a1){};
				\vertex[below left=1cm of a1](a2);
				\vertex[below left=0.8cm of a2](a3){\(b\)};
				\vertex[above left=1cm of a1](a4);
				\vertex[above left=0.8cm of a4](a5){\(s,d\)};
				\vertex[right=1.5cm of a1](a6){\(\gamma\)};
				
				\diagram* {
					(a3) --[fermion] (a2) --[fermion, very thick, edge label'=\(u_i\)] (a1) --[fermion, very thick, edge label'=\(u_j\)] (a4) --[fermion] (a5),
					(a2) --[boson, edge label=\(W\)] (a4),
					(a1) --[red, boson, very thick] (a6),
				};
			\end{feynman}
		\end{tikzpicture}
	}
	\hspace{1cm}
	\subfloat[]{
		\begin{tikzpicture}
			\begin{feynman}\label{fig:rad_2}
				\vertex[square dot, blue, minimum size=4pt, inner sep=2pt](a1){};
				\vertex[below left=1cm of a1](a2);
				\vertex[below left=0.8cm of a2](a3){\(b\)};
				\vertex[above left=1cm of a1](a4);
				\vertex[above left=0.8cm of a4](a5){\(s,d\)};
				\vertex[right=1.5cm of a1](a6){\(g\)};
				
				\diagram* {
					(a3) --[fermion] (a2) --[fermion, very thick, edge label'=\(u_i\)] (a1) --[fermion, very thick, edge label'=\(u_j\)] (a4) --[fermion] (a5),
					(a2) --[boson, edge label=\(W\)] (a4),
					(a1) --[blue, gluon, very thick] (a6),
				};
			\end{feynman}
		\end{tikzpicture}
	}\\
	\subfloat[]{
		\begin{tikzpicture}
			\begin{feynman}\label{fig:b2sll_1}
				\vertex[square dot, draw=red, fill=red, minimum size=4pt, inner sep=2pt](a1){};
				\vertex[below left=1cm of a1](a2);
				\vertex[below left=0.8cm of a2](a3){\(b\)};
				\vertex[above left=1cm of a1](a4);
				\vertex[above left=0.8cm of a4](a5){\(s,d\)};
				\vertex[right=1.cm of a1](a6);
				\vertex[above right=1.5cm of a6](a7){\(\ell\)};
				\vertex[below right=1.5cm of a6](a8){\(\ell\)};
				\diagram* {
					(a3) --[fermion] (a2) --[fermion, very thick, edge label'=\(u_i\)] (a1) --[fermion, very thick, edge label'=\(u_j\)] (a4) --[fermion] (a5),
					(a2) --[boson, edge label=\(W\)] (a4),
					(a1) --[red, boson, very thick, edge label=\(\gamma/Z\)] (a6),
					(a8) --[fermion](a6) --[fermion](a7),
				};
			\end{feynman}
		\end{tikzpicture}
	}
	\subfloat[]{
		\begin{tikzpicture}
			\begin{feynman}\label{fig:b2sll_2}
				\vertex[square dot, draw=violet, fill=violet, minimum size=4pt, inner sep=2pt](a1){};
				\vertex[below left=1cm of a1](a2);
				\vertex[below left=0.8cm of a2](a3){\(b\)};
				\vertex[above left=1cm of a1](a4);
				\vertex[above left=0.8cm of a4](a5){\(s,d\)};
				\vertex[right=1.cm of a1](a6);
				\vertex[above right=1.5cm of a6](a7){\(\ell\)};
				\vertex[below right=1.5cm of a6](a8){\(\ell\)};
				\diagram* {
					(a3) --[fermion] (a2) --[fermion, very thick, edge label'=\(u_i\)] (a1) --[fermion, very thick, edge label'=\(u_j\)] (a4) --[fermion] (a5),
					(a2) --[boson, edge label=\(W\)] (a4),
					(a1) --[violet, scalar, very thick, edge label=\(H\)] (a6),
					(a8) --[fermion](a6) --[fermion](a7),
				};
			\end{feynman}
		\end{tikzpicture}
	}
	\subfloat[]{
		\begin{tikzpicture}
			\begin{feynman}\label{fig:Inv_1}
				\vertex[square dot, draw=red, fill=red, minimum size=4pt, inner sep=2pt](a1){};
				\vertex[below left=1cm of a1](a2);
				\vertex[below left=0.8cm of a2](a3){\(b\)};
				\vertex[above left=1cm of a1](a4);
				\vertex[above left=0.8cm of a4](a5){\(s,d\)};
				\vertex[right=1.cm of a1](a6);
				\vertex[above right=1.5cm of a6](a7){\(\nu\)};
				\vertex[below right=1.5cm of a6](a8){\(\nu\)};
				\diagram* {
					(a3) --[fermion] (a2) --[fermion, very thick, edge label'=\(u_i\)] (a1) --[fermion, very thick, edge label'=\(u_j\)] (a4) --[fermion] (a5),
					(a2) --[boson, edge label=\(W\)] (a4),
					(a1) --[red, boson, very thick, edge label=\(Z\)] (a6),
					(a8) --[fermion](a6) --[fermion](a7),
				};
			\end{feynman}
		\end{tikzpicture}
	}
	\caption{Feynman diagrams for various FCNC processes receiving contributions from top FCNC interactions. Here, $u_i$ and $u_j$ represent up-type quarks, where one is the top quark ($t$) and the other can be either an up ($u$) or charm ($c$) quark.}
	
	\label{fig:FCNC_processes}
\end{figure}
Radiative decays, such as $b \to s(d) \gamma$, are loop-induced processes in the SM and proceed via electromagnetic penguin diagrams. In the SM, these decays predominantly occur via loops involving heavy virtual particles, especially the top quark, along with $W$ bosons. In our analysis, alongside the SM penguin loop contributions, we also obtain additional NP effects from top FCNC operators, as shown in figs.~\ref{fig:rad_1} and \ref{fig:rad_2}. We can match the contributions of these diagrams to the most general Hamiltonian describing radiative decays $(b \to d_i \,\gamma)$ which can be written as (in WET basis):
\begin{equation}\label{eq:Heff_rad}
\mathcal{H}_{\rm eff}^{b \to d_i \gamma} = -\frac{4G_F}{\sqrt{2}} V_{tb} V_{td_i}^* \sum_{i=7,8} \left( C_i(\mu) O_i + C_i'(\mu) O_i' \right) ,
\end{equation}
Here, $d_i = b, s$ are the down-type quarks involved in the radiative transition, and the explicit expressions for the operators appearing in eq.~\eqref{eq:Heff_rad} are given below:
\begin{subequations}
\begin{align}\label{eq:opro7o8}
	O_{7} &= \frac{e}{16 \pi^2} m_b
	(\bar{d_j} \sigma_{\mu \nu} P_R b) F^{\mu \nu} ,&
	O_{7}^\prime &= \frac{e}{16 \pi^2} m_b
	(\bar{d_j} \sigma_{\mu \nu} P_L b) F^{\mu \nu} , \\
	O_{8} &= \frac{g_s}{16\pi^2} m_b
	(\bar{d_j} \sigma_{\mu \nu} T^a P_R b) G^{\mu \nu \, a} ,&
	O_{8}^\prime &= \frac{g_s}{16 \pi^2} m_b
	(\bar{d_j} \sigma_{\mu \nu} T^a P_L b) G^{\mu \nu \, a}.
	\end{align}
    \end{subequations}
In the SM, both the coefficients $C_7$ and $C_8$ contribute to the radiative decay $b \to s(d)\gamma$. In addition to the SM contributions, the WCs defined above may include NP contributions. To disentangle the NP effects from the SM ones, we redefine the low-energy WCs as
\begin{align}
	C_i(\mu) = C_i^{\rm SM}(\mu) + C_i^{\rm NP}(\mu)\,.
\end{align}
The SM predictions for the WCs $C_i^{\rm SM}(\mu_b)$, incorporating both QED and QCD corrections, are available in the literature~\cite{Mahmoudi:2024zna}.

In our analysis, we obtain NP contributions to the operators $O_7$ and $O_8$. It is important to note that the contribution from $O_8$ to the $b \to d_i \gamma$ process arises through RGE mixing with the $O_7$ operator ~\cite{Chetyrkin:1996vx,Buras:1998raa}, and 
\begin{align}\label{eq:RGEC78}
\begin{pmatrix}
	        C_7\\
	        C_8
	     \end{pmatrix}_{\mu_b}=\begin{pmatrix}
	 0.66301 & 0.09259 \\
	 0.00326 & 0.69877
	 \end{pmatrix}\begin{pmatrix}
	         C_7\\
	         C_8
	     \end{pmatrix}_{\mu_{\rm{EW}}} \,.
 \end{align}
After computing the one-loop diagrams and applying the appropriate matching conditions, we express the $C_i^{\rm NP}$ in terms of the top-quark FCNC couplings and the relevant mass scales. These matching relations are obtained at the electroweak scale. It is important to note that the WET basis defined above is not identical to the LEFT basis; the mapping between these two sets of couplings can be found in ref~\cite{Dekens:2019ept}. Using these relations, we rewrite the NP contributions to the $C_{7,8}^{\rm NP}(\mu)$ in terms of the corresponding LEFT coefficients as follows:
\begin{align}
	C_7^{\rm NP}(\mu) &= N_{d_i b}^{\rm rad}\frac{e}{m_b\, }\, L^{d\gamma}_{d_i b}(\mu)\,, \quad 
	C_8^{\rm NP}(\mu) =N_{d_i b}^{\rm rad} \frac{e^2}{g_s\, m_b\, }\, L^{dG}_{d_i b}(\mu)\,.
\end{align} 
\begin{subequations}
\begin{align}\label{eq:Append_FCNC_rad}
    L^{d\gamma}_{d_i b}(\mu)&=-i\frac{e\, m_b\,g_2^2}{16\pi^2 M_W^2}\Bigg[\frac{V_{tb} V^*_{u_is}(1+x_t)\lambda_R^{u_i t}(\mu)-2V_{ts}^* V_{u_i b}(1-x_t)\lambda_R^{tu_i\,*}(\mu)}{8(x_t-1)}\nonumber\\
    &+\frac{V_{tb} V_{u_is}^*\lambda_R^{u_i t}(\mu)}{4}\log\left(\frac{\mu^2}{M_W^2}\right)-\frac{V_{tb}V^*_{u_i s} x_t^2 \lambda_R^{u_it}(\mu)}{4(x_t-1)^2}\log x_t\Bigg]
\end{align}
\begin{align}\label{eq:Append_FCNC_rad1}
    L^{dG}_{d_ib}(\mu)&=-i \frac{g_s\, m_b\,g_2^2}{16\pi^2 M_W^2}\Bigg[\frac{V_{tb} V^*_{u_is}(1+x_t)\xi_R^{u_i t}(\mu)-2V_{ts}^* V_{u_i b}(1-x_t)\xi_R^{tu_i\,*}(\mu)}{8(x_t-1)}\nonumber\\&+\frac{V_{tb} V_{u_is}^*\xi_R^{u_i t}(\mu)}{4}\log\left(\frac{\mu^2}{M_W^2}\right)
    -\frac{V_{tb}V^*_{u_i s} x_t^2 \xi_R^{u_it}(\mu)}{4(x_t-1)^2}\log x_t\Bigg]
\end{align}
\end{subequations}
The mathematical expressions for $L^{d\gamma}_{d_i b}(\mu)$ and $L^{dG}_{d_i b}(\mu)$ are given in eqs.~(\ref{eq:Append_FCNC_rad}-\ref{eq:Append_FCNC_rad1}).
Here, the prefactor $(N_{d_i b}^{\rm rad})^{-1} = \frac{4 G_F}{\sqrt{2}}\, \frac{\alpha_{\rm em}}{4\pi}\, V_{ts}^* V_{tb}$, where $\alpha_{\rm em}$ denotes the fine-structure constant, approximately equal to $1/127$. Finally, we obtain $C_{7,8}^{\rm NP}(\mu_b)$ by evolving these expressions using the RGEs defined in eq.~\ref{eq:RGEC78}.

In our numerical analysis, we include measurements of both inclusive and exclusive radiative decays provided by various experimental collaborations~\cite{Belle:2017hum,Belle-II:2024tru,Belle:2014sac,HeavyFlavorAveragingGroupHFLAV:2024ctg}. A list of different radiative processes, along with their corresponding SM predictions and experimental values, is tabulated in Table~\ref{tab:Inv}. The branching ratio of the inclusive radiative decay $B \to X_s \gamma$, in the presence of NP WCs, can be expressed as~\cite{Misiak:2020vlo}:
\begin{equation}
	\mathcal{B}(B \to X_{s} \gamma) \times 10^{4} = (3.40 \pm 0.17) - 8.25\, C_{7}^{\rm NP}(\mu_{\rm EW}) - 2.10\, C_{8}^{\rm NP}(\mu_{\rm EW}) \,. 
\end{equation}
On the other hand, the branching ratios for exclusive decays into a vector meson can be written as~\cite{Paul:2016urs, Beneke:2001at}:
\begin{equation}
	\mathcal{B}(B_q \to V\gamma) = \tau_{B_q} \frac{G_F^2 \alpha_{\text{em}} m_{B_q}^3 m_b^2}{32 \pi^3} 
	\left( 1 - \frac{m_V^2}{m_{B_q}^2} \right)^3 
	|\lambda_t|^2 \left( |C_7(\mu_b)|^2 + |C_7'(\mu_b)|^2 \right) T_1(0),
\end{equation}
where $\lambda_{t} = V_{tq}^{\ast} V_{tb}$, and $T_1(0)$ is the tensor form factor, whose values can be found in~\cite{Bharucha:2015bzk}.

The direct asymmetry $B\to X_{s}\gamma$ \cite{Kagan:1998bh}\,can be written as,
\begin{align}\label{eq:acpbsgamma}
&A_{\text{CP}}(b \to s\gamma) \equiv \frac{\Gamma(B \to X_{\bar{s}}\gamma) - \Gamma(\bar{B} \to X_s\gamma)}{\Gamma(B \to X_{\bar{s}}\gamma) + \Gamma(\bar{B} \to X_s\gamma)} \notag \\
&\simeq - \frac{1}{|C_7(\mu_b)|^2 + |C_7^\prime(\mu_b)|^2} \Big( 1.23 \, \text{Im}[C_2(\mu_b) C_7^*(\mu_b)] - 9.52 \, \text{Im}[C_8(\mu_b) C_7^*(\mu_b) + C_8^\prime(\mu_b) C_7^{\prime *}(\mu_b)] \notag \\
&+ 0.10 \, \text{Im}[C_2(\mu_b) C_8^*(\mu_b)] \Big) 
 - 0.5 \quad (\text{in } \%) , 
\end{align}
For the exclusive radiative channels $B\to V\gamma$, we adopt the expressions for the direct CP asymmetry $A_{\mathcal{CP}}$ given in ref.~\cite{Bosch:2001gv}, which are used throughout our numerical analysis. In addition, the mixing-induced CP asymmetry in $B^0\to K^{*0}\gamma$ can be expressed at leading order as
\begin{equation}\label{eq:Sbsgamma}
S_{K^{*}\gamma}
\simeq
\frac{2\,\mathrm{Im}\left(e^{-i\phi_d}\,C_7(\mu_b)\,C_7'(\mu_b)\right)}
{|C_7(\mu_b)|^2 + |C_7'(\mu_b)|^2}\,,
\end{equation}
where $\phi_d$ denotes the phase of the $B_d$--$\bar B_d$ mixing amplitude. Using the value reported by the PDG~\cite{ParticleDataGroup:2024cfk} $\sin\phi_d = 0.710 \pm 0.011$.
we evaluate the corresponding NP contributions to $S_{K^{*}\gamma}$. In the SM, the asymmetry is suppressed by the ratio $m_s/m_b$, and the corresponding corrections can be found in ref.~\cite{Ball:2006eu}. The resulting prediction is
\begin{align}
S_{K^{*}\gamma}^{\rm SM}=(-2.3\pm1.6)\%,.
\end{align}

\paragraph{\underline{\textbf{Semileptonic $b\to d_i\ell\ell$ decays:}}}
The general Low-energy Lagrangian describing the $b \to d_i \ell^+\ell^-$ transition is expressed as:
\begin{equation} \label{eq:Heff_b2di}
	{\cal H}^{b\to d_i\ell\ell}_{eff} = - \frac{4\,G_F}{\sqrt{2}} V_{tb}V_{td_j}^\ast
	\left[  \sum_{i=1}^{6} C_i (\mu)
	O_i(\mu) + \sum_{i=7,8,9,10,P,S} \biggl(C_i (\mu)  O_i + C'_i (\mu) 
	O'_i\biggr)\right] \,,
\end{equation}
Among this large set of operators, in our analysis, the dipole-type operators $O_7$, $O_8$, the vector-type operators $O_9$, $O_{10}$ and the scalar-type operator $O_S$ receive additional contributions to their couplings from top-quark FCNC interactions, as illustrated in the diagrams shown in fig.~\ref{fig:b2sll_1}, \ref{fig:b2sll_2}. We are using the notation for $O_i$ from refs.~\cite{ Becirevic:2012fy, Bobeth:1999mk, Altmannshofer:2008dz}, where operator structure of $O_7$ and $O_8$ are already mentioned in eq.~\eqref{eq:opro7o8}.
\begin{subequations}
\begin{align}
    O_{9} &= \frac{e^2}{16 \pi^2} 
		(\bar{d_j} \gamma_{\mu} P_L b)(\bar{\ell} \gamma^\mu \ell) ,&
		O_{9}^\prime &= \frac{e^2}{16 \pi^2} 
		(\bar{d_j} \gamma_{\mu} P_R b)(\bar{\ell} \gamma^\mu \ell) , \\
		O_{10} &=\frac{e^2}{16 \pi^2}
		(\bar{d_j}  \gamma_{\mu} P_L b)(  \bar{\ell} \gamma^\mu \gamma_5 \ell) ,&
		O_{10}^\prime &=\frac{e^2}{16 \pi^2}
		(\bar{d_j}  \gamma_{\mu} P_R b)(  \bar{\ell} \gamma^\mu \gamma_5 \ell) , \\
		O_{S} &=\frac{e^2}{16\pi^2} m_b
		(\bar{d_j} P_R b)(  \bar{\ell} \ell) ,&
		O_{S}^\prime &=\frac{e^2}{16\pi^2} m_b
		(\bar{d_j} P_L b)(  \bar{\ell} \ell) .
\end{align}
\end{subequations}
where $\ell = e$, $\mu$, $\tau$, but we restrict our analysis to $\ell = \mu$, as only upper bounds exist for other lepton flavors. Similar to the radiative case, we follow the same procedure to incorporate NP effects within the low-energy WCs. Specifically, we split the contributions to the WCs into two parts: one arising solely from SM corrections (QCD and QED), and the other from NP effects. Again, one can find the SM contribution $(C_i^{\rm SM})$ from the ref.~\cite{Mahmoudi:2024zna}. Like the SM, NP contributions also arise via penguin diagrams shown in fig.~\ref{fig:FCNC_processes}. Among the possible NP diagrams, the one proportional to $V_{tb} V_{cs}^*$ provides the dominant contribution. We have performed the NP loop calculation and subsequently obtained the low-energy WCs in terms of top FCNC couplings at the electroweak scale following the appropriate matching conditions.

Like the radiative decay, we draw a connection between LEFT WCs and low-energy WCs $C_9^{\rm NP}, C_{10}^{\rm NP}$ and $C_S^{\rm NP}$ \cite{Jenkins:2017jig},
\begin{subequations}
	\begin{align}
		C_9^{\rm NP}(\mu)&=\lambda_1\left(L_{\underset{\mu\mu d_i b}{ed}}^{V,LL}+L_{\underset{d_i b \mu\mu}{de}}^{V,LR}
		\right)\,, & C_9^{\prime\rm NP}(\mu)&=\lambda_1\left(L_{\underset{\mu\mu d_i b}{ed}}^{V,LR}+L_{\underset{\mu\mu d_i b}{ed}}^{V,RR}
		\right)\,, \\
		C_{10}^{\rm NP}(\mu)&=\lambda_1\left(-L_{\underset{\mu\mu d_i b}{ed}}^{V,LL}+L_{\underset{d_i b \mu\mu}{de}}^{V,LR}\right)\,,& C_{10}^{\prime\rm NP}(\mu)&=\lambda_1\left(-L_{\underset{\mu\mu d_i b}{ed}}^{V,LR}+L_{\underset{ \mu\mu d_i b}{ed}}^{V,RR}\right)\,,\\
		C_S^{\rm NP}(\mu)&=\frac{\lambda_1}{ m_b}\left(L_{\underset{\mu\mu d_i b}{ed}}^{S,RR}+L_{\underset{\mu\mu b d_i}{ed}}^{S,RL\,*}\right)\,, &C_S^{\prime \rm NP}(\mu)&=\frac{\lambda_1}{ m_b}\left(L_{\underset{\mu\mu d_i b}{ed}}^{S,RL}+L_{\underset{\mu\mu b d_i}{ed}}^{S,RR\,*}\right)\,, \\ (\lambda_1)^{-1}&=-\frac{4 G_F}{\sqrt{2}}V_{tb} V_{ts}^* \frac{e^2}{16\pi^2}\,.
	\end{align}
\end{subequations}
where $L_i$ are the different LEFT WCs. We present the expressions of all these LEFT WCs in eqs.~(\ref{eq:Append_FCNC_rare}-\ref{eq:Append_FCNC_rare1}). \footnote{A discussion on one-loop SMEFT and LEFT matching for a subset of our chosen SMEFT operator sets relevant to the $b \to s \ell \ell$ process is already available in the literature~\cite{Aebischer:2015fzz}. We have verified our results against these existing studies.}

\begin{subequations}
\begin{align}\label{eq:Append_FCNC_rare}
    &L_{\underset{\mu\mu d_i b}{ed}}^{V,LL}(\mu)=-\frac{1}{2}\frac{e^2\,g_2^2}{16\pi^2 M_W^2}\Bigg[\frac{V_{tb}V_{u_i s}^* \lambda_R^{u_it}(\mu)-V_{ts}^*V_{ub}\lambda_R^{tu_i\,*}(\mu)}{2(x_t-1)}\log x_t\Bigg]\\
    &-\frac{g_2^4 g_{Z_L}}{16\pi^2M_Z^2 c_W^2}\Bigg[\left(V_{tb}V_{u_is}^* X_L^{u_i t}(\mu)+V_{ts}^*V_{u_i b}X_L^{t u_i*}(\mu)\right)\left\{\frac{2x_t+3}{16}+\frac{x_t}{8}\log\left(\frac{\mu^2}{m_t^2}\right)\right\}\Bigg]\nonumber
\end{align}
\begin{align}
    &L_{\underset{\mu\mu d_i b}{ed}}^{V,LR}(\mu)=-\frac{1}{2}\frac{e^2\,g_2^2}{16\pi^2 M_W^2}\Bigg[\frac{V_{tb}V_{u_i s}^* \lambda_R^{u_it}(\mu)-V_{ts}^*V_{ub}\lambda_R^{tu_i\,*}(\mu)}{2(x_t-1)}\log x_t\Bigg]\\
    &-\frac{g_2^4 g_{Z_R}}{16\pi^2M_Z^2 c_W^2}\Bigg[\left(V_{tb}V_{u_is}^* X_L^{u_i t}(\mu)+V_{ts}^*V_{u_i b}X_L^{t u_i*}(\mu)\right)\left\{\frac{2x_t+3}{16}+\frac{x_t}{8}\log\left(\frac{\mu^2}{m_t^2}\right)\right\}\Bigg]\nonumber
\end{align}
\begin{align}\label{eq:Append_FCNC_rare1}
    &L_{\underset{\mu\mu d_i b}{ed}}^{S,RR}(\mu)=L_{\underset{\mu\mu b d_i }{ed}}^{S,RL\,*}(\mu)=\frac{1}{2}\frac{m_b\,m_{\mu}\,m_t\, g_2^2}{16\pi^2 v \,M_H^2\,M_W^2}\Bigg[\frac{V_{tb}V_{u_i s}^*\eta_R^{u_i t}(\mu)+2 V_{ts}^*V_{u_ib}\eta_R^{tu_i\,*}(\mu)}{4\sqrt{2}}\log\left(\frac{\mu^2}{m_t^2}\right)\nonumber\\
    &+\frac{3 V_{tb}V_{u_i s}^* \eta_R^{u_i t}}{4\sqrt{2}(x_t-1)^2}\log x_t+\frac{V_{tb}V_{u_i s}^*(x_t-7)\eta_R^{u_i t}+4(x_t-1)V_{ts}^*V_{u_ib}\eta_R^{tu_i\,*}}{8\sqrt{2}(x_t-1)}\Bigg]
\end{align}
\end{subequations}
Furthermore, we have to run down our low-energy WCs to the scale of $\mu_b$ to utilise the experimental inputs related to $b\to d_i\ell\ell$ observables, and the governing RGEs connecting low-energy WCs at the scale $\mu_b$  are given by:
\begin{align}\label{eq:FCNC_EW_to_mb}
	\begin{pmatrix}
		C_9\\
		C_{10}\\
		C_S
	\end{pmatrix}_{{\mu_b}}&=\begin{pmatrix}
		 0.99522 & 0.00716 & 0 \\
		 0.00716 & 1.0& 0 \\
	 0& 0 &1.37433 
	\end{pmatrix}   \begin{pmatrix}
		C_9\\
		C_{10}\\
		C_S
\end{pmatrix}_{\mu_{\rm{EW}}}\,.
\end{align}


\paragraph{\underline{\textbf{Leptonic rare decays:}}}
Dileptonic decays of neutral mesons are another important class of decay channels for probing NP, as they are rare and highly suppressed within the SM. In our analysis, we have accounted for the following processes: $B_s^0 \to \ell^+ \ell^-$, $B^0 \to \ell^+ \ell^-$,  $K_{L,S} \to \ell^+ \ell^-$. For the $B$ meson decays, we consider only $\ell = \mu$, as for other lepton flavours, only upper bounds on the branching fractions are currently available. Rare leptonic decays can also be described using the effective Hamiltonian given in eq.~\eqref{eq:Heff_b2di}. For $K$-meson decays, the same effective Hamiltonian applies with the replacements $b \leftrightarrow s$ and $s \leftrightarrow d$. 

Based on the Hamiltonian mentioned above the expressions for the branching fraction for the rare dileptonic decays can be written as  
\begin{equation}
\begin{split} \label{eq:BR_Bqll}
\mathcal{B}(B_q \rightarrow \mu^+ \mu^-) = & \tau_{B_q} f_{B_q}^2  m_{B_q}^3 \frac{G_F^2 \alpha^2}{64 \pi^3} |V^*_{tq}V_{tb}|^2 \beta_{\mu}(m_{B_q}^2) \left[   \frac{m_{B_q}^2}{m_b^2} |C_s - C'_s|^2 \left(1-\frac{4m_{\mu}^2}{m_{B_q}^2}\right) \right.\\& \left.  + \bigg|\frac{m_{B_q}}{m_b}(C_p - C'_p) + 2\frac{m_{\mu}}{m_{B_q}} (C_{10} - C'_{10})\bigg|^2 \right]\,,
\end{split}
\end{equation}
where $\beta_{\mu}(q^2) = \sqrt{1 - \frac{4 m_{\mu}^2}{q^2}}$, and the $B_q$ meson decay constant is defined via the matrix element 
\begin{equation}
	\langle 0| \bar{s} \gamma_{\mu} P_L b | B_q(p) \rangle = \frac{i}{2} f_{B_q} p_\mu,
\end{equation}
where $q = (d, s)$. We have used the values of the decay constants: $f_{B} = (190.0 \pm 1.3)\, \text{MeV}$ and $f_{B_s} = (230.3 \pm 1.3)\, \text{MeV}$~\cite{FlavourLatticeAveragingGroupFLAG:2024oxs}, whereas for other input we have used the values from PDG24 \cite{ParticleDataGroup:2024cfk}. In the SM, the branching ratio for $B_q \to \ell^+ \ell^-$ receives contributions solely from the coupling $C_{10}$. However, in the presence of physics beyond the SM, additional contributions can arise from the other WCs $C_S, C_S^{\prime}, C_P, C_P^{\prime},$ and $C_{10}^{\prime}$. In our analysis, the NP contributions enter through $C_S^{\rm NP}$ and $C_{10}^{\rm NP}$ WCs.

\paragraph{ \bf\underline{$K_{L} \to \mu^+\mu^-$ decays}:}
For the $K_L\to \mu^+\mu^-$ decays, there will be the short and the complex long-distance contributions in the decay amplitude. Only the short-distance (SD) contributions can be reliably calculated, as they arise from perturbative electroweak loops. For conservative bounds on the long-distance contribution, the readers may look at the refs.~\cite{DAmbrosio:1997eof, Isidori:2003ts} which suffer from significant hadronic uncertainties. The short-distance contribution in this dileptonic kaon decay can be written as \cite{Buras:2013rqa}:
\begin{equation}\label{eq:BR_K2mumu}
\begin{split}
    \mathcal{B}(K_L\to \mu^+\mu^-)_{\rm SD}=& \tau_{K_L} f_K^2 m_K \frac{G_F^2 \alpha^2}{8 \pi^3}m_{\mu}^2 \beta_{\mu}(m_K^2)\\
    & \times \left\{\Bigg[\mathfrak{R}\left(V_{ts}^* V_{td}\hat{P}\right)\Bigg]^2+\Bigg[\mathfrak{I}\left(V_{ts}^* V_{td}\hat{S}\right)\Bigg]^2\right\}
    \end{split}
\end{equation}
with 
\begin{align*}
    \hat{P}(K)&\equiv\left(C_{10}-C_{10}^{\prime}\right)+\frac{m_K^2}{2 m_{\mu}}\frac{m_s}{m_d+m_s}\left(C_P-C_P^{\prime}\right)\\
    \hat{S}(K)&\equiv \beta_{\mu}(m_K^2)\frac{m_K^2}{2 m_{\mu}}\frac{m_s}{m_d+m_s}\left(C_s-C_s^{\prime}\right)
\end{align*}


\paragraph{\underline{\textbf{Invisible decays:}}}
For the decay of $B$ mesons (both charged and neutral), the underlying quark-level transition is $b \to s(d) \, \nu \bar{\nu}$, while for $K$ meson decays, it is governed by the transition $s \to d \, \nu \bar{\nu}$. With the assumptions that the right-handed neutrinos are absent and the neutrinos are Dirac types, the most general effective Hamiltonian for $b\to d_i\nu\bar{\nu}$ can be written as
\begin{align}\label{eq:effbtosnunu}
    \mathcal{H}^{b\to d_i\nu\bar{\nu}}=\frac{4 G_F}{\sqrt{2}}V_{tb} V_{td_i}^* (\mathcal{C}_{L}^{\nu} \mathcal{O}_{L}^{\nu} + \mathcal{C}_{R}^{\nu} \mathcal{O}_{R}^{\nu} )
\end{align}
where 
\begin{align}
    \mathcal{O}_{L}^{\nu} = \frac{e^2}{16 \pi^2}  \left( \bar{d}_{j} \gamma_{\mu} P_{L}b \right) \left(\bar{\nu} \gamma^{\mu}P_{L}\nu\right) , & \ \ \ \ &  \mathcal{O}_{R}^{\nu}=  \frac{e^2}{16 \pi^2}  \left( \bar{d}_{j} \gamma_{\mu} P_{R}b \right) \left(\bar{\nu} \gamma^{\mu}P_{L}\nu\right)\,,
\end{align}
Using the above Hamiltonian, we obtain the differential decay rates of various $B$ meson decay channels, which are given by \cite{Becirevic:2023aov, Chen:2024jlj}:
\begin{align}
	\begin{split}
		\frac{d \mathcal{B}(B^+\to K^+\nu\bar{\nu})}{dq^2}&=\tau_B \frac{G_F^2 \alpha^2}{256 \pi^5}\frac{\lambda(q^2,m_B^2,m_K^2)^{3/2}}{m_B^3}\left|V_{tb}V_{ts}^*\right|^2\bigg[f_+(q^2)\bigg]^2\left|C_L^{\nu} +C_R^{\nu}\right|^2\,,
	\end{split}
\end{align}
where $\lambda$ is the Källén function, defined as $\lambda(a^2,b^2,c^2) = (a^2 - (b - c)^2)(a^2 - (b + c)^2)$, $q^2$ is the dineutrino invariant mass squared ranging over $0 < q^2 \leq (m_B - m_K)^2$, and $f_+(q^2)$ denotes the $B \to K$ transition form factor. For $B\to K^*\nu\bar{\nu}$ decay, the differential decay rate distribution is given by  
\begin{align}
	\frac{d \mathcal{B}(B\to K^*\nu\bar{\nu})}{dq^2}&=\tau_B \frac{G_F^2 \alpha^2}{128 \pi^5}\frac{\lambda(q^2,m_B^2,m_{K^*}^2)^{1/2}\,q^2}{m_B^3}(m_B+m_{K^*})^2\left|V_{tb}V_{ts}^*\right|^2 \nonumber\\
	&\times \left\{\left(\bigg[A_1(q^2)\bigg]^2+ \frac{32 m_B^2 m_{K^*}^2}{q^2 (m_B+m_{K^*})^2}\bigg[A_{12}(q^2)\bigg]^2\right)\left|C_L^{\nu}-C_R^{\nu}\right|^2\right.\\& \left. +\frac{\lambda(q^2,m_B^2,m_{K^*}^2)}{(m_B+m_{K^*})^4}\bigg[V(q^2)\bigg]^2\left|C_L^{\nu}+C_R^{\nu}\right|^2\right\}\nonumber
\end{align}
Here, $A_1(q^2)$, $A_{12}(q^2)$, and $V(q^2)$ are the form factors associated with the $B \to K^*$ transition. We have used the values of the decay constants $f_K = (155.7 \pm 0.3)\,\text{MeV}$ and $f_{K^*} = (205 \pm 6)\,\text{MeV}$~\cite{FlavourLatticeAveragingGroupFLAG:2024oxs}. All form factors relevant to the $B \to K$ and $B \to K^*$ decays are adopted from ref.~\cite{Becirevic:2023aov}, where a combination of inputs from Lattice QCD (HPQCD \cite{Parrott:2022rgu}, FNAL/MILC\cite{Bailey:2015dka}) and Light-Cone Sum Rules (LCSR)\cite{Bharucha:2015bzk} has been utilised.

In our scenario, the NP contributions in these modes will arise via the diagrams shown in fig.~\ref{fig:Inv_1}, and contribute to the left-handed vector-type WC $C_L^{\nu}$. In the SM as well, only the left-handed vector current contributes to these processes. These low-energy couplings can again be decomposed into SM and NP parts, where the NP WCs can be expressed in terms of the LEFT coefficients as:
\begin{align}
    C_L^{\nu\,\rm{NP}}&=-\lambda_1\, L_{\underset{\nu\nu d_i b}{\nu d}}^{V,LL}\,,&
    C_R^{\nu\,\rm{NP}}&= 0\,.
\end{align}
We have used the following SM values of the WCs: $C^{\nu\,\rm{SM}}_L = (-12.64 \pm 0.14)$ and $C^{\nu\,\rm{SM}}_R = 0$. In order to utilise the experimental measurements of invisible decays, RGEs of the WCs are required. However, it is important to note that the coefficients $C_{L,R}^{\nu}$ do not evolve under the RGEs \cite{Aebischer:2017gaw}. Hence, we can approximate the NP contributions at low energy as $C_{L}^{\nu, \rm NP}(\mu_b) \approx C_{L}^{\nu, \rm NP}(\mu_{\rm EW})$. The expression for $C^{\nu\,\rm{NP}}_L$ in terms of top FCNC couplings is presented in eq.~\ref{eq:Append_FCNC_inv}.
\begin{align}\label{eq:Append_FCNC_inv}
    L_{\underset{\nu\nu d_i b}{\nu d}}^{V,LL}(\mu)&=-\frac{g_2^4 g_{Z_L}}{16\pi^2M_Z^2 c_W^2}\Bigg[\left(V_{tb}V_{u_is}^* X_L^{u_i t}(\mu)+V_{ts}^*V_{u_i b}X_L^{t u_i*}(\mu)\right)\left\{\frac{2x_t+3}{16}+\frac{x_t}{8}\log\left(\frac{\mu^2}{m_t^2}\right)\right\}\Bigg]
\end{align}
The effective Hamiltonian and the operator structure describing the $s \to d \nu\bar{\nu}$ transition are the same as in eq.~\ref{eq:effbtosnunu} with the replacements $b \leftrightarrow s$ and $s \leftrightarrow d$. The expression for the branching ratio of $K^+ \to \pi^+ \nu \bar{\nu}$ can be written as given in ref.~\cite{Li:2019fhz, Buras:2024ewl}.
\begin{align}\label{eq:BR_K2pinunu}
    \mathcal{B}(K^+\to \pi\nu\bar{\nu})= 3\lambda_2^{-2} J_{V}^{K^+}\left|C_L^{\nu}+C_R^{\nu}\right|^2\,.
\end{align}
with
\begin{subequations}
\begin{align}
    J_V^{K^+}&=\frac{1}{\Gamma_{K^+}^{\rm Exp}}\frac{1}{3\cdot 2^9 \pi^3 m_{K^+}^3}\int ds \,\lambda^{3/2}(s,m_{K^+}^2,m_{\pi^+}^2)\left|f_+^{K^+}(s)\right|^2=0.23 G_F^{-2}\\
    (\lambda_2)^{-1}&=\frac{4 G_F}{\sqrt{2}}V_{td} V_{ts}^* \frac{e^2}{16\pi^2}\,,\,\,\,\, C_L^{\nu\,\rm{SM}}=\lambda_2\times (1.30 \times 10^{-10}\,\,\mathrm{GeV}^{-2})\,,\,\,C_R^{\nu\,\rm{SM}}=0
\end{align}
\end{subequations}
Note that the branching ratio expression in eq.~\eqref{eq:BR_K2pinunu} contains only the vector-type current. However, it can be further parametrised in terms of scalar and tensor currents~\cite{Li:2019fhz, Buras:2024ewl}. Also, for this decay we obtain $C_R^{\nu\,\rm{NP}}= 0$. Since the SM contribution is purely vectorial and our NP analysis also yields only vector-type currents, we restrict our discussion to this case.

\subsection{FCCC Processes}\label{sec:appFCCC}
We now turn to the theoretical evaluation of FCCC observables within our effective framework. Because these processes are unsuppressed in the SM, they serve as an exceptionally sensitive probe for NP; any deviation from SM predictions would be a clear signal of NP. Given the relatively high-precision measurements and the relatively clean theoretical predictions in these channels, they provide a powerful means to impose stringent constraints on NP scenarios that might contribute to these processes, constraints that are expected to become even stronger with future data.
\begin{figure}[t]
	\centering
	\subfloat[]{
		\begin{tikzpicture}
			\begin{feynman}
				\vertex[](a1);
				\vertex[below left=1cm of a1](a2);
				\vertex[below left=0.8cm of a2](a3){\(b\)};
				\vertex[square dot, draw=blue, fill=blue, minimum size=4pt, inner sep=2.5pt,above left=1cm of a1](a4){};
				\vertex[above left=1.1cm of a4](a5){\(c,u\)};
				\vertex[right=1.cm of a1](a6);
				\vertex[above right=1.5cm of a6](a7){\(\ell\)};
				\vertex[below right=1.5cm of a6](a8){\(\nu\)};
				\diagram* {
					(a3) --[fermion] (a2) --[fermion,  edge label'=\(b\)] (a1) --[fermion, very thick, edge label'=\(t\)] (a4) --[fermion,very thick] (a5),
					(a2) --[blue,gluon, very thick, edge label=\(g\)] (a4),
					(a1) --[ boson,  edge label=\(W\)] (a6),
					(a8) --[fermion](a6) --[fermion](a7),
				};
			\end{feynman}
		\end{tikzpicture}
	}
	\subfloat[]{
		\begin{tikzpicture}
			\begin{feynman}
				\vertex (a1){\( b\)};
				\vertex [above right=1.5 cm of a1](a2);
				\vertex [ above left=0.5cm of a2](a3);
				\vertex [square dot, draw=blue, fill=blue, minimum size=4pt, inner sep=2.5pt,above left=0.7 cm of a3](a4){};
				\vertex [above left=1cm of a4](a5){\( c,u\)};
				\vertex [right=1.cm of a2](a6);
				\vertex [above right=1.5cm of a6](a7){\(\ell\)};
				\vertex [below right=1.5cm of a6](a8){\(\nu\)};

				\diagram* {
					(a1) --[fermion, arrow size=1.5pt] (a2),
					(a2) --[fermion, , arrow size=1.5pt,edge label={\(t\)}] (a3) --[fermion, very thick, arrow size=1.5pt,edge label={\(t \)}] (a4) --[fermion,very thick, arrow size=1.5pt] (a5),
					(a4) --[blue, very thick, gluon, half left, looseness=2, edge label = \(g\)] (a3),
					(a2) --[boson, edge label=\( W\)] (a6),
					(a8) --[fermion, arrow size=1.5pt](a6) --[fermion, arrow size=1.5pt](a7),
				};
			\end{feynman}
	\end{tikzpicture}}\\
	\subfloat[]{
		\begin{tikzpicture}
			\begin{feynman}
				\vertex[](a1);
				\vertex[below left=1cm of a1](a2);
				\vertex[below left=0.8cm of a2](a3){\(b\)};
				\vertex[square dot, draw=red, fill=red, minimum size=4pt, inner sep=2.5pt,above left=1cm of a1](a4){};
				\vertex[above left=1.1cm of a4](a5){\(c,u\)};
				\vertex[right=1.cm of a1](a6);
				\vertex[above right=1.5cm of a6](a7){\(\ell\)};
				\vertex[below right=1.5cm of a6](a8){\(\nu\)};
				\diagram* {
					(a3) --[fermion] (a2) --[fermion,  edge label'=\(b\)] (a1) --[fermion, very thick, edge label'=\(t\)] (a4) --[fermion,very thick] (a5),
					(a2) --[red,boson, very thick, edge label=\(\gamma/Z\)] (a4),
					(a1) --[ boson,  edge label=\(W\)] (a6),
					(a8) --[fermion](a6) --[fermion](a7),
				};
			\end{feynman}
		\end{tikzpicture}
	}
	\subfloat[]{
		\begin{tikzpicture}
			\begin{feynman}
				\vertex[](a1);
				\vertex[below left=1cm of a1](a2);
				\vertex[below left=0.8cm of a2](a3){\(b\)};
				\vertex[square dot, draw=red, fill=red, minimum size=4pt, inner sep=2.5pt,above left=1cm of a1](a4){};
				\vertex[above left=1.1cm of a4](a5){\(c,u\)};
				\vertex[right=1.cm of a1](a6);
				\vertex[above right=1.5cm of a6](a7){\(\ell\)};
				\vertex[below right=1.5cm of a6](a8){\(\nu\)};
				\diagram* {
					(a3) --[fermion] (a2) --[fermion, very thick, edge label=\(t\)] (a4) --[fermion, very thick] (a5) ,
					(a4) --[red,boson, very thick, edge label=\(\gamma/Z\)] (a1) --[boson,edge label=\(W\)](a2),
					(a1) --[ boson,  edge label=\(W\)] (a6),
					(a8) --[fermion](a6) --[fermion](a7),
				};
			\end{feynman}
		\end{tikzpicture}
	}
	\subfloat[]{
		\begin{tikzpicture}
			\begin{feynman}
				\vertex (a1){\( b\)};
				\vertex [above right=1.5 cm of a1](a2);
				\vertex [ above left=0.5cm of a2](a3);
				\vertex [square dot, draw=red, fill=red, minimum size=4pt, inner sep=2.5pt,above left=0.7 cm of a3](a4){};
				\vertex [above left=1cm of a4](a5){\( c,u\)};
				\vertex [right=1.cm of a2](a6);
				\vertex [above right=1.5cm of a6](a7){\(\ell\)};
				\vertex [below right=1.5cm of a6](a8){\(\nu\)};

				\diagram* {
					(a1) --[fermion, arrow size=1.5pt] (a2),
					(a2) --[fermion, , arrow size=1.5pt,edge label={\(t\)}] (a3) --[fermion, very thick, arrow size=1.5pt,edge label={\(t \)}] (a4) --[fermion,very thick, arrow size=1.5pt] (a5),
					(a4) --[red, very thick, boson, half left, looseness=2, edge label = \(\gamma/Z\)] (a3),
					(a2) --[boson, edge label=\( W\)] (a6),
					(a8) --[fermion, arrow size=1.5pt](a6) --[fermion, arrow size=1.5pt](a7),
				};
			\end{feynman}
	\end{tikzpicture}}\\
	\subfloat[]{
		\begin{tikzpicture}
			\begin{feynman}
				\vertex[](a1);
				\vertex[below left=1cm of a1](a2);
				\vertex[below left=0.8cm of a2](a3){\(b\)};
				\vertex[square dot, draw=violet, fill=violet, minimum size=4pt, inner sep=2.5pt,above left=1cm of a1](a4){};
				\vertex[above left=1.1cm of a4](a5){\(c,u\)};
				\vertex[right=1.cm of a1](a6);
				\vertex[above right=1.5cm of a6](a7){\(\ell\)};
				\vertex[below right=1.5cm of a6](a8){\(\nu\)};
				\diagram* {
					(a3) --[fermion] (a2) --[fermion,  edge label'=\(b\)] (a1) --[fermion, very thick, edge label'=\(t\)] (a4) --[fermion,very thick] (a5),
					(a2) --[violet,scalar, very thick, edge label=\(H\)] (a4),
					(a1) --[ boson,  edge label=\(W\)] (a6),
					(a8) --[fermion](a6) --[fermion](a7),
				};
			\end{feynman}
		\end{tikzpicture}
	}
	\subfloat[]{
		\begin{tikzpicture}
			\begin{feynman}
				\vertex[](a1);
				\vertex[below left=1cm of a1](a2);
				\vertex[below left=0.8cm of a2](a3){\(b\)};
				\vertex[square dot, draw=violet, fill=violet, minimum size=4pt, inner sep=2.5pt,above left=1cm of a1](a4){};
				\vertex[above left=1.1cm of a4](a5){\(c,u\)};
				\vertex[right=1.cm of a1](a6);
				\vertex[above right=1.5cm of a6](a7){\(\ell\)};
				\vertex[below right=1.5cm of a6](a8){\(\nu\)};
				\diagram* {
					(a3) --[fermion] (a2) --[fermion, very thick, edge label=\(t\)] (a4) --[fermion, very thick] (a5) ,
					(a4) --[violet,scalar, very thick, edge label=\(H\)] (a1) --[boson,edge label=\(W\)](a2),
					(a1) --[ boson,  edge label=\(W\)] (a6),
					(a8) --[fermion](a6) --[fermion](a7),
				};
			\end{feynman}
		\end{tikzpicture}
	}
	\subfloat[]{
		\begin{tikzpicture}
			\begin{feynman}
				\vertex (a1){\( b\)};
				\vertex [above right=1.5 cm of a1](a2);
				\vertex [ above left=0.5cm of a2](a3);
				\vertex [square dot, draw=violet, fill=violet, minimum size=4pt, inner sep=2.5pt,above left=0.7 cm of a3](a4){};
				\vertex [above left=1cm of a4](a5){\( c,u\)};
				\vertex [right=1.cm of a2](a6);
				\vertex [above right=1.5cm of a6](a7){\(\ell\)};
				\vertex [below right=1.5cm of a6](a8){\(\nu\)};

				\diagram* {
					(a1) --[fermion, arrow size=1.5pt] (a2),
					(a2) --[fermion, , arrow size=1.5pt,edge label={\(t\)}] (a3) --[fermion, very thick, arrow size=1.5pt,edge label={\(t \)}] (a4) --[fermion,very thick, arrow size=1.5pt] (a5),
					(a4) --[violet, very thick, scalar, half left, looseness=2, edge label = \(H\)] (a3),
					(a2) --[boson, edge label=\( W\)] (a6),
					(a8) --[fermion, arrow size=1.5pt](a6) --[fermion, arrow size=1.5pt](a7),
				};
			\end{feynman}
	\end{tikzpicture}}
	\caption{Feynman diagrams contributing to the FCCC processes $(b \to c(u)\, \ell \nu)$. Similarly, top FCNC couplings also affect the quark-level transitions $c \to s(d)\, \ell \nu$ and $s \to u\, \ell \nu$, although those diagrams are not explicitly shown here.}
	\label{fig:FCCC_processes_1}
\end{figure}

\paragraph{\underline{\textbf{Leptonic and Semi-leptonic decays:}}}
The contributions in the FCCC semileptonic and leptonic decays will be via the one-loop diagrams shown in fig.~\ref{fig:FCCC_processes_1}. The resulting effective Hamiltonian relevant to the processes $d_i \to u_j \ell \nu_{\ell}$ can be written as:
\begin{align}\label{eq:FCCC}
	\mathcal{H}^{d_i \to u_j \ell \nu_{\ell}}=\frac{4 G_F}{\sqrt{2}}V_{u_j d_i}\bigg[\left(1+C_{V_L}\right)O_{V_L}+C_{V_R}O_{V_R}\bigg]
\end{align}
with the operator defined as 
\begin{align}
	O_{V_L}&=\left(\bar{u}_j\gamma^{\mu}P_L d_i\right)\left(\bar{\ell}\gamma_{\mu}P_L \nu_{\ell}\right)\,,& O_{V_R}&=\left(\bar{u}_j\gamma^{\mu}P_R d_i\right)\left(\bar{\ell}\gamma_{\mu}P_L \nu_{\ell}\right). 
\end{align}
Note that the most general effective Hamiltonian for $d_i \to u_j \ell \nu_{\ell}$ transition also contains scalar- and tensor-type four fermion currents. 
However, the diagrams in fig.~\ref{fig:FCCC_processes_1} will contribute only in $C_{V_L}$ and $C_{V_R}$, respectively. 
\begin{subequations}\label{eq:Append_FCCC}
\begin{align}
    &L^{V,LR}_{\nu eb u_i}=\frac{\,g_2^2}{16\sqrt{2}\pi^2 M_W^2 G_F}\left|\frac{V_{tb}}{V_{u_i b}}\right|\Bigg[\frac{m_b}{m_t}\left(\frac{g_s^2}{12}\left(-1+6\log\frac{\mu^2}{m_t^2}\right)\xi_L^{u_it}+\frac{e^2}{48}\left(1-6\log\frac{\mu^2}{m_t^2}\right)\lambda_L^{u_it}\right.\nonumber\\
    &\left.+\frac{g_1^2}{96}\left(\frac{6(x_t X_R^{u_it}-2x_Z \kappa_L^{u_it})}{(x_t-x_Z)}\log\frac{x_t}{x_Z}+\frac{x_t X_R^{u_it}-2 x_Z \kappa_L^{u_it}}{x_Z}+\frac{2(x_t X_R^{u_i t}+6 x_Z \kappa_L^{u_i t})}{x_Z}\log\frac{\mu^2}{m_t^2}\right)\right)\nonumber\\
    &+\frac{m_b}{16\sqrt{2}v}\left(\frac{2 x_t}{(x_H-x_t)}\log\frac{x_H}{x_t}-\left(1+2\log\frac{\mu^2}{M_H^2}\right)\right)\eta_L^{u_it}+\frac{g_1^2}{12}\left(\frac{x_t-x_Z}{x_Z}X_R^{u_i t}+2\frac{x_t+x_Z}{x_t}\kappa_L^{u_i t}\right.\nonumber\\
    &\left. +3\frac{x_Z(x_t X_R^{u_i t}-2 x_Z \kappa_L^{u_i t})}{x_t(x_t-x_Z)}\log\frac{x_t}{x_Z}+\left(\frac{x_t-3 x_Z}{x_Z}X_R^{u_it}+6\frac{x_Z+x_t}{x_t}\kappa_L^{u_i t}\right)\log\frac{\mu^2}{m_t^2}\right)\Bigg]
\end{align}

\begin{align}
    &L^{V,LL}_{\nu ebu_i}=\frac{g_2^2}{16\sqrt{2}\pi^2 M_W^2 G_F}\left|\frac{V_{tb}}{V_{u_i b}}\right|\Bigg[g_s^2\left(\frac{5}{6}+\log \frac{\mu^2}{m_t^2}\right) \xi_R^{u_i t}\nonumber\\
    &+e^2\left(\frac{3}{8(x_t-1)}\log x_t+\frac{15 x_t-34}{96}+\frac{3 x_t-14}{16}\log\frac{\mu^2}{m_t^2}\right)\lambda_R^{u_it}\nonumber\\
    &+\frac{3g_2^2}{82\sqrt{2}}\frac{v m_t}{M_W^2}\left(\frac{1}{2}-\frac{\log x_H}{(x_H-1)(x_t-1)}-\frac{1}{3}\frac{x_t(x_t-4)}{(x_H-x_t)(x_t-1)}\log\frac{x_H}{x_t}+\frac{1}{3}\log\frac{\mu^2}{M_H^2}\right)\eta_R^{u_i t}\nonumber\\
    &+\frac{m_t}{4\sqrt{2}v}\left(1-\frac{x_H}{(x_H-x_t)}\log\frac{x_H}{x_t}+\log\frac{\mu^2}{m_t^2}\right)\eta_R^{u_it}\\
    &-\frac{g_2^2}{16}\left(2 X_L^{u_i t}-3\kappa_R^{u_i t}\right)\left(\frac{5}{2}+\frac{3\log x_t}{(x_t-1)(x_Z-1)}+\frac{3 x_Z^2}{(x_Z-1)(x_Z-x_t)}\log\frac{x_t}{x_Z}+3\log\frac{\mu^2}{m_t^2}\right)\nonumber\\
    &-\frac{(g_1^2-3g_2^2)}{16}\left(\frac{3 x_Z(x_t X_L^{u_i t}-2 x_Z \kappa_R^{u_i t})}{x_t(x_t-x_Z)}\log\frac{x_t}{x_Z}+\left(\frac{x_t-x_Z}{x_Z}+\frac{x_t-3x_Z}{x_Z}\log\frac{\mu^2}{m_t^2}\right)X_L^{u_i t}\right.\nonumber\\
    &\left.2 \frac{x_t+x_Z}{x_t}\left(1+3\log\frac{\mu^2}{m_t^2}\right)\kappa_R^{u_it}\right)+\frac{(g_1^2+3 g_2^2)}{2}\left(-6 \kappa_R^{u_i t}+\left(2\frac{x_t}{x_Z}\left(1+\log\frac{\mu^2}{m_t^2}\right)+3 \right)X_L^{u_it}\right)\Bigg]\nonumber
\end{align}

\end{subequations}
In the SM, both $C_{V_L}$ and $C_{V_R}$ are exactly zero. However, anomalous NP interactions induce non-zero contributions at the one-loop level, as detailed in eq.~\eqref{eq:Append_FCCC}. To connect these high-scale effects to low-energy observables, the corresponding WCs are evolved down to the scale $\mu_b$ using the appropriate RGEs, which are given by:
\begin{align}\label{eq:FCCC_EW_to_mb}
	\begin{pmatrix}
		C_{V_L}\\
		C_{V_R}
	\end{pmatrix}_{{\mu_b}}&=\begin{pmatrix}
		1.00716 &0 \\
		0 & 1.00358 
	\end{pmatrix} \begin{pmatrix}
		C_{V_L}\\
		C_{V_R}
	\end{pmatrix}_{\mu_{\rm{EW}}}\,.
\end{align}
Using the effective Hamiltonian in eq.~\eqref{eq:FCCC}, we obtain the differential decay rate for the process $P \to M \ell \nu$, where $P$ and $M$ are pseudoscalar mesons, corresponding to the quark-level transition $d_i~(u_i) \to u_j~(d_j)\, \ell \nu_{\ell}$ \cite{Sakaki:2013bfa}, as:
\begin{align}
\frac{d\Gamma (P \to M \ell \nu_\ell)}{dq^2} &= \frac{G_F^2 |V_{u_j d_i}|^2}{192 \pi^3 m_P^3} \, q^2 \sqrt{\lambda(q^2,m_P^2,m_M^2)} \left(1 - \frac{m_\ell^2}{q^2}\right)^2 \left|1 + C_{V_L} + C_{V_R}\right|^2 \nonumber \\
&\quad \times \left\{ \left(1 + \frac{m_\ell^2}{2q^2} \right) H_{V,0}^{s\,2} + \frac{3}{2} \frac{m_\ell^2}{q^2} H_{V,t}^{s\,2} \right\},
\label{eq:br_semileptonic_FCCC}
\end{align}
All the relevant parameters in eq.~\eqref{eq:br_semileptonic_FCCC}, including the helicity amplitudes $H_V$, are taken from ref.~\cite{Sakaki:2013bfa, Biswas:2021pic}. Similarly, the branching fraction for the leptonic decay $P \to \ell \nu_{\ell}$ can be written as:
\begin{equation}\label{eq:br_lep_FCCC}
	\begin{split}
		\mathcal{B}(P\to\ell\nu_{\ell}) = & \frac{\tau_P}{8\pi}\, m_P \, m_\ell^2\,  f_P^2\,  G_F^2 \left(1-\frac{m_\ell^2}{m_P^2}\right)^2 \left|V_{u_j d_i}(1+C_{V_{1}}^{\ell}-C_{V_{2}}^{\ell}) \right|^2.
	\end{split}
\end{equation}
NP contributions alter the decay rates and kinematic distributions, directly impacting the extraction of CKM matrix elements. This effect is captured by the shift:
\begin{equation}
	\left| V_{u_j d_i} \right| \longrightarrow \left| V_{u_j d_i}\,(1 + C_{V_L} \pm C_{V_R}) \right|,
\end{equation}
where the $+(-)$ sign corresponds to semileptonic (leptonic) decays. Consequently, CKM determinations from $P \to M \ell \nu_\ell$ transitions serve as stringent constraints on the NP parameter space.

\subsection*{Anomalous $Wtb$ coupling}\label{Append:Anomalous_Wtb}
In this subsection, we present the loop amplitudes associated with the anomalous $Wtb$ couplings, as discussed in the section \ref{sec:FCCC}.
\begin{align}\label{eq:Append_anomalous_Wtb}
    &C_{V_L}=\frac{V_{u_i b}}{8 \pi^2}\Bigg[\left(C_F\,g_s^2 \xi_R^{u_i t}+\frac{e^2}{3}\lambda_R^{u_i t}\right)\left(1+\frac{3 x_t+1}{4(x_t-1)}\log\left(-\frac{\mu^2}{m_t^2}\right)-\frac{1}{(x_t-1)}\log\left(-\frac{\mu^2}{M_W^2}\right)\right)\nonumber\\
    &+\frac{(g_1^2+3 g_2^2)}{192\pi^2}\left(\left(X_L^{u_i t}+3\kappa_R^{u_i t}\right)\log\frac{\mu^2}{M_Z^2}-\frac{(x_t+2)(x_t-x_Z-1)}{(x_t-1)^2}X_L^{u_it}-\frac{4 x_t^2-x_t(3 x_Z+4)-x_Z}{x_t(x_t-1)}\kappa_R^{u_i t}\right.\nonumber\\
    &\left.\left(+\frac{(2 x_t^2-4 x_t x_Z-2 x_Z+x_t-3)}{(x_t-1)^3} X_L^{u_i t}+4\frac{x_t-2 x_Z-1}{(x_t-1)^2}\kappa_R^{u_it}\right)\log(-x_Z)\right)\\
    &+e^2\left(\frac{3x_t+5}{8(x_t-1)}-\frac{x_t^2+6 x_t+1}{4(x_t-1)^2}\log(-x_t)\right)\lambda_R^{u_i t}+\frac{g_2^2}{4\sqrt{2}}\frac{m_t v}{M_W^2}\left(\frac{1}{x_t-1}+\frac{x_t-x_H+1}{(x_t-1)^2}\log x_H\right.\nonumber\\
    &\left.+\frac{M_W^2}{m_t}\frac{(x_t-x_H)(x_t-1)}{(x_t-1)^2}\log\frac{x_H}{x_H-x_t} \right)\eta_R^{u_i t}+\frac{g_2^2}{8}\left(\frac{(2 x_t^2 X_L^{u_i t}+(3 x_t^2+x_t(5-6 x_Z)+2x_Z)\kappa_R^{u_i t}}{x_t(x_t-1)}\right.\nonumber\\
    &\left. -2\frac{(x_t^2-x_t-x_Z+2)X_L^{u_i t}+(1+x_t^2+x_t(6-4 x_Z)-4 x_Z+2 x
    _Z^2)\kappa_R^{u_i t}}{(x_t-1)^2}\log x_Z\right.\nonumber\\
    &\left.+\frac{2(x_t-x_Z)(X_L^{u_it} x_t(2-x_t+x_t^2)+(x_t+x_t^3-3 x_t^2(x_Z-2)-x_Z)\kappa_R^{u_i t}}{x_t^2 (x_t-1)^2}\log\frac{x_Z}{x_Z-x_t}\right)\Bigg]\nonumber
\end{align}
 $C_{V_R}$ is light quark mass dependent and gives a small contribution.

\subsection{Neutron EDM}\label{sec:appEDM}
\begin{figure}[h!]
	\centering
	\subfloat[]{\label{fig:EDM_double_a}
		\begin{tikzpicture}
			\begin{feynman}
				\vertex[](a1){\(u_i\)};
				\vertex[right=1cm of a1](a2);
				\vertex[right=1.3cm of a2](a3);
				\vertex[square dot,blue,right=0.8cm of a3](a4){};
				\vertex[right=1cm of a4](a5){\(u_i\)};
				\vertex[below=1.3cm of a4](a6){\(\gamma,g\)};
				\diagram* {
					(a1) --[thick, fermion](a2) --[thick, fermion, edge label'=\(d_j\)](a3) --[thick, fermion, edge label'=\(t\)](a4) --[thick, fermion](a5),
					(a4) --[thick,blue,boson](a6),
					(a2) --[thick, boson, half left, looseness=2.5, edge label = \(W\)](a3),
				};
			\end{feynman}
		\end{tikzpicture}
	}
	\subfloat[]{\label{fig:EDM_double_b}
		\begin{tikzpicture}
			\begin{feynman}
				\vertex[](a1){\(u_i\)};
				\vertex[square dot,blue,right=1cm of a1](a2){};
				\vertex[right=0.8cm of a2](a3);
				\vertex[right=1.3cm of a3](a4);
				\vertex[right=1cm of a4](a5){\(u_i\)};
				\vertex[below =1.3 cm of a2](a6){\(\gamma,g\)};
				\diagram* {
					(a1) --[thick, fermion](a2) --[thick, fermion,edge label'=\(t\)](a3) --[thick, fermion, edge label'=\(d_j\)](a4) --[thick, fermion](a5),
					(a3) --[thick, boson, half left, looseness=2.5,edge label=\(W\)](a4),
					(a2) --[blue,thick,boson](a6),
				};
			\end{feynman}
	\end{tikzpicture}}
	\subfloat[]{\label{fig:EDM_double_c}
		\begin{tikzpicture}
			\begin{feynman}
				\vertex[](a1){\(d_i\)};
				\vertex[right=1cm of a1](a2);
				\vertex[square dot, blue,right=1cm of a2](a3){};
				\vertex[right=1cm of a3](a4);
				\vertex[right=0.6cm of a4](a5){\(d_i\)};
				\vertex[below= 1.3cm of a3](a6){\(\gamma,g\)};
				\diagram*{
					(a1) --[thick, fermion](a2) --[thick, fermion, edge label'=\(t\)](a3) --[thick, fermion, edge label'=\(u/c\)](a4) --[thick, fermion](a5),
					(a3) --[blue,thick,boson](a6),
					(a2) --[thick, boson, half left, looseness=1.5, edge label=\(W\)](a4)
				};
			\end{feynman}
		\end{tikzpicture}
	}\\
	\subfloat[]{\label{fig:EDM_double_d}
		\begin{tikzpicture}
			\begin{feynman}
				\vertex[](a1){\(u_i\)};
				\vertex[square dot,blue,right=1cm of a1](a2){};
				\vertex[right=1cm of a2](a3);
				\vertex[square dot,blue,right=1cm of a3](a4){};
				\vertex[right=1cm of a4](a5){\(u_i\)};
				\vertex[below= 1cm of a3](a6){\(\gamma,g\)};
				\diagram*{
					(a1) --[thick,fermion](a2) --[thick, fermion, edge label'=\(t\)](a3) --[thick, fermion, edge label'=\(t\)](a4) --[thick, fermion](a5),
					(a3) --[thick,boson](a6),
					(a2) --[blue,gluon,thick, half left, looseness=1.5, edge label=\(g\)](a4)
				};
			\end{feynman}
		\end{tikzpicture}
	}
	\subfloat[]{\label{fig:EDM_double_e}
		\begin{tikzpicture}
			\begin{feynman}
				\vertex[](a1){\(u_i\)};
				\vertex[square dot,red,right=1cm of a1](a2){};
				\vertex[right=1cm of a2](a3);
				\vertex[square dot,red,right=1cm of a3](a4){};
				\vertex[right=1cm of a4](a5){\(u_i\)};
				\vertex[below= 1cm of a3](a6){\(\gamma,g\)};
				\diagram*{
					(a1) --[thick,fermion](a2) --[thick,fermion, edge label'=\(t\)](a3) --[thick,fermion, edge label'=\(t\)](a4) --[thick,fermion](a5),
					(a3) --[thick,boson](a6),
					(a2) --[red,boson,thick, half left, looseness=1.5, edge label=\(\gamma/Z\)](a4)
				};
			\end{feynman}
		\end{tikzpicture}
	}
	\subfloat[]{\label{fig:EDM_double_f}
		\begin{tikzpicture}
			\begin{feynman}
				\vertex[](a1){\(u_i\)};
				\vertex[square dot,violet,right=1cm of a1](a2){};
				\vertex[right=1cm of a2](a3);
				\vertex[square dot,violet,right=1cm of a3](a4){};
				\vertex[right=1cm of a4](a5){\(u_i\)};
				\vertex[below= 1cm of a3](a6){\(\gamma,g\)};
				\diagram*{
					(a1) --[thick,fermion](a2) --[thick,fermion, edge label'=\(t\)](a3) --[thick,fermion, edge label'=\(t\)](a4) --[thick,fermion](a5),
					(a3) --[thick,boson](a6),
					(a2) --[violet,scalar,thick, half left, looseness=1.5, edge label=\(H\)](a4)
				};
			\end{feynman}
		\end{tikzpicture}
	}\\
	\subfloat[]{\label{fig:EDM_double_g}
		\begin{tikzpicture}
			\begin{feynman}
				\def\r{1.0 cm}
				\coordinate (O) at (0,0);
				\draw[thick] (O) circle (\r);
				\draw[fermion, thick] (360:\r) arc (360:180:\r);
				\vertex (A) at (90:\r)  ;   
				\vertex (B) at (210:\r) ;   
				\vertex (C) at (330:\r) ;   
				\vertex (D) at (0:\r)   {}; 
				\vertex (E) at (180:\r) {};
				\vertex[above=0.5 of A](F){\(g\)};
				\vertex[below left=0.5cm of B](G){\(g\)};
				\vertex[below right=0.5cm of C](H){\(g\)};
				\vertex[below=1.1cm of O](a1) {\(u_i\)};
				\vertex[above right=1.0 cm of O](a2) {\(t\)};
				\vertex[square dot, red, at=(D)] {};
				\vertex[square dot, red, at=(E)] {};
				\diagram*{
					(A) --[thick, gluon](F),
					(B) --[thick, gluon](G),
					(C) --[thick, gluon](H),
					(D) --[boson, thick, red, edge label'=\(\gamma/Z\)](E),
				};
			\end{feynman}
		\end{tikzpicture}
	}
	\subfloat[]{\label{fig:EDM_double_h}
		\begin{tikzpicture}
			\begin{feynman}
				\def\r{1.0 cm}
				\coordinate (O) at (0,0);
				\draw[thick] (O) circle (\r);
				\draw[fermion, thick] (360:\r) arc (360:180:\r);
				\vertex (A) at (90:\r)  ;   
				\vertex (B) at (210:\r) ;   
				\vertex (C) at (330:\r) ;   
				\vertex (D) at (0:\r)   {}; 
				\vertex (E) at (180:\r) {};
				\vertex[above=0.5 of A](F){\(g\)};
				\vertex[below left=0.5cm of B](G){\(g\)};
				\vertex[below right=0.5cm of C](H){\(g\)};
				\vertex[below=1.1cm of O](a1) {\(u_i\)};
				\vertex[above right=1.0 cm of O](a2) {\(t\)};
				\vertex[square dot, violet, at=(D)] {};
				\vertex[square dot, violet, at=(E)] {};
				\diagram*{
					(A) --[thick, gluon](F),
					(B) --[thick, gluon](G),
					(C) --[thick, gluon](H),
					(D) --[scalar, thick, violet, edge label'=\(H\)](E),
				};
			\end{feynman}
		\end{tikzpicture}
	}
	\caption{Dipole moment of light quarks affected by the top FCNC operators. Here $u_i$ denotes $u$ and $c$-quark respectively. The diagrams in the last row denote the contribution of NP in the Weinberg operator.}
	\label{fig:dipole_moment_light_quark}
\end{figure}
The electric dipole moments of the light quarks $(u,d,s)$ contribute directly to the EDM of the nucleon. These effects are parametrised by hadronic matrix elements that connect fundamental dipole operators to measurable neutron and proton EDMs. For quark dipole operators, this hadronic input is captured by the tensor charges, which have been computed using Lattice QCD~\cite{Gupta:2018lvp,FlavourLatticeAveragingGroupFLAG:2024oxs}. The theoretical estimate, for the neutron EDM, is expressed in terms of the light-quark contributions as follows~\cite{Pospelov:2005pr}:
\begin{align}\label{eq:neutron_EDM}
	\frac{d_n}{e}&=\left(g_T^d\frac{d_u}{e}+g_T^u\frac{d_d}{e}+g_T^s\frac{d_s}{e}\right)+(1\pm0.5)\times 1.1 \left(\tilde{d}_d+0.5 \tilde{d}_u\right)\,,\nonumber\\
	&+(22\pm10)\times 10^{-3}\mathrm{GeV}\cdot w\,.
\end{align}
In this expression, the first parenthesis captures the direct light-quark EDM contributions, the second accounts for the CEDMs, and the final term arises from the Weinberg operator defined in eq.~\ref{eq:lag_EDM}. The tensor charges, evaluated at the hadronic renormalisation scale $\mu_{\rm had}=2\,\mathrm{GeV}$, are~\cite{FlavourLatticeAveragingGroupFLAG:2024oxs}:
\begin{align}
	g_T^u=0.784(28)(10)\,,\,\,\,\,g_T^d=-0.204(11)(10)\,,\,\,\,\, g_T^s=-0.0027(16)\,.
\end{align}

Fig.~\ref{fig:dipole_moment_light_quark} illustrates the loop diagrams through which the top-quark FCNC couplings induce light-quark EDMs and CEDMs. In addition to diagrams with single insertions of the NP couplings, we include topologies with double insertions. These are particularly important because the corresponding loop amplitudes are dominated by the top-quark contribution and can therefore be numerically significant. Note that the contribution of the Weinberg operator, $w(\mu_t)$, is numerically negligible for light-quark EDMs and is therefore omitted from their evaluation~\cite{Gorbahn:2014sha}\footnote{The contribution from the Weinberg operator becomes relevant when computing the EDMs of the heavier $c$- and $b$-quarks, which are not considered in this work. Since the effective operators defined in eq.~\ref{eq:lag_EDM} are evaluated at the scale $\mu = 2\,\mathrm{GeV}$, we work within an effective theory with three active light flavours, effectively integrating out the $c$ and $b$ quarks.}. 

Matching the one-loop amplitudes in fig.~\ref{fig:dipole_moment_light_quark} to the effective operators in eq.~\ref{eq:lag_EDM} yields the expressions for the light-quark EDMs ($d_q$) and CEDMs ($\tilde{d}_q$) detailed in eq.~\eqref{Loop:up_EDM},\eqref{Loop:charm_weinberg} and \eqref{Loop:CEDM} respectively. In terms of the LEFT coefficients, these are given by:
\begin{align}\label{eq:Matching_EDM_LEFT}
	d_q(\mu)=2\,\mathrm{Im}[L^{q\gamma}_{ii}](\mu)\,,\,\,\,\, \tilde{d}_q(\mu)=\frac{2}{g_s(\mu)}\,\mathrm{Im}[L^{qG}_{ii}](\mu)\,.
\end{align}
Because we define $L_{ii}/\Lambda \equiv L_{ii}$ with dimension $\mathrm{GeV}^{-1}$, we apply a conversion factor $\left(1\,\mathrm{GeV}^{-1}=1.97 \times 10^{-14}\,\mathrm{cm}\right)$ for consistency with conventional EDM normalisation. Substituting these coefficients into the nucleon EDM expression yields:
\begin{align}\label{eq:light_EDM}
	\frac{d_{n}}{e}(\mu)&=-\frac{3.94\times 10^{-14}}{e}\left(g_T^d \,\mathrm{Im}[L^{u\gamma}_{11}] (\mu_{\rm had})+g_T^u\,\mathrm{Im}[L^{d\gamma}_{11}](\mu_{\rm had})+g_T^s\,\mathrm{Im}[L^{d\gamma}_{22}](\mu_{\rm had})\right)\,,\nonumber\\
   & -\frac{4.3\times 10^{-14}}{g_s(\mu)}(1\pm0.5)\left\{\mathrm{Im}[L^{dG}_{11}](\mu_{\rm had})+0.5\,\mathrm{Im}[L^{uG}_{11}](\mu_{\rm had})\right\}\,.
\end{align}

All low-energy couplings in eq.~\eqref{eq:neutron_EDM} are defined at the hadronic scale, $\mu_{\rm had}$. The running of these couplings from the electroweak scale $(\mu_{\rm EW})$ down to $\mu_{\rm had}$ is performed through the LEFT framework in the leading-logarithmic approximation, as described in ref.~\cite{Jenkins:2017dyc}:
\begin{align}\label{eq:RGE_mat_EDM}
	\begin{pmatrix}
		L^{u\gamma}\\
		L^{uG}\\
		L^{d\gamma}\\
		L^{dG}
	\end{pmatrix}_{\mu_{\rm had}}=\begin{pmatrix}
		0.87 & -0.064 &0 &0\\
		-0.05 & 1.23 & 0 &0 \\
		0 & 0 & 0.88 & 0.03 \\
		0 & 0 & 0.024 & 1.23
	\end{pmatrix}\begin{pmatrix}
		L^{u\gamma}\\
		L^{uG}\\
		L^{d\gamma}\\
		L^{dG}
	\end{pmatrix}_{\mu_{\rm EW}}
\end{align}
While the matrix in eq.~\eqref{eq:RGE_mat_EDM} applies to the evolution of both $d_q$ and $\tilde{d}_q$, the matching relation for $\tilde{d}_q$ in eq.~\eqref{eq:Matching_EDM_LEFT} explicitly contains the strong coupling $g_s(\mu)$. Because $g_s(\mu)$ varies significantly toward low energies, an additional factor of $\left(g_s(\mu_{\rm EW})/g_s(\mu_{\rm had})\right)$ must be included when relating $\tilde{d}_q(\mu_{\rm had})$ to $\tilde{d}_q(\mu_{\rm EW})$ via this RGE matrix.

Beyond the direct contributions of light quarks, the $b$ and $c$ quarks influence the nucleon EDM indirectly. Because they are integrated out at their corresponding mass thresholds, we do not compute their non-perturbative matrix elements directly~\cite{Alexandrou:2019brg}; instead, their effects are captured through RGE-induced mixing and matching~\cite{Ema:2022pmo, Hou:2025bjy, Brod:2023wsh}. Specifically, at the hadronic scale, the light-quark dipole moments receive leading-logarithmic contributions from the charm-quark CEDM~\cite{Gorbahn:2014sha}:
\begin{align}\label{eq:charm_RGE_EDM}
	\frac{d_d(\mu_{\rm had})}{e} &= 2.3 \times 10^{-8}\,\tilde{d}_c(\mu_t) 
	+ 1.0 \times 10^{-4}~\mathrm{GeV}\cdot w(\mu_t)\,, \nonumber\\[4pt]
	\frac{d_u(\mu_{\rm had})}{e} &= -2.1 \times 10^{-8}\,\tilde{d}_c(\mu_t) 
	- 9.1 \times 10^{-5}~\mathrm{GeV}\cdot w(\mu_t)\,, \nonumber\\[4pt]
	\tilde{d}_d(\mu_{\rm had}) &= 1.8 \times 10^{-6}\,\tilde{d}_c(\mu_t) 
	+ 7.0 \times 10^{-4}~\mathrm{GeV}\cdot w(\mu_t)\,, \\[4pt]
	\tilde{d}_u(\mu_{\rm had}) &= 8.2 \times 10^{-7}\,\tilde{d}_c(\mu_t) 
	+ 3.1 \times 10^{-4}~\mathrm{GeV}\cdot w(\mu_t)\,, \nonumber\\[4pt]
	w(\mu_{\rm had}) &= 1.7 \times 10^{-2}~\mathrm{GeV}^{-1}\,\tilde{d}_c(\mu_t) 
	+ 0.41\,w(\mu_t)\,.\nonumber
\end{align}
Because the Weinberg operator mixes into the quark EDMs and CEDMs (but reverse mixing does not occur~\cite{Gorbahn:2014sha}), inserting eq.~\eqref{eq:charm_RGE_EDM} into eq.~\eqref{eq:neutron_EDM} isolates the contribution to the neutron EDM arising purely from the charm-quark CEDM at scale $\mu_t$:
\begin{align}\label{eq:charm_nuetron_EDM}
	\left|\frac{d_n}{e}\right|=\left|(3.7 \pm 1.7)\times 10^{-4}\cdot\tilde{d}_c(\mu_t)+(9.3 \pm 4.1)\times 10^{-3}\,\mathrm{GeV}\cdot w(\mu_t)\right|\,.
\end{align}

In summary, the dominant theoretical contributions to $d_n$ depend strongly on the active FCNC transitions. If NP effects are present only in the $t \to u$ transition, or simultaneously in both $t \to u$ and $t \to c$ transitions, the dominant contributions to $d_n$ originate solely from the light-quark EDMs and CEDMs in eq.~\ref{eq:light_EDM}, including those generated via RGE mixing as in eq.~\ref{eq:RGE_mat_EDM}. In this case, the contributions from charm dipole operators, evaluated at the electroweak scale $\mu_{\mathrm{EW}}$ and evolved down to the hadronic scale $\mu_{\mathrm{had}}$, are numerically negligible.

In contrast, if NP effects are assumed to affect only the $t \to c$ FCNC vertices, the dominant contribution to the neutron EDM arises from the charm-quark CEDM through RGE-induced mixing, as given in eq.~\ref{eq:charm_nuetron_EDM}, all the other contributions are negligible. 

\paragraph{Loop contribution to light quark EDM}
\begin{subequations}
	\begin{align}\label{Loop:up_EDM}
		&\frac{d_{u_i}}{e}(\mu)=\sum_{j} \frac{ g_2^2 V_{td_j}V^*_{u_i d_j}}{32\pi^2}\frac{m_{u_i}}{m_t^2}\Bigg[\frac{5 x_{d_j}^2-5 x_{d_j}-6}{2(x_{d_j}-1)}+3 x_{d_j}\log\left(\frac{\mu^2}{m_{d_j}^2}\right)+3 \frac{x_{d_j} \log(x_{d_j})}{(x_{d_j}-1)^2}\Bigg]\mathrm{Re}\left(\lambda_R^{u_it}\right)\nonumber\,,\\
		&+\frac{q_{u_i}}{16 \pi^2 m_t}\left\{C_F\,g_s^2(\mu)\,\mathrm{Im}\left(\xi_L^{u_it}\xi_R^{tu_i\,*}\right)\log\left(\frac{\mu^2}{m_t^2}\right)+e^2\,\mathrm{Im}\left(\lambda_L^{u_it}\lambda_R^{tu_i\,*}\right)\log\left(\frac{\mu^2}{m_t^2}\right)\right\}\,,\\
		&+\frac{q_{u_i}}{32\pi^2}\frac{m_t}{M_H^2}\left\{\frac{(x_{t/H}-3)}{2(x_{t/H}-1)^2}+\frac{\log x_{t/H}}{(x_{t/H}-1)^3}\right\}\mathrm{Im}\left(\eta_L^{u_it}\eta_R^{tu_i\,*}\right)\,,\nonumber\\
		&+\frac{g_2^2}{16 \pi^2 c_W^2}\frac{q_{u_i}}{m_t}\Bigg[\mathrm{Im}\left(\kappa_L^{u_it}X_L^{tu_i\,*}+X_R^{u_it}\kappa_R^{tu_i\,*}+\kappa_L^{u_it}\kappa_R^{tu_i\,*}\right)\log\left(\frac{\mu^2}{m_t^2}\right)\,,\nonumber\\
		&+\frac{\log x_t}{4(x_{t/Z}-1)^2}\left\{x_{t/Z}^2\,\mathrm{Im}\left(2 X_L^{tu_i\,*}X_R^{u_it}+X_L^{u_it}\kappa_L^{tu_i\,*}+X_R^{tu_i\,*}\kappa_R^{u_it}\right)\,,\right.\nonumber\\
		&\left.+2\,\mathrm{Im}\left(X_L^{u_it}\kappa_L^{tu_i\,*}+X_R^{tu_i\,*}\kappa_R^{u_it}+\kappa_L^{u_it}\kappa_R^{tu_i\,*}\right)\right.\nonumber\\
		&\left.+x_{t/Z}\,\mathrm{Im}\left(2 X_L^{tu_i\,*}X_R^{u_it}+9(X_L^{tu_i\,*}\kappa_L^{u_it}+X_R^{u_it}\kappa_R^{tu_i\,*})+10 \kappa_L^{u_it}\kappa_R^{tu_i\,*}\right)\right\}\,,\nonumber\\
		&+\frac{\log x_{t/Z}}{2 (x_{t/Z}-1)^3}\left\{2 x_{t/Z}^2\,\mathrm{Im}\left(X_L^{tu_i\,*}X_R^{u_it}\right)+x_{t/Z}\,\mathrm{Im}\left(5(X_L^{tu_i\,*}\kappa_L^{tu_i\,*}+X_R^{tu_i\,*}\kappa_R^{u_it})+8 \kappa_L^{u_it}\kappa_R^{tu_i\,*}\right)\right.\,,\nonumber\\
		&\left.+2\,\mathrm{Im}\left(X_L^{u_it}\kappa_L^{tu_i\,*}+X_R^{tu_i\,*}\kappa_R^{u_it}+\kappa_L^{tu_i\,*}\kappa_R^{u_it}\right)\right\}\Bigg]\nonumber
	\end{align}
	
	\begin{align}\label{Loop:charm_weinberg}
		w(\mu)&=\frac{g_s^3(\mu)}{(32\pi^2)^2}\frac{1}{M_H^2}\left\{-\frac{x_{t/H}^2-5 x_{t/H}-2}{3(x_{t/H}-1)^2}-\frac{2 x_{t/H}}{(x_{t/H}-1)^4}\log x_{t/H}\right\}\mathrm{Im}\left(\eta_L^{ct}\eta_R^{tc\,*}\right)
	\end{align}
\end{subequations}

Here, $d_j$ represents the down-type quarks of the three generations, and the factors 
$x_i$ are defined as $x_{d_j} = m_{d_j}^2/M_W^2$, $x_{t/Z} = m_t^2/M_Z^2$, and 
$x_{t/H} = m_t^2/M_H^2$, respectively. 
The loop contribution of the Weinberg operator in eq.~\eqref{Loop:charm_weinberg} 
is only relevant for the charm-quark CEDM. 
The expression for the CEDM, $\tilde{d}_{u_i}$, of up- and charm-type quarks is structurally analogous to that of the EDM, $\left(d_{u_i}/e\right)$, apart from a few key modifications. Specifically, for the topologies shown in fig.~\ref{fig:EDM_double_d}–\ref{fig:EDM_double_f}, corresponding to two insertions of the NP couplings, the electric charge of the up quark $(q_{u_i}=2/3)$ does not contribute. In the case of fig.~\ref{fig:EDM_double_d}, the color factor $C_F=4/3$ in the EDM expression is replaced by $-1/(2N)$ for the CEDM. Moreover, for the single-insertion diagrams depicted in fig.~\ref{fig:EDM_double_a}–\ref{fig:EDM_double_b}, the factor $\mathrm{Re}\,\left(\lambda_R^{u_it}\right)$ appearing in the EDM expression must be replaced by $\mathrm{Re}\,\left(\xi_R^{u_it}\right)$ in the CEDM case.
\begin{subequations}\label{Loop:CEDM}
	\begin{align}
		\frac{d_{d_i}}{e}(\mu)=\frac{g_2^2 m_{d_i}}{32\pi^2 M_W^2}V_{td_i}\left(V_{cd_i}^*\,\mathrm{Im}\left(\lambda_R^{ct}\right)+V_{ud_i}^*\,\mathrm{Im}\left(\lambda_R^{ut}\right)\right)\left\{\frac{1}{(x_t-1)}+\frac{x_t}{(x_t-1)^2}\log x_t\right\}
	\end{align}
	\begin{align}
		\frac{\tilde{d}_{d_i}}{g_s}(\mu)=\frac{g_2^2 m_{d_i}}{32\pi^2 M_W^2}V_{td_i}\left(V_{cd_i}^*\,\mathrm{Im}\left(\xi_R^{ct}\right)+V_{ud_i}^*\,\mathrm{Im}\left(\xi_R^{ut}\right)\right)\left\{\frac{1}{(x_t-1)}+\frac{x_t}{(x_t-1)^2}\log x_t\right\}
	\end{align}
\end{subequations}
In the above equation, the loop calculations for the down and strange quarks have been performed at the electroweak scale, $\mu_{\rm EW}$.
\subsubsection*{Loop contribution to top quark EDM \& CMDM}\label{sec:appTopCMDM}

Top quark CMDM in the presence of NP modified as,
\begin{subequations}\label{eq:Append_top_CMDM}
    \begin{align}
       \hat{d}_t^{\rm NP}&=  \sum_{i,j}\frac{g_2^2 m_t}{64\pi^2 M_W^2} V_{td_j}V_{u_i d_j}\left(-m_{u_i}\,\mathrm{Im}(\xi_L^{u_i t})+m_t\,\mathrm{Im}(\xi_R^{u_i t})\right)\left(\frac{2x_t^2-x_t-2}{x_t^2}+\log\frac{\mu^2}{M_W^2}\right.\nonumber\\
       &+\left.\frac{(x_t-1)^2(x_t+2)}{x_t^3}\log\frac{M_W^2}{M_W^2-m_t^2}\right)
    \end{align}

    \begin{align}
       \hat{\mu}_t^{\rm NP}&= \sum_{i,j}\frac{g_2^2 m_t}{64\pi^2 M_W^2} V_{td_j}V_{u_i d_j}\left(m_{u_i}\,\mathrm{Re}(\xi_L^{u_i t})+m_t\,\mathrm{Re}(\xi_R^{u_i t})\right)\left(\frac{2x_t^2-x_t-2}{x_t^2}+\log\frac{\mu^2}{M_W^2}\right.\nonumber\\
       &+\left.\frac{(x_t-1)^2(x_t+2)}{x_t^3}\log\frac{M_W^2}{M_W^2-m_t^2}\right)
    \end{align}
\end{subequations}
Here, $d_j$ represents the down-type quarks of the three generations and $u_i$ corresponds to up and charm quark respectively.
\subsection{EWPO}\label{sec:appEWPO}
\paragraph{\underline{\textbf{Oblique Parameters:}}}
The precision extraction of the oblique parameters $S$, $T$, and $U$ provide a model-independent framework to quantify the effects of NP on EWPOs via corrections to the vacuum polarisations (self-energies) of the electroweak gauge bosons \cite{Peskin:1991sw,Peskin:1990zt,Maksymyk:1993zm}. These parameters are extracted by comparing precise experimental measurements, primarily from the $Z$-pole observables and the W boson mass, with their SM predictions. These oblique EWPOs represent a crucial phenomenological sector, offering exceptional sensitivity to the NP, particularly at future lepton colliders such as FCC-ee
\cite{Baak:2014ora,Asadi:2022xiy,Bagnaschi:2022whn,ALEPH:2013dgf,Altmann:2025feg}. Among the three oblique parameters, $T$ measures the difference between NP contributions to neutral and charged current interactions at low energies, making it sensitive to custodial isospin violation. Parameters $S$ and $S+U$ capture NP effects in neutral and charged current processes across different energy scales. The parameter $U$ is mainly constrained by the $W$ boson mass and width. In practice, $S$ and $T$ typically provide stronger constraints on NP than $U$.
\begin{figure}[h!]
    \centering
    \begin{tikzpicture}
        \begin{feynman}
            \vertex (a1){\(V_i\)};
            \vertex[square dot,blue,right=1.8cm of a1](a2){};
            \vertex[square dot,blue,right=1.7cm of a2](a3){};
            \vertex[right=1.8cm of a3](a4){\(V_j\)};
            
            \diagram*{
                (a1) --[boson,thick](a2) --[fermion,thick, half left, looseness=1.5,arrow size=1pt,edge label={\(t\)}](a3) --[boson,thick](a4),
                (a3) --[fermion,thick, half left, looseness=1.5,arrow size=1.pt,edge label={\(c/u\)}](a2)
            }; 
        \end{feynman}
    \end{tikzpicture}
     \caption{Gauge boson $(V_i=\gamma,Z)$ self-energy correction affected due to the presence of top-FCNC processes. 
 }
    \label{fig:gauge_boson_self_energy}
\end{figure}
The general expression describing the renormalized one-particle irreducible (1PI) two-point functions is defined through $\Pi_{V_i V_j}(q^2)$, where the NP contributions are encoded in $\delta \Pi_{V_i V_j}(q^2)$. Here, $V_i = \gamma, W^{\pm}, Z$ denotes the gauge bosons involved.
\begin{subequations}
\begin{align}\label{eq:gauge_self}
\Pi_{V_i V_j}(q^2)=\Pi_{V_i V_j}^{\rm SM}(q^2)+\delta \Pi_{V_i V_j}(q^2)
\end{align}
\begin{align}
    \left(\Pi_{V_i V_j}\right)_{\mu\nu}(q^2)=-i g_{\mu\nu}\left(q^2-m_{V_i}^2\right)-i\left(g_{\mu\nu}-\frac{q_{\mu}q_{\nu}}{q^2}\right)\Sigma^T_{V_i V_j}(q^2)-i\frac{q_{\mu}q_{\nu}}{q^2}\Sigma^L_{V_i V_j}(q^2)
\end{align}
\end{subequations}
In our analysis, various top-quark FCNC processes induce corrections to the self-energies of the $\gamma$ and $Z$ bosons as shown in fig.~\ref{fig:gauge_boson_self_energy}, which lead to modifications of the oblique parameters $S$, $T$, and $U$. Oblique parameters $(S,T, \mathrm{and}\,U)$ originally defined by Peskin and Takeuchi \cite{Peskin:1991sw}.
\begin{align}\label{eq:oblique_param}
    S&=\left(\frac{4 s_W^2 c_W^2}{\alpha_e}\right) \left(\Bigg[\frac{\delta \Sigma_{ZZ}^T(m_Z^2)-\delta \Sigma^T_{ZZ}(0)}{m_Z^2}\Bigg]-\frac{c_W^2-s_W^2}{c_W s_W}\frac{\delta \Sigma^T_{\gamma Z}(m_Z^2)}{m_Z^2}-\frac{\delta \Sigma^T_{\gamma\gamma}(m_Z^2)}{m_Z^2} \right)\,, \nonumber \\
    T&=\frac{1}{\alpha_e}\left(\frac{\delta \Sigma^T_{WW}(0)}{m_W^2} -\frac{\delta \Sigma^T_{ZZ}(0)}{m_Z^2}\right)\,,\\
    U&=\frac{4 s_W^2}{\alpha_e} \left(\Bigg[\frac{\delta \Sigma^T_{WW}(m_W^2)-\delta \Sigma^T_{WW}(0)}{m_W^2}\Bigg]-\frac{c_W}{s_W}\frac{\delta \Sigma^T_{Z\gamma}(m_Z^2)}{m_Z^2}-\frac{\delta \Sigma^T_{\gamma\gamma}(m_Z^2)}{m_Z^2}\right)-S\,. \nonumber
\end{align}
where $\alpha_e$ is defined at the electro-weak scale $(\mu_{\rm EW})$. For our fitting procedure, we have used the global fit results provided by the Gfitter group \cite{Baak:2014ora}, including the correlations among the oblique parameters. The self-energy expressions of the gauge bosons corresponding to fig.~\ref{fig:gauge_boson_self_energy} are provided in eqs.~\eqref{eq:Append_gauge_WR}.
\begin{subequations}\label{eq:Append_gauge_WR}
\begin{align}
    \Sigma_{\gamma\gamma}^T(q^2)&=-\frac{e^2}{8\pi^2}\left(|\lambda_L^{u_it}|^2+|\lambda_R^{u_it}|^2\right)q^2\\
    \Sigma_{\gamma Z}^T(q^2)&=-\frac{e g_2}{64\pi^2 c_W}\Bigg[\left(X_L^{tu_i\,*}\lambda_L^{u_i t}+X_R^{tu_i\,*}\lambda_R^{u_i t}+X_R^{u_i t}\lambda_L^{tu_i\,*}+X_L^{u_i t}\lambda_R^{tu_i\,*}\right)\left(1+2\log\frac{\mu^2}{m_t^2}\right)\nonumber\\
    &+8 \mathrm{Re}\left(\kappa_L^{tu_i*}\lambda_L^{u_it}+\kappa_R^{tu_i*}\lambda_R^{u_it}\right)\Bigg]q^2
\end{align}
\begin{align}
    &\Sigma_{ZZ}^T(q^2)=-\frac{g_2^2 m_t^2}{64\pi^2c_W^2}\left(|X_L^{u_it}|^2+|X_R^{u_it}|^2\right)\left(1+2\log\frac{\mu^2}{m_t^2}\right)\nonumber\\
    &+\frac{g_2^2}{576\pi^2 c_W^2}\Bigg[4\left(|X_L^{u_it}|^2+|X_R^{u_i t}|^2\right)\left(1+3\log\frac{\mu^2}{m_t^2}\right)-72\left(|\kappa_L^{u_it}|^2+|\kappa_R^{u_it}|^2\right)\nonumber\\
    &-18\,\mathrm{Re}\left(X_L^{tu_i\,*}\kappa_L^{u_i t}+X_R^{tu_i*}\kappa_L^{u_i t}+X_L^{tu_i\,*}\kappa_R^{u_i t}+X_R^{tu_i*}\kappa_R^{u_i t}\right)\left(1+2\log\frac{\mu^2}{m_t^2}\right)\Bigg]q^2
\end{align}
\end{subequations}
\paragraph{\underline{\textbf{$Z$ Pole Observables:}}}

\begin{figure}[h!]
	\centering
	\subfloat[]{
		\begin{tikzpicture}
			\begin{feynman}
				\vertex[square dot, draw=red, fill=red, minimum size=4pt, inner sep=2.5pt](a1){};
				\vertex[left=1.5cm of a1](a2){\(Z\)};
				\vertex[above right=1cm of a1](a3);
				\vertex[above right=0.8cm of a3](a4){\(b(s)\)};
				\vertex[below right=1cm of a1](a5);
				\vertex[below right=0.8cm of a5](a6){\(\bar{b}(\bar{s})\)};
				\diagram*{
					(a2) --[red,very thick,boson] (a1),
					(a6) --[fermion](a5) --[fermion, very thick, edge label=\(t\)](a1) --[fermion, very thick, edge label=\(u/c\)](a3) --[fermion](a4),
					(a5) --[boson, edge label'=\(W\)](a3),
				};
			\end{feynman}
		\end{tikzpicture}
	}
	\subfloat[]{
		\begin{tikzpicture}
			\begin{feynman}
				\vertex[square dot, draw=red, fill=red, minimum size=4pt, inner sep=2.5pt](a1){};
				\vertex[left=1.5cm of a1](a2){\(Z\)};
				\vertex[above right=1cm of a1](a3);
				\vertex[above right=0.8cm of a3](a4){\(b(s)\)};
				\vertex[below right=1cm of a1](a5);
				\vertex[below right=0.8cm of a5](a6){\(\bar{b}(\bar{s})\)};
				\diagram*{
					(a2) --[red,very thick,boson] (a1),
					(a6) --[fermion](a5) --[fermion, very thick, edge label=\(u/c\)](a1) --[fermion, very thick, edge label=\(t\)](a3) --[fermion](a4),
					(a5) --[boson, edge label'=\(W\)](a3),
				};
			\end{feynman}
		\end{tikzpicture}
	}
	\caption{Feynman diagrams of $Z\to d_i \bar{d}_i$ decay, modified by the top FCNC operators.}
	\label{fig:Z_pole}
\end{figure}

The $Z$-pole observables are among the most precise and sensitive measurements within EWPOs. They play a crucial role in testing the SM and probing potential NP effects. In our analysis, these observables are modified by corrections to the processes $Z \to f \bar{f}$.

Among these, the decays $Z \to b\bar{b}$ and $Z \to s\bar{s}$ receive the dominant contributions, primarily from single insertions of top-quark FCNC operators, as illustrated in fig.~\ref{fig:Z_pole}. We also consider processes such as $Z \to c\bar{c}$, $Z \to \ell\bar{\ell}$, and $Z \to \nu\bar{\nu}$. These channels are affected through double insertions of NP operators, with contributions arising from diagrams involving self-energy corrections to the $Z$-boson, as discussed earlier (see fig.~\ref{fig:gauge_boson_self_energy}). However, these effects are subleading, being suppressed by $\mathcal{O}(1/\Lambda^4)$, compared to the leading $\mathcal{O}(1/\Lambda^2)$ contributions in the $Z \to b\bar{b}$ and $Z \to s\bar{s}$ channels.

The Lagrangian describing the $Z \to f \bar{f}$ interaction can be written as:
\begin{equation}
    \mathcal{L}_{Zf\bar{f}} = \frac{g_2}{2\cos\theta_W} \sum_f \bar{f} \gamma^\mu \left( g_{Z,\,v}^{f} - g_{Z,\,a}^{f} \gamma^5 \right) f \, Z_\mu \,,
\end{equation}
The vertices can be written in terms of the NP correction as: 
\begin{align}\label{eq:Z_couplings}
    g_{Z,\,a}^{f}&\to g_{Z,\,a}^{f,\,\rm SM}+\delta g_{Z,\,a}^{f,\, \rm NP}\nonumber\\
    g_{Z,\,v}^{f}&\to g_{Z,\,v}^{f,\,\rm SM}+\delta g_{Z,\,v}^{f,\,\rm NP}
\end{align}\\
where $g_{Z,\,a}^{f,\,\rm SM}$ and $g_{Z,\,v}^{f,\,\rm SM}$ are the axial and vector couplings of the $Z$ boson to fermions respectively and $g_{Z,\,v}^{f,\,\rm SM} =(I^3-2 Q_f \sin^2\theta)$ and $g_{Z,\,a}^{f,\,\rm SM} =  I^3$, where $I^3$ and $Q_f$ represent the third component of the isospin and the charge of the fermion, respectively. The expression for NP couplings $\delta g_{Z,\,a}^{f,\, \rm NP},\delta g_{Z,\,v}^{f,\,\rm NP}$ are mentioned in the eq.~\eqref{eq:Append_Z_coupling}.

\begin{align}\label{eq:Append_Z_coupling}
    &\delta g_{Z,v}^{b\,,\rm NP}=-\delta g_{Z,a}^{b\,,\rm NP}=\frac{g_2^2}{32\pi^2}\sum_{u_i=u,c}V_{tb} V_{u_ib}^*\left\{\,\mathrm{Re}\left(X_L^{u_i t}\right)\Bigg[\frac{2(5 x_Z+2-x_t(1- x_Z))}{x_Z}\right. \nonumber\\
    &+2 x_t \log\frac{\mu^2}{m_t^2} +\frac{2(2+3x_Z-2x_t(1+x_Z))}{(x_t-1)x_Z}\log x_t
    +\frac{2(x_t-x_Z)(x_t-3x_Z-2)}{x_Z^2} \log\left(\frac{x_t}{x_t-x_z}\right)\Bigg]\nonumber\\
    &\left.-8 \,i\mathrm{Im}\left(\kappa_R^{u_i t}\right)\Bigg[1-\log x_t-\frac{x_t-x_Z}{x_Z}\log\left(\frac{x_t}{x_t-x_Z}\right)\Bigg]\right\}
\end{align}
The expression for $\delta g_Z^{s,\rm NP}$ will be same $\delta g_Z^{b,\rm NP}$ with the replacement in the CKM, $V_{tb}V_{u_i b}^*\Leftrightarrow V_{ts} V_{u_i s}^*$. The partial decay width of $Z\to f\bar{f}$ process can be written as:
\begin{align}
\label{eq:decaywidth_Zpole}
\Gamma(Z \to f \bar{f}) = \frac{N_c^b}{48} \frac{\alpha}{s_W^2 c_W^2} m_Z \sqrt{1-\mu_{f}^2}& \biggl( |g_{Z\,,a}^{f}|^2 (1-\mu_{f}^2) + |g_{Z\,,v}^{f}|^2(1+\frac{\mu_{f}^2}{2})\biggr)(1+\delta_{f}^0)(1+\delta_{b}) \nonumber \\ &  (1+\delta_{\rm QCD})(1+\delta_{\rm QED}) (1+\delta^{f}_{\mu}) \,,
\end{align}
The parameter $\mu_f^2 = \frac{4 m_f^2}{m_Z^2}$ is relevant primarily for the $Z \to b\bar{b}$ decay channel. The corrections $\delta_f^0$, $\delta_b$, $\delta_{\text{QCD}}$, $\delta_{\text{QED}}$, and $\delta_f^{\mu}$ account for various radiative effects beyond the leading-order contribution. A detailed discussion can be found in refs.~\cite{Soni:2010xh,Bernabeu:1990ws} and the references therein. After performing the loop integration, various $Z$ boson decay channels receive corrections, which subsequently affect various $Z$-pole observables. These include ratio observables, asymmetry parameters, and the total decay width of the $Z$ boson. The corresponding values of these observables are listed in table.~\ref{tab:Z_pole}. In our analysis, we have followed the methodology developed in our previous works~\cite{Kolay:2024wns,Kolay:2025jip,Kala:2025srq}, which simplifies the expressions for $Z$-pole observables in terms of the modified vector and axial-vector couplings.

\paragraph{\underline{\textbf{$W$ Pole Observables:}}}

\begin{figure}[h!]
	\centering
	\subfloat[]{
		\begin{tikzpicture}
			\begin{feynman}
				\vertex[](a1);
				\vertex[left=1.cm of a1](a2){\(W\)};
				\vertex[square dot, draw=red, fill=red, minimum size=4pt, inner sep=2.5pt,above right=1cm of a1](a3){};
				\vertex[above right=1.2cm of a3](a4){\(c(u)\)};
				\vertex[below right=1cm of a1](a5);
				\vertex[below right=1.cm of a5](a6){\(\bar{s}(\bar{d})\)};
				\diagram*{
					(a1) --[boson] (a2),
					(a6) --[fermion] (a5) --[fermion, very thick, edge label'=\(t\)] (a3) --[fermion, very thick] (a4),
					(a1) --[red, boson, very thick, edge label=\(\gamma/Z\)] (a3),
					(a1) --[boson, edge label'=\(W\)] (a5),
				};
			\end{feynman}
		\end{tikzpicture}
	}
	\subfloat[]{
		\begin{tikzpicture}
			\begin{feynman}
				\vertex[](a1);
				\vertex[left=1.cm of a1](a2){\(W\)};
				\vertex[square dot, draw=red, fill=red, minimum size=4pt, inner sep=2.5pt,above right=1cm of a1](a3){};
				\vertex[above right=1.2cm of a3](a4){\(c(u)\)};
				\vertex[below right=1cm of a1](a5);
				\vertex[below right=1.cm of a5](a6){\(\bar{s}(\bar{d})\)};
				\diagram*{
					(a1) --[boson] (a2),
					(a6) --[fermion] (a5) --[fermion, edge label=\(s(d)\)] (a1) --[fermion, very thick, edge label=\(t\)] (a3) --[fermion, very thick](a4),
					(a5) --[red, boson, very thick, edge label'=\(\gamma/Z\)] (a3),
				};
			\end{feynman}
		\end{tikzpicture}
	}
	\subfloat[]{
		\begin{tikzpicture}
			\begin{feynman}
				\vertex[](a1);
				\vertex[left=1.cm of a1](a2){\(W\)};
				\vertex[square dot, draw=blue, fill=blue, minimum size=4pt, inner sep=2.5pt,above right=1cm of a1](a3){};
				\vertex[above right=1.2cm of a3](a4){\(c(u)\)};
				\vertex[below right=1cm of a1](a5);
				\vertex[below right=1.cm of a5](a6){\(\bar{s}(\bar{d})\)};
				\diagram*{
					(a1) --[boson] (a2),
					(a6) --[fermion] (a5) --[fermion, edge label=\(s(d)\)] (a1) --[fermion, very thick, edge label=\(t\)] (a3) --[fermion, very thick](a4),
					(a5) --[blue, gluon, very thick, edge label'=\(g\)] (a3),
				};
			\end{feynman}
		\end{tikzpicture}
	}\\
	\subfloat[]{
		\begin{tikzpicture}
			\begin{feynman}
				\vertex[](a1);
				\vertex[left=1.cm of a1](a2){\(W\)};
				\vertex[square dot, draw=violet, fill=violet, minimum size=4pt, inner sep=2.5pt,above right=1cm of a1](a3){};
				\vertex[above right=1.2cm of a3](a4){\(c(u)\)};
				\vertex[below right=1cm of a1](a5);
				\vertex[below right=1.cm of a5](a6){\(\bar{s}(\bar{d})\)};
				\diagram*{
					(a1) --[boson] (a2),
					(a6) --[fermion] (a5) --[fermion, edge label=\(s(d)\)] (a1) --[fermion, very thick, edge label=\(t\)] (a3) --[fermion, very thick](a4),
					(a5) --[violet, scalar, very thick, edge label'=\(H\)] (a3),
				};
			\end{feynman}
		\end{tikzpicture}
	}
	\subfloat[]{
		\begin{tikzpicture}
			\begin{feynman}
				\vertex[](a1);
				\vertex[left=1.cm of a1](a2){\(W\)};
				\vertex[square dot, draw=violet, fill=violet, minimum size=4pt, inner sep=2.5pt,above right=1cm of a1](a3){};
				\vertex[above right=1.2cm of a3](a4){\(c(u)\)};
				\vertex[below right=1cm of a1](a5);
				\vertex[below right=1.cm of a5](a6){\(\bar{s}(\bar{d})\)};
				\diagram*{
					(a1) --[boson] (a2),
					(a6) --[fermion] (a5) --[fermion, very thick, edge label'=\(t\)] (a3) --[fermion, very thick] (a4),
					(a1) --[violet, scalar, very thick, edge label=\(H\)] (a3),
					(a1) --[boson, edge label'=\(W\)] (a5),
				};
			\end{feynman}
		\end{tikzpicture}
	}
	\caption{Feynman diagrams of $W\to u_i \bar{d}_j$ decay, modified by the top FCNC operators.}
	\label{fig:W_pole}
\end{figure}

In addition to the $Z$-pole observables, the $W$-pole observables are also affected in the presence of top quark FCNC operators, as shown in fig.~\ref{fig:W_pole}. These operators induce one-loop level corrections to the $Wc\bar{s}$ and $Wu\bar{d}$ vertex, which lead to shifts in observables such as $R_{Wc}$ and $\Gamma_{\rm hadrons}$.
As a result, precision measurements of $W$-pole observables provide complementary constraints on the NP effects in our framework. The list of observables along with their corresponding values is provided in table.~\ref{tab:W_pole}.

\subsection{Higgs Observables}\label{sec:appHiggs}
\begin{figure}[t]
    \centering
    \subfloat[]{
        \begin{tikzpicture}
          \begin{feynman}
            \vertex[square dot, draw=violet, fill=violet, minimum size=4pt, inner sep=2.5pt](a1){};
            \vertex[left=1.5cm of a1](a2){\(H\)};
            \vertex[above right=1cm of a1](a3);
            \vertex[above right=0.8cm of a3](a4){\(b\)};
            \vertex[below right=1cm of a1](a5);
            \vertex[below right=0.8cm of a5](a6){\(\bar{b}\)};
            \diagram*{
                (a2) --[violet,very thick,scalar] (a1),
                (a6) --[fermion](a5) --[fermion, very thick, edge label=\(t\)](a1) --[fermion, very thick, edge label=\(u/c\)](a3) --[fermion](a4),
                (a5) --[boson, edge label'=\(W\)](a3),
            };
          \end{feynman}
        \end{tikzpicture}
    }
    \subfloat[]{
        \begin{tikzpicture}
          \begin{feynman}
            \vertex[square dot, draw=violet, fill=violet, minimum size=4pt, inner sep=2.5pt](a1){};
            \vertex[left=1.5cm of a1](a2){\(H\)};
            \vertex[above right=1cm of a1](a3);
            \vertex[above right=0.8cm of a3](a4){\(b\)};
            \vertex[below right=1cm of a1](a5);
            \vertex[below right=0.8cm of a5](a6){\(\bar{b}\)};
            \diagram*{
                (a2) --[violet,very thick,scalar] (a1),
                (a6) --[fermion](a5) --[fermion, very thick, edge label=\(u/c\)](a1) --[fermion, very thick, edge label=\(t\)](a3) --[fermion](a4),
                (a5) --[boson, edge label'=\(W\)](a3),
            };
          \end{feynman}
        \end{tikzpicture}
    }
    \caption{Modification of the $H \to b\bar{b}$ decay channel due to the presence of top FCNC operators.}
    \label{fig:Hbbbar_decay}
\end{figure}
\paragraph{\underline{\textbf{$H\to b\bar{b}$ decay:}}}
In the presence of top FCNC interactions, the $Hb\bar{b}$ vertex receives loop-level corrections, as depicted in fig.~\ref{fig:Hbbbar_decay}. While the vertex is purely scalar in the SM, these corrections introduce deviations that modify its chiral structure. In our analysis, we incorporate these effects through modification of the chiral current.

\begin{align}\label{eq:Hbb_vertex}
\mathcal{L}_{Hbb}=C_L^H(\bar{b} P_{L} b )+C_R^H(\bar{b} P_{R} b )
\end{align}
The modified $Hb\bar{b}$ vertex can be expressed as $C^H_{L(R)} = C_{L(R)}^{H\,,\rm{SM}} + C_{L(R)}^{H\,,\rm{NP}}$, where the SM contribution is given by $C_{L, R}^{H\,,\rm{SM}} = \frac{m_b}{v}$, with $v$ being the vacuum expectation value of the Higgs field. The expressions for the new physics (NP) contributions $C_{L, R}^{H\,,\rm{NP}}$ are provided in eqs.~\eqref{eq:Append_Higgs_b}-\eqref{eq:Append_Higgs_b1}.
\begin{subequations}
    \begin{align}\label{eq:Append_Higgs_b}
    C_L^{H\,,\rm NP}&=\sum_{u_i= u,c}\frac{g_2^2 m_b m_t}{128\sqrt{2}\pi^2M_H^2}V_{tb} V_{u_i b}^*\left(-8(x_H+1)\eta_R^{u_it}+\frac{8-x_H(x_t-3)-8 x_t}{x_t-1}\eta_R^{tu_i\,*}\right.\nonumber\\
    &-2x_H(2 \eta_R^{u_i t}+\eta_R^{tu_i\,*})\log\frac{\mu^2}{m_t^2}+2\left(4 \eta_R^{u_i t}-\frac{4 x_t+x_H-4}{(x_t-1)^2}\eta_R^{tu_i\,*}\right)\log x_t\\
    &\left. -\frac{4(x_H+x_t)((2+x_H)\eta_R^{u_i t}+2 \eta_R^{tu_i\,*})}{x_H}\log\frac{x_t}{x_t+x_H}\right)\nonumber
\end{align}

\begin{align}\label{eq:Append_Higgs_b1}
    C_R^{H\,,\rm NP}&=\sum_{u_i= u,c}\frac{g_2^2 m_b m_t}{128\sqrt{2}\pi^2M_H^2}V_{tb} V_{u_i b}^* \left(\left(8+ x_H\frac{x_t-3}{x_t-1}\right)\eta_R^{u_i t}+8(1+x_H)\eta_R^{tu_i*} \right.\nonumber\\
    &+2 x_H(\eta_R^{u_i t}+2 \eta_R^{tu_i\,*})\log\frac{\mu^2}{m_t^2}-2\left(4 \eta_R^{t u_i \,*}-\frac{4 x_t+x_H-4}{(x_t-1)^2}\eta_R^{u_it}\right)\log x_t\\
    &\left. +\frac{4(x_H+x_t)((2+x_H)\eta_R^{t u_i\,*}+2 \eta_R^{u_i t})}{x_H}\log\frac{x_t}{x_t+x_H}\right)\nonumber
\end{align}
\end{subequations}

The decay width can then be written in terms of the modified $Hb\bar{b}$ current as:
\begin{equation}
    \Gamma(H \to b \bar{b}) = \frac{N_{c}}{16 \pi m_{H}^2 }\sqrt{m_{H}^2 - 2 m_{b}^2}  \left[  m_{H}^2 \left( (C_{L}^H)^2 + (C_{R}^H)^2 \right)  - 2 m_{b}^2 \left(C_{L}^H+ C_{R}^H \right)^2 \right]\,.
\end{equation}
Here, the color factor $N_c = 3$ accounts for the three possible color states of the $b\bar{b}$ pair. We take the running bottom quark mass at the Higgs mass scale, $m_b = m_b(M_H) \approx 2.7$ GeV. The experimental value of the partial decay width for the process $H \to b\bar{b}$ is given by:
$$
\Gamma(H \to b\bar{b}) = \left(1.961 \pm 0.923\right) \times 10^{-3}~\text{GeV} \,.
$$
This value is derived using the total Higgs decay width $\Gamma_H^{\rm tot} = 3.7^{+1.9}_{-1.4}$ MeV, which corresponds to the PDG average~\cite{ParticleDataGroup:2024cfk}, based on measurements reported by both ATLAS~\cite{ATLAS:2023dnm} and CMS~\cite{CMS:2022ley,CMS:2023vzh}. We also checked the process $H \to WW^* \to W f \bar{f}^{\prime}$, which is, in principle, affected by top FCNC operators at one-loop. However, due to the unitarity properties of the CKM matrix, the loop correction effectively vanishes at zero light quark mass limit.\\

\paragraph{\underline{\textbf{$H \to \gamma \gamma(Z)$ decay:}}}
In the SM, the Higgs boson does not couple to photons at tree level, making the $H \to \gamma\gamma$ and $H \to \gamma Z$ decay channels rare with small branching ratios. Despite this, the diphoton channel played a crucial role in the discovery of the Higgs boson due to its clean experimental signature at colliders.
\begin{figure}[h!]
    \centering
\subfloat[]{
        \begin{tikzpicture}
          \begin{feynman}
            \vertex[square dot, draw=violet, fill=violet, minimum size=4pt, inner sep=2.5pt](a1){};
            \vertex[left=1.5cm of a1](a2){\(H\)};
            \vertex[square dot, red,above right=1cm of a1](a3){};
            \vertex[above right=1.2cm of a3](a4){\(V_i\)};
            \vertex[below right=1cm of a1](a5);
            \vertex[below right=0.8cm of a5](a6){\(V_j\)};
            \diagram*{
                (a2) --[violet,very thick,scalar] (a1),
                (a5) --[fermion, very thick, edge label=\(t\)](a1) --[fermion, very thick, edge label=\(u/c\)](a3) --[fermion, very thick, edge label=\(t\)](a5),
                (a5) --[boson](a6),
                (a3) --[red, very thick,boson](a4),

            };
          \end{feynman}
        \end{tikzpicture}
    }
\subfloat[]{
        \begin{tikzpicture}
          \begin{feynman}
            \vertex[square dot, draw=violet, fill=violet, minimum size=4pt, inner sep=2.5pt](a1){};
            \vertex[left=1.5cm of a1](a2){\(H\)};
            \vertex[square dot, red,above right=1cm of a1](a3){};
            \vertex[above right=1.2cm of a3](a4){\(V_i\)};
            \vertex[below right=1cm of a1](a5);
            \vertex[below right=0.8cm of a5](a6){\(V_j\)};
            \diagram*{
                (a2) --[violet,very thick,scalar] (a1),
                (a5) --[fermion, very thick, edge label=\(u/c\)](a1) --[fermion, very thick, edge label=\(t\)](a3) --[fermion, very thick, edge label=\(u/c\)](a5),
                (a5) --[boson](a6),
                (a3) --[red, very thick,boson](a4),

            };
          \end{feynman}
        \end{tikzpicture}
    }
\subfloat[]{
        \begin{tikzpicture}
          \begin{feynman}
            \vertex[square dot, draw=violet, fill=violet, minimum size=4pt, inner sep=2.5pt](a1);
            \vertex[left=1.cm of a1](a2){\(H\)};
            \vertex[square dot, red,above right=1cm of a1](a3){};
            \vertex[above right=1.2cm of a3](a4){\(V_i\)};
            \vertex[square dot, red,below right=1cm of a1](a5){};
            \vertex[below right=1.2cm of a5](a6){\(V_j\)};
            \diagram*{
                (a2) --[scalar] (a1),
                (a5) --[fermion, very thick, edge label=\(t\)](a1) --[fermion, very thick, edge label=\(t\)](a3) --[fermion, very thick, edge label=\(u/c\)](a5),
                (a5) --[red, very thick,boson](a6),
                (a3) --[red, very thick,boson](a4),

            };
          \end{feynman}
        \end{tikzpicture}
    }
\subfloat[]{
        \begin{tikzpicture}
          \begin{feynman}
            \vertex[square dot, draw=violet, fill=violet, minimum size=4pt, inner sep=2.5pt](a1);
            \vertex[left=1.cm of a1](a2){\(H\)};
            \vertex[square dot, red,above right=1cm of a1](a3){};
            \vertex[above right=1.2cm of a3](a4){\(V_i\)};
            \vertex[square dot, red,below right=1cm of a1](a5){};
            \vertex[below right=1.2cm of a5](a6){\(V_j\)};
            \diagram*{
                (a2) --[scalar] (a1),
                (a5) --[fermion, very thick, edge label=\(u/c\)](a1) --[fermion, very thick, edge label=\(u/c\)](a3) --[fermion, very thick, edge label=\(t\)](a5),
                (a5) --[red, very thick,boson](a6),
                (a3) --[red, very thick,boson](a4),

            };
          \end{feynman}
        \end{tikzpicture}
    }
    \caption{Feynman diagrams depicted $H \to V_i V_j$ decay channels, modified by the top FCNC operators. Here, $V_i, V_j = \gamma, Z, g$. Additionally, these diagrams correspond to the $H \to \gamma\gamma$, $H \to \gamma Z$ and $H \to g g$ decays. 
 }
    \label{fig:Hdiphoton}
\end{figure}\\
In the SM, the decays $H \to \gamma\gamma$ and $H \to \gamma Z$ are mediated by loop diagrams involving closed fermion and gauge boson loops, with the dominant contribution arising from the top quark. In our analysis, we obtain additional fermion loop contributions induced by top FCNC operators as depicted in fig.~\ref{fig:Hdiphoton}, which modify the amplitude beyond the SM prediction. The effective vertices can be written as
\begin{subequations}\label{eq:Lag_HphotonZ}
\begin{align}
\delta\mathcal{L}_{h\gamma\gamma}&=-\frac{\delta_{\gamma\gamma}}{2\Lambda^2}e^2 v h F_{\mu\nu}F^{\mu\nu}\,,& 
    \delta\mathcal{L}_{h\gamma Z}&=-\frac{\delta_{\gamma Z}}{2 \Lambda^2}e^2 v h F_{\mu\nu}Z^{\mu\nu}\,.
\end{align}
\end{subequations}
Thus, the resulting modified decay width can be expressed as \cite{Manohar:2006gz} :
\begin{subequations}
\begin{align}
    \frac{\Gamma(h \to \gamma \gamma)}{\Gamma^{\rm{SM}}(h\to \gamma \gamma)}&\simeq \left| 1- \frac{4 \pi^2 v^2 \delta_{\gamma\gamma}}{\Lambda^2 I^{\gamma\gamma}}\right|^2\\
    \frac{\Gamma(h \to \gamma Z)}{\Gamma^{\rm{SM}}(h\to \gamma Z)}&\simeq \left| 1- \frac{4 \pi^2 v^2 \delta_{\gamma Z}}{\Lambda^2 I^{\gamma Z}}\right|^2
\end{align}
\end{subequations}
Here, $I^{\gamma\gamma(Z)}$ denotes the Feynman parameter integrals with $I^{\gamma\gamma}\approx-1.65$ and $I^{\gamma Z}\approx -2.87$. The analytic and numerical expressions of these Feynman integrals are taken from refs.~\cite{Manohar:2006gz,Alonso:2013hga,Bergstrom:1985hp}. The new physics coupling modifiers $\delta_{\gamma\gamma(Z)}$ are provided in eqs.~\eqref{eq:Append_kappa_gamma}-~\eqref{eq:Append_kappa_gammaZ}.
\begin{subequations}
\begin{align}\label{eq:Append_kappa_gamma}
 &\delta_{\gamma\gamma}=\frac{1}{12\sqrt{2}\pi^2 m_t v}\left(2i\,\mathrm{Im}\left(\eta_L^{tu_i\,*}\lambda_L^{u_i t}+\eta_R^{tu_i\,*}\lambda_R^{u_i t}\right)\left(\log\frac{\mu^2}{m_t^2}+\log\frac{m_t^2}{m_t^2-M_H^2}\right)\right.\nonumber\\
 &\left.-2i\,\,\mathrm{Im}\left(\eta_L^{u_it}\lambda_L^{tu_i\,*}+\eta_R^{u_it}\lambda_R^{tu_i\,*}\right)+\frac{3\sqrt{2} m_t}{v}\left(|\lambda_L^{u_it}|^2+|\lambda_R^{u_i t}|^2\right)\right)
\end{align}

\begin{align}\label{eq:Append_kappa_gammaZ}
    &\delta_{\gamma Z}=\frac{1}{192 \pi^2 v^2 m_t}\Bigg[\left(-2\sqrt{2}\,iv\frac{5 g_1^2-3g_2^2}{g_1 g_2}\mathrm{Im}\left(\eta_L^{tu_i\,*}\lambda_L^{u_it}+\eta_R^{tu_i\,*}\lambda_R^{u_it}\right)+6\,i\frac{m_t}{g_1 g_2}\mathrm{Im}\left(X_R^{tu_i\,*}\lambda_L^{u_i t}+X_L^{tu_i\,*}\lambda_R^{u_i t}\right)\right.\nonumber\\
    &\left.+4\sqrt{2}e^2 g_1 g_2\left(2i\mathrm{Im}\left(\eta_L^{tu_i\,*}\kappa_L^{u_i t}+\eta_R^{tu_i*}\kappa_R^{u_i t}\right)-\left(\eta_L^{u_i t}\kappa_L^{tu_i *}+\eta_R^{u_i t}\kappa_R^{tu_i *}\right)\right)\right)\left(1
    +\log\frac{\mu^2}{m_t^2}\right)\nonumber\\
    &+\frac{24m_t}{g_1 g_2}\mathrm{Re}\left(\kappa_L^{tu_i\,*}\lambda_L^{u_i t}+\kappa_R^{tu_i\,*}\lambda_R^{u_i t}\right)+\left(2\sqrt{2}i \,v\frac{(5 g_1^2-3 g_2^2)}{g_1 g_2}\mathrm{Im}\left(\eta_L^{tu_i*}\lambda_L^{u_i t}+\eta_R^{tu_i*}\lambda_R^{u_i t}\right)\right.\nonumber\\\ &\left.-4\sqrt{2} e^2 g_1 g_2\left(2 i \mathrm{Im}\left(\eta_L^{tu_i*}\kappa_L^{u_i t}+\eta_R^{tu_i*}\kappa_R^{u_i t}\right)-\left(\eta_L^{u_it}\kappa_L^{tu_i *}+\eta_R^{u_i t}\kappa_R^{t u_i *}\right)\right)\right)\log\frac{m_t^2}{m_t^2-M_H^2}\Bigg]
\end{align}
\end{subequations}

\paragraph{\underline{\textbf{$g g\to H$:}}}
Like the $H \to \gamma\gamma\,(Z)$ channels, the $gg \to H$ process is another important probe, as it is loop-mediated in the SM. In the SM, the dominant contribution arises from a top-quark triangle loop. In our analysis, similar types of NP triangular loop contributions arise, as mentioned in fig.~\ref{fig:Hdiphoton} (mirror symmetric). The effective Lagrangian describing the NP Higgs-gluon production channel is given by:
\begin{align}\label{eq:gluon_Higgs}
   \delta \mathcal{L}_{ggh}=-\frac{\delta_{gg}}{2 \Lambda^2}g_s^2 v h G^a_{\mu\nu}G^{a\,\mu\nu} 
\end{align}
The corresponding changes in the Higgs production and decay cross sections with respect to the SM predictions can be expressed as~\cite{Manohar:2006gz}:
\begin{align}
    \frac{\sigma(gg\to h)}{\sigma^{\rm SM}(gg\to h)}\simeq \frac{\Gamma(h\to gg)}{\Gamma^{\rm SM}(h\to g g)}\simeq\left|1-\frac{8\pi^2 v^2 \delta_{gg}}{\Lambda^2 I^{gg}}\right|^2
\end{align}
where $I^{gg} \approx 0.37$~\cite{Manohar:2006gz,Alonso:2013hga,Bergstrom:1985hp} is the Feynman parameter integral arising from the $gg \to h$ loop diagrams in the SM.
\paragraph{} In our analysis, we adopt the $\kappa$-framework to confront the new physics scenario, focusing specifically on the Higgs boson production and decay channels: $H \to \gamma\gamma$, $H \to Z\gamma$, and $H \to gg$. The general expression of a $\kappa_i$ parameter is given by
\begin{equation}
	\kappa_i^2 = \frac{\Gamma_i^{\rm Total}}{\Gamma_i^{\rm SM}} \,,
\end{equation}
where $\Gamma_i^{\rm SM}$ is the partial decay width in the SM and $\Gamma_i^{\rm Total}$ is the same width including the contribution from NP. In the SM, these ratios $\kappa_i = 1$, and any observed deviation from this value would indicate the presence of physics beyond the SM. 
We have presented the resulting expressions for $\delta_{gg}$, for different top-FCNC operator insertions, in eq~.\eqref{eq:Append_kappa_gg}.
\begin{align}\label{eq:Append_kappa_gg}
    &\delta_{gg}=\frac{1}{8\sqrt{2}\pi^2 m_t v}\left(i\,\mathrm{Im}\left(\eta_L^{tu_i\,*}\xi_L^{u_i t}+\eta_R^{tu_i\,*}\xi_R^{u_i t}\right)\left(\log\frac{\mu^2}{m_t^2}+\log\frac{m_t^2}{m_t^2-M_H^2}\right)\right.\nonumber\\
 &\left.-i\,\,\mathrm{Im}\left(\eta_L^{u_it}\xi_L^{tu_i\,*}+\eta_R^{u_it}\xi_R^{tu_i\,*}\right)+\frac{\sqrt{2} m_t}{v}\left(|\xi_L^{u_it}|^2+|\xi_R^{u_i t}|^2\right)\right)
\end{align}

\section{Correlation Matrices}\label{Append:correlation_mat}
In this section, we present the correlation matrices among the various top FCNC couplings obtained from the global fit, which is presented in the section \ref{sec:Result}.
\begin{table}[htb!]
\centering
\footnotesize
\renewcommand{\arraystretch}{1.4}
\setlength{\tabcolsep}{6pt}
\begin{tabular}{|c|c c c c||c|c c c c|}
\hline
\rowcolor{gray!20}
\textbf{Coupling} &
$\mathrm{Re}(\xi_L^{ct})$ &
$\mathrm{Im}(\xi_L^{ct})$ &
$\mathrm{Re}(\xi_R^{ct})$ &
$\mathrm{Im}(\xi_R^{ct})$ &
\textbf{Coupling} &
$\mathrm{Re}(\xi_L^{ut})$ &
$\mathrm{Im}(\xi_L^{ut})$ &
$\mathrm{Re}(\xi_R^{ut})$ &
$\mathrm{Im}(\xi_R^{ut})$ \\
\hline
$\mathrm{Re}(\xi_L^{ct})$ & $1.0$ & $0.0$ & $-0.001$ & $-0.007$ &
$\mathrm{Re}(\xi_L^{ut})$ & $1.0$ & $0.120$ & $-0.376$ & $0.020$ \\
$\mathrm{Im}(\xi_L^{ct})$ &     & $1.0$   & $0.0$ & $0.0$ &
$\mathrm{Im}(\xi_L^{ut})$ &     & $1.0$   & $0.183$ & $-0.010$ \\
$\mathrm{Re}(\xi_R^{ct})$ &     &       & $1.0$   & $-0.005$ &
$\mathrm{Re}(\xi_R^{ut})$ &     &       & 1.0 & $0.210$ \\
$\mathrm{Im}(\xi_R^{ct})$ &     &       &       & $1.0$ &
$\mathrm{Im}(\xi_R^{ut})$ &     &       &       & 1.0 \\
\hline
\end{tabular}
\caption{Correlation matrices of the complex top–gluon FCNC dipole couplings at the scale $\mu_{\rm EW}$.}
\label{tab:corr_top_gluon}
\end{table}

\begin{table}[htb!]
\centering
\footnotesize
\renewcommand{\arraystretch}{1.4}
\setlength{\tabcolsep}{6pt}
\begin{tabular}{|c|c c c c||c|c c c c|}
\hline
\rowcolor{gray!20}
\textbf{Coupling} &
$\mathrm{Re}(\lambda_L^{ct})$ &
$\mathrm{Im}(\lambda_L^{ct})$ &
$\mathrm{Re}(\lambda_R^{ct})$ &
$\mathrm{Im}(\lambda_R^{ct})$ &
\textbf{Coupling} &
$\mathrm{Re}(\lambda_L^{ut})$ &
$\mathrm{Im}(\lambda_L^{ut})$ &
$\mathrm{Re}(\lambda_R^{ut})$ &
$\mathrm{Im}(\lambda_R^{ut})$ \\
\hline
$\mathrm{Re}(\lambda_L^{ct})$ & 1.0 & $-0.001$ & $-0.001$ & 0.005 &
$\mathrm{Re}(\lambda_L^{ut})$ & 1.0 & 0.009 & 0.013 & $-0.014$ \\
$\mathrm{Im}(\lambda_L^{ct})$ &       & 1.0    & $-0.003$ & 0.019 &
$\mathrm{Im}(\lambda_L^{ut})$ &       & 1.0   & $-0.027$ & 0.028 \\
$\mathrm{Re}(\lambda_R^{ct})$ &       &          & 1.0    & 0.918 &
$\mathrm{Re}(\lambda_R^{ut})$ &       &         & 1.0    & 0.934 \\
$\mathrm{Im}(\lambda_R^{ct})$ &       &          &          & 1.0 &
$\mathrm{Im}(\lambda_R^{ut})$ &       &         &          & 1.0 \\
\hline
\end{tabular}
\caption{Correlation matrices of the complex top–photon FCNC dipole couplings
$\lambda_{L,R}^{ct}$ and $\lambda_{L,R}^{ut}$ at the scale $\mu_{\rm EW}$.}
\label{tab:corr_top_photon}
\end{table}

\begin{table}[h!]
\centering
\footnotesize
\renewcommand{\arraystretch}{1.4}
\setlength{\tabcolsep}{6pt}
\begin{tabular}{|c|c c c c||c|c c c c|}
\hline
\rowcolor{gray!20}
\textbf{Coupling} &
$\mathrm{Re}(\kappa_L^{ct})$ &
$\mathrm{Im}(\kappa_L^{ct})$ &
$\mathrm{Re}(\kappa_R^{ct})$ &
$\mathrm{Im}(\kappa_R^{ct})$ &
\textbf{Coupling} &
$\mathrm{Re}(\kappa_L^{ut})$ &
$\mathrm{Im}(\kappa_L^{ut})$ &
$\mathrm{Re}(\kappa_R^{ut})$ &
$\mathrm{Im}(\kappa_R^{ut})$ \\
\hline
$\mathrm{Re}(\kappa_L^{ct})$ & 1.000 & $-0.123$ & $-0.231$ & 0.375 &
$\mathrm{Re}(\kappa_L^{ut})$ & 1.000 & 0.181 & 0.213 & $-0.044$ \\
$\mathrm{Im}(\kappa_L^{ct})$ &       & 1.000    & $-0.055$ & 0.090 &
$\mathrm{Im}(\kappa_L^{ut})$ &       & 1.000 & $-0.297$ & 0.171 \\
$\mathrm{Re}(\kappa_R^{ct})$ &       &          & 1.000 & 0.168 &
$\mathrm{Re}(\kappa_R^{ut})$ &       &       & 1.000 & 0.122 \\
$\mathrm{Im}(\kappa_R^{ct})$ &       &          &       & 1.000 &
$\mathrm{Im}(\kappa_R^{ut})$ &       &       &       & 1.000 \\
\hline
\end{tabular}
\caption{Correlation matrices of the complex top--$Z$ boson FCNC dipole couplings
$\kappa_{L,R}^{ct}$ and $\kappa_{L,R}^{ut}$ at the scale $\mu_{\rm EW}$.}
\label{tab:corr_top_Z_dipole}
\end{table}

\begin{table}[htb!]
\centering
\footnotesize
\renewcommand{\arraystretch}{1.4}
\setlength{\tabcolsep}{6pt}
\begin{tabular}{|c|c c c c||c|c c c c|}
\hline
\rowcolor{gray!20}
\textbf{Coupling} &
$\mathrm{Re}(\eta_L^{ct})$ &
$\mathrm{Im}(\eta_L^{ct})$ &
$\mathrm{Re}(\eta_R^{ct})$ &
$\mathrm{Im}(\eta_R^{ct})$ &
\textbf{Coupling} &
$\mathrm{Re}(\eta_L^{ut})$ &
$\mathrm{Im}(\eta_L^{ut})$ &
$\mathrm{Re}(\eta_R^{ut})$ &
$\mathrm{Im}(\eta_R^{ut})$ \\
\hline
$\mathrm{Re}(\eta_L^{ct})$ & 1.0 & $-0.007$ & $-0.068$ & $-0.003$ &
$\mathrm{Re}(\eta_L^{ut})$ & 1.0 & 0.012 & 0.033 & 0.022 \\
$\mathrm{Im}(\eta_L^{ct})$ &       & 1.0    & $-0.150$ & $-0.005$ &
$\mathrm{Im}(\eta_L^{ut})$ &       & 1.0    & $-0.182$ & $-0.170$ \\
$\mathrm{Re}(\eta_R^{ct})$ &       &          & 1.0    & $-0.056$ &
$\mathrm{Re}(\eta_R^{ut})$ &       &          & 1.0    & $-0.341$ \\
$\mathrm{Im}(\eta_R^{ct})$ &       &          &          & 1.0 &
$\mathrm{Im}(\eta_R^{ut})$ &       &          &          & 1.0 \\
\hline
\end{tabular}
\caption{Correlation matrices of the top--Higgs ($\eta^{ct}, \eta^{ut}$) FCNC couplings at the scale $\mu_{\rm EW}$.}
\label{tab:corr_eta_ct_ut}
\end{table}
\section{SMEFT Couplings}\label{Append:SMEFT_couplings}
In this section, we present the fitted values of the SMEFT couplings 
$\mathcal{C}_{3i}$, which correspond to the left-chiral components of the 
top--FCNC effective interactions.
\begin{table}[htb!]
    \centering
    \footnotesize
    \renewcommand{\arraystretch}{2.}
    \begin{tabular}{|c|c|c|c|}
    \hline
\multicolumn{4}{|c|}{\textbf{ Real SMEFT Couplings} $(\mu_{\rm EW})$} \\
    \hline
    \rowcolor{gray!20}
       \textbf{Coupling}  & \textbf{Value}~$(\mathrm{GeV}^{-2})$ &  \textbf{Coupling}  & \textbf{Value}~$(\mathrm{GeV}^{-2})$  \\
\hline
\hline
$\mathcal{C}^{uG}_{32}\times 10^7 $ & $(-0.27 \pm 3.89)$ &$\mathcal{C}^{uG}_{31}\times 10^6 $& $(-0.07 \pm 2.63)$ \\
\hline
$\mathcal{C}^{uB}_{32}\times 10^7 $ & $(0.07 \pm 0.76)$ &$\mathcal{C}^{uB}_{31}\times 10^6 $ & $(-0.16 \pm 0.62)$\\
\hline
$\mathcal{C}^{uW}_{32}\times 10^7 $ & $(-0.16 \pm 0.98)$ &$\mathcal{C}^{uW}_{31}\times 10^7 $& $(-0.16 \pm 0.64)$ \\
\hline
$\mathcal{C}^{u\phi }_{32}\times 10^6$ & $(-0.03 \pm 2.89)$ &$\mathcal{C}^{u\phi }_{31}\times 10^6$ & $(16.37\pm 58.08)$ \\
\hline
\end{tabular}
\caption{Fit result for the top FCNC SMEFT WCs in the real scenario.}
\label{tab:SMEFT_C3i} 
\end{table}

\bibliographystyle{JHEP}
\bibliography{biblio.bib}

\end{document}